\def\lae{\mathrel{<\kern-1.0em\lower0.9ex\hbox{$\sim$}}}
\def\gae{\mathrel{>\kern-1.0em\lower0.9ex\hbox{$\sim$}}}

\newcommand{\be}{\begin{equation}}
\newcommand{\ee}{\end{equation}}

\documentclass[apj]{emulateapj}
\usepackage{apjfonts,lscape}

\slugcomment{Accepted for Publication in ApJS}
\shortauthors{JORD\'AN ET AL.}
\shorttitle{GLOBULAR CLUSTER LUMINOSITY FUNCTIONS}

\begin{document}
 
\title{The ACS Virgo Cluster Survey. XII. \\
The Luminosity Function of Globular Clusters in Early Type Galaxies
\altaffilmark{1}}

\author{
Andr\'es Jord\'an\altaffilmark{2},
Dean E. McLaughlin\altaffilmark{3},
Patrick C\^ot\'e\altaffilmark{4},
Laura Ferrarese\altaffilmark{4},
Eric W. Peng\altaffilmark{4},
Simona Mei\altaffilmark{5},
Daniela Villegas\altaffilmark{2,6},
David Merritt\altaffilmark{7},
John L. Tonry\altaffilmark{8},
Michael J. West\altaffilmark{9,10}
}

\begin{abstract}
We analyze the luminosity function of the globular clusters (GCs) belonging to 
the early-type galaxies observed in the ACS Virgo Cluster Survey.  We have
obtained maximum likelihood estimates for a Gaussian representation of the 
globular cluster luminosity function (GCLF) for 89 galaxies. We have also
fit the luminosity functions with an ``evolved Schechter function'', 
which is meant to reflect the preferential depletion
of low-mass GCs, primarily by evaporation due to two-body
relaxation, from an initial Schechter
mass function similar to that of young massive clusters in local
starbursts and mergers. We  find a highly significant trend of the GCLF
dispersion $\sigma$ with galaxy luminosity, in the sense that the GC systems
in smaller galaxies have narrower luminosity functions. The GCLF dispersions
of our Galaxy
and M31 are quantitatively in keeping with this trend, and thus the
correlation between $\sigma$ and galaxy luminosity would seem more fundamental
than older notions that the GCLF dispersion depends on Hubble type. 
We show that this narrowing of the GCLF in a Gaussian description is driven
by a steepening of the cluster mass function above the classic turnover
mass, as one moves to lower-luminosity host galaxies. In a Schechter-function
description, this is 
reflected by a steady decrease in the value of the exponential cut-off mass
scale. We argue that this behavior at the high-mass end of the GC
mass function is most likely a consequence of systematic variations of the
initial cluster mass function rather than long-term dynamical evolution.
The GCLF turnover mass $M_{\rm TO}$ is roughly constant, at
$M_{\rm TO}\simeq(2.2\pm0.4)\times10^5\,M_\odot$ in bright galaxies, but it
decreases slightly (by $\sim\!35\%$ on average, with significant
scatter) in dwarf galaxies with $M_{B, {\rm gal}} \ga -18$. It could be
important to allow for this effect when using the GCLF as a distance
indicator. We show that part, though perhaps not all, of the variation could
arise from the shorter dynamical friction timescales in less massive
galaxies. We probe the variation of the GCLF to projected galactocentric radii
of 20--35~kpc in the Virgo  giants M49 and M87, finding that the turnover
point is essentially constant over these spatial scales.
Our fits of evolved Schechter functions imply average dynamical mass losses
($\Delta$) over a Hubble time that vary more than $M_{\rm TO}$, and
systematically but non-monotonically as a function of galaxy luminosity.
If the initial GC mass distributions rose steeply towards low masses as we
assume, then these losses fall in the range
$2\times10^5\,M_\odot\la \Delta < 10^6\,M_\odot$ per GC for all of our
galaxies. The trends in $\Delta$ are broadly consistent with observed, small
variations of the mean GC half-light radius in ACSVCS galaxies, and with
rough estimates of the expected scaling of average evaporation rates (galaxy
densities) versus total luminosity. We agree with previous suggestions
that if the full GCLF is to be understood in more detail, especially
alongside other properties of GC systems, the next generation of GCLF
models will have to include self-consistent treatments of dynamical
evolution inside time-dependent galaxy potentials.
\end{abstract}


\altaffiltext{1}{Based on observations with the NASA/ESA
{\it Hubble Space Telescope}
obtained at the Space Telescope Science Institute, which is operated
by the Association
of Universities for Research in Astronomy, Inc., under
NASA contract NAS 5-26555}
\altaffiltext{2}{European Southern Observatory,
Karl-Schwarzschild-Stra{\ss}e 2, 85748 Garching bei M\"unchen, Germany;
ajordan@eso.org}
\altaffiltext{3}{Department of Physics \& Astronomy, University
of Leicester, Leicester, LE1 7RH, UK; dean.mclaughlin@astro.le.ac.uk}
\altaffiltext{4}{Herzberg Institute of Astrophysics, Victoria, 
BC V9E 2E7, Canada}
\altaffiltext{5}{GEPI, Observatoire de Paris, Section de Meudon, 5 place Jules 
Janssen,  92195 Meudon Cedex, France}
\altaffiltext{6}{Departamento de Astronom\'{\i}a y Astrof\'{\i}sica, 
P. Universidad Cat\'olica de Chile, 
Avenida Vicu\~na Mackenna 4860, Casilla 306, Santiago 22, Chile}
\altaffiltext{7}{Department of Physics, Rochester Institute of Technology,
84 Lomb Memorial Drive, Rochester, NY 14623}
\altaffiltext{8}{Institute of Astronomy, University of Hawaii,
2680 Woodlawn Drive, Honolulu, HI 96822}
\altaffiltext{9}{Department of Physics \& Astronomy, University of Hawaii,
Hilo, HI 96720}
\altaffiltext{10}{Gemini Observatory, Casilla 603, La Serena, Chile}

\section{Introduction}
\label{sec:intro}
One of the remarkable features of the systems of globular clusters (GCs)
found around most galaxies is the shape of their luminosity function, or the
relative number of GCs with any given magnitude. Historically most important
has been the fact 
that these distributions always appear to peak, or turn over, at a GC absolute
magnitude around $M_{V,{\rm TO}} \approx -7.5$ (e.g., Harris 2001),
corresponding roughly to a mass of $M_{\rm TO}\sim 2\times10^5\,M_\odot$.
The near
universality of this magnitude/mass scale for GCs has motivated the widespread
use of the globular cluster luminosity function (GCLF) as a distance
indicator (see Harris 2001; also Ferrarese et al.~2000), and it has also posed
one of the longest-standing challenges to theories of GC formation and
evolution.

In recent years, some significant amount of attention has also been paid to
the way that GCs are distributed in {\it mass} around the peak of the
GCLF. Traditionally, the full GCLF has most often been modeled as a Gaussian
distribution in magnitude, corresponding to a lognormal distribution of GC
masses. However, if one focuses only on the distribution of GCs above the
point where the magnitude distribution turns over, it is found that the mass
function can usually be described by a power law (Harris \& Pudritz
1994), or perhaps a Schechter (1976) function (Burkert \& Smith 2000), which is
very similar to the mass distributions of giant molecular clouds and the young
massive star clusters forming in starbursts and galaxy mergers in the local
universe (e.g., Zhang \& Fall 1999). The main difference between
ancient GCs and the present-day sites of star-cluster formation is then
that the mass functions of the latter rise steeply upwards towards masses much
less than $M_{\rm TO}\sim\,10^5\,M_\odot$, far exceeding the observed
frequency of such low-mass GCs.

There are two main possibilities to explain this fundamental
difference. The first is that the conditions of star cluster formation in the
early universe when GCs were assembling may have favored the formation of
objects with masses in a fairly narrow range around
$\sim\! 10^5$--$10^6\,M_\odot$ (to the exclusion, in particular, of much
smaller masses). These conditions would no longer prevail in the environments
forming young clusters in the nearby universe. Some theoretical models along
these lines invoke the $\sim\! 10^6\,M_\odot$ Jeans mass at the epoch of
recombination (Peebles \& Dicke 1968), the detailed properties of
$\sim\! 10^6\,M_\odot$ cold clouds in a two-phase protogalactic medium (Fall
\& Rees 1985), and reionization-driven compression of the gas in
subgalactic ($\la\! 10^7\,M_\odot$) dark-matter halos (Cen 2001).

The second possibility is that GCs were in fact born with a wide
spectrum of masses, like that observed for young
star clusters, extending from $\! 10^6$--$10^7\,M_\odot$ down to
$\sim\! 10^3$--$10^4\,M_\odot$ or below. A subsequent transformation to the
characteristic mass function of GCs today could then be effected mainly by
dynamical processes (relaxation and tidal shocking) that are particularly
efficient at destroying low-mass clusters over the lifetime of a GC system
(e.g., Fall \& Rees 1977; Ostriker \& Gnedin 1997; Fall \& Zhang 2001).
Some observational evidence has been reported for such an evolution in the
mass functions of young and intermediate-age star clusters
(e.g., de Grijs, Bastian \& Lamers 2003, Goudfrooij et~al. 2004).

If we take the Occam's-razor view that indeed GCs formed through substantially
the same processes as star clusters today, then the picture offered by
observations of old GCLFs is unavoidably one of survivors. There has been
some debate as to whether it was in fact the long-term dynamical
mechanisms just mentioned that were mainly responsible for destroying large
numbers of low-mass globulars, or whether processes more related
to cluster formation strongly depleted many low-mass protoclusters
on shorter timescales (Fall \& Zhang 2001; Vesperini \& Zepf
2003). Even the most massive Galactic GCs have rather low binding
energies $E_b \la 10^{52}$~erg (McLaughlin 2000), so that if conditions were
not just right, very many protoglobular clusters could have been easily
destroyed in the earliest $\sim\! 10^7$~yr of their evolution, through the
catastrophic mass loss induced by massive-star winds and supernova explosions
(see, e.g., Kroupa \& Boily 2002; Fall, Chandar \& Whitmore 2005). Furthermore,
any clusters that survive this earliest mass-loss phase intact but with too low
a concentration could potentially still dissolve within a relatively short
time of $\sim\! 10^8$--$10^9$~yr (Chernoff \& Weinberg 1990). Homogeneous
observations of large samples of old GCLFs can help clarify the relative
importance of such early evolution versus longer-term dynamical mass loss in
the lives of star clusters generally.

The largest previous studies of GCLFs in early-type galaxies were performed
with archival HST/WFPC2 data. Kundu \& Whitmore (2001a, b) studied the
GCLF for 28 elliptical and 29 S0 galaxies. They concluded that the turnover
magnitude of the GCLF is an excellent distance indicator, and that the
difference in the turnover luminosity between the $V$ and $I$ bands increases
with the mean metallicity of the GCs essentially as expected 
if the GC systems in most galaxies have similar age and mass distributions.
Larsen et~al. (2001) studied the GCLF for 17 nearby early-type galaxies. 
They fitted Student's $t$ distributions separately to the subpopulations of
metal-rich and metal-poor GCs in each galaxy, and found
that any difference in the derived turnovers was consistent
with these subpopulations having similar mass and age distributions and the
same GCLF turnover {\it mass} scale. Larsen et al. also fitted
power laws to the mass distributions of GCs in the range
$M\simeq 10^5$--$10^6\,M_\odot$ and found they were well described by
power-law exponents similar to those that fit the mass
functions of young cluster systems.

In this paper, we study the GCLFs of 89 early-type galaxies observed by HST as
part of the ACS Virgo Cluster Survey (C\^ot\'e et al.~2004). This represents
the most comprehensive and homogeneous study of its kind to date. Some of the
results in this paper are also presented in a companion paper 
(Jord\'an et~al. 2006).
In the next section, we briefly describe our data and present our observed
GCLFs in a machine-readable table available
for download from the electronic edition of the {\it Astrophysical Journal}.
In \S\ref{sec:models} we discuss two different models that
we fit to the GCLFs, and in \S~\ref{sec:method} we 
describe our (maximum-likelihood) fitting methodology. Section
\ref{sec:results} presents the fits themselves, while \S\ref{sec:trends}
discusses a number of trends
for various GCLF parameters as a function of host galaxy luminosity and
touches briefly on the issue of GCLF variations within galaxies.
In \S\ref{sec:disc} we discuss some aspects of our results in the light of
ideas about GC formation and dynamical evolution, focusing in particular on
the relation between our data and a model of evaporation-dominated GCLF
evolution. In \S\ref{sec:conclusions} we conclude.

\section{Data}
\label{sec:obs}

A sample of 100 early-type galaxies in the Virgo cluster 
was observed for the ACS
Virgo Cluster Survey (ACSVCS; C\^ot\'e et~al. 2004). Each galaxy
was imaged in the F475W ($\simeq$ Sloan $g$) and F850LP
($\simeq$ Sloan $z$) bandpasses for a total of 
750 s and 1210 s respectively, and
reductions were performed as described in Jord\'an et~al. (2004a).
These data have been used previously to analyze the surface-brightness 
profiles of the galaxies and their nuclei 
(Ferrarese et~al. 2006ab, C\^ot\'e et~al. 2006), 
their surface brightness fluctuations (Mei et~al. 2005ab; 2007),
and the properties of their populations of star clusters, mainly
GCs (Jord\'an et~al. 2004b, Jord\'an et~al. 2005, Peng et~al. 2006a)
but also dwarf-globular transition objects (or UCDs, Ha\c{s}egan et~al. 2005)
and diffuse star clusters (Peng et~al. 2006b).

One of the main scientific objectives of the ACSVCS is the study of
the GC systems of the sample galaxies. We have developed
a procedure by which we select GC candidates from the totality of
observed sources around each galaxy, discarding the inevitable
foreground stars and background galaxies that are contaminants for our
purposes. This GC selection
uses a statistical clustering method, described in
detail in another paper in this series (Jord\'an et~al. 2007, in 
preparation), in which each source in the field of view of each galaxy
is assigned a probability $p_{\rm GC}$ that it is a GC.
Our samples of GC candidates are then constructed by selecting all sources
that have $p_{\rm GC} \ge 0.5$. The results of our classification 
method are illustrated in Figure~1 of Peng et~al. (2006a). For every GC
candidate we record the background surface brightness $I_b$ of the host
galaxy at the position of the candidate, and we measure $z$- and $g$-band
magnitudes and a half-light radius $R_h$ by fitting PSF-convolved
King (1966) models to the local light distribution of the cluster
(Jord\'an et~al. 2005).
Photometric zeropoints are taken from Sirianni et~al. 2005 (see also
Jord\'an et~al. 2004a), and  aperture corrections are applied as
described by Jord\'an et~al. (2007, in
preparation). 

Note that, as part of the ACSVCS we have measured the distances to
most of our target galaxies using the method of surface brightness
fluctuations (SBF; Tonry \& Schneider 1988). The reduction procedures for
SBF measurements, feasibility simulations for our observing configuration, 
and calibration have been presented in Mei et~al.\ (2005ab) and the
distance catalog is presented in another paper in this series (Mei et~al.
2007). We use these distances in this paper\footnote{We use the distances 
obtained using the polynomial calibration presented in Mei et al. 2007.} to
transform observed GC magnitudes into absolute ones on a per galaxy basis
whenever we wish to assess GCLF properties in physical (i.e., mass-based)
terms or need to compare the GCLFs of two or more galaxies. While some
galaxies have larger distances, the average distance modulus that we employ is
$(m-M)_0=31.09\pm0.03 \rm{(random)} \pm0.15 \rm{(systematic)}$, corresponding
to $D=16.5\pm0.1 \rm{(random)} \pm 1.1 \rm{(systematic)}$~Mpc.

\subsection{GCLF Histograms}
\label{ssec:gclfhists}

There are three main ingredients we need to construct a GCLF for any galaxy.
First, we have sets of magnitudes, in both the $z$ and $g$ bands,
for all GC candidates. As mentioned above, we generally isolate
GC candidates from a list of all detected objects by requiring that
$p_{\rm GC}\ge 0.5$. Note that here and throughout, we use $g$ as
shorthand to refer to the F475W filter, and $z$ denotes F850LP. Also, all
GC magnitudes in this paper have already been de-reddened (see \S~2.7
in Jord\'an et~al. 2004a for details).

Second, we have the (in)completeness functions in both bandpasses. Our
candidate GCs are marginally resolved with the ACS, and thus these
completeness functions depend not only on the GC apparent magnitude
$m$ and its position in its parent galaxy (through the local
background surface brightness $I_b$), but also on the GC projected
half-light radius $R_h$. Separate $z$- and $g$-band completeness functions
$f(m,R_h,I_b) \le 1$ have therefore been calculated from simulations in which
we first added simulated GCs with sizes $R_h=1,3,6,10$ pc and King (1966)
concentration parameter $c=1.5$ to actual images from the ACSVCS (making sure
to avoid sources already present), and then reduced the simulated images in an
identical fashion to the survey data. We next found the fraction of artificial
sources that were recovered, as a function of input magnitude and half-light
radius, in each of ten separate bins of background light intensity.
The final product is a three-dimensional look-up table on which we interpolate
to obtain $f$ for any arbitrary values of $(m,R_h,I_b)$.

Last, we have the expected density of contaminants as a function of magnitude
for each galaxy, obtained from analysis of archival ACS images (unassociated
with the Virgo Cluster Survey) of 17 blank, high-latitude control fields,
each observed with both $g$ and $z$ filters to depths greater than in the
ACSVCS. We ``customized'' these data to our survey galaxies
by performing object detections on every control field as if it contained each
galaxy in turn. This procedure is described in more detail in
Peng et al.~(2006a, their \S2.2). The net result is 17 separate estimates of
the number of foreground and background objects, as a function of $g$ and $z$
magnitude, expected to contaminate the list of candidate GCs in every ACSVCS
field.

Of the 100 galaxies in the ACSVCS, we restrict our analysis to those that have
more than 5 probable GCs, as estimated by subtracting the total number
of expected contaminants from the full list of GC candidates for
each galaxy. We additionally eliminated two galaxies
for which we could not usefully constrain the GCLF parameters. 
This results in a final sample of 89 galaxies. The
GCLF data for these are presented in Table \ref{tab:gclf_hists}.

\placetable{\ref{tab:gclf_hists}}

The first column of Table \ref{tab:gclf_hists} is the galaxy ID in the
Virgo Cluster Catalogue (VCC: Binggeli, Sandage \& Tammann 1985; see Table~1
in C\^ot\'e et~al. 2004 for
NGC and Messier equivalents). Column (2) contains an apparent $z$-band
magnitude 
defining the midpoint of a bin with width $h_z$ given in column (3). This
binwidth was chosen to be 0.4 for all galaxies.
Columns (4)--(6) of the table then give the total number
$N_{z,{\rm tot}}$ of observed sources in this bin; the number
$N_{z,{\rm cont}}$ of contaminants in the bin as estimated from the average of
our 17 control fields; and the average completeness fraction $f_z$ in the
bin---all applying to the candidate-GC sample defined on the basis of our GC
probability threshold, $p_{\rm GC}\ge 0.5$.  Columns (7)--(11) repeat
this information for the galaxy's GC candidates identified
in the $g$ band. Columns (12)--(21) are the corresponding $z$- and
$g$-band data for an alternate GC sample defined strictly by magnitude
cuts and an upper limit of $R_h<0\farcs064\simeq 5$~pc (which will include the
large majority of real GCs; Jord\'an et al.~2005), rather than by relying
on our $p_{\rm GC}$ probabilities. This provides a way of checking that
selecting GC candidates by $p_{\rm GC}$ does not introduce any subtle biases
into the GCLFs (see also \S\ref{sec:method} below).

The data in Table \ref{tab:gclf_hists} can be converted to distributions of
absolute GC magnitude by applying the individual galaxy distances
given in Mei et al.~(2007). If they are used to fit
model GCLFs, it should be by comparing the observed $N_{\rm tot}$ against a
predicted $(f \times N_{\rm model} + N_{\rm cont})$ as a function of
magnitude. This is essentially what we will do here, although we employ
maximum-likelihood techniques rather than using the binned data.
However, before describing our model-fitting methodology, we pause first
to discuss in some detail the models
themselves. We work with two different distributions in this paper: one
completely standard, and one that is meant to
elucidate the connections between observed GC mass distributions and plausible
initial conditions and dynamical evolution histories.

\section{Two GCLF Models}
\label{sec:models}

The term ``globular cluster luminosity
function''  is customarily used to refer to a directly
observed histogram of the number of GCs per unit magnitude. We
follow this standard useage here, and in addition whenever
we refer simply to a ``luminosity function,'' we in fact mean the GCLF, i.e.,
the distribution of magnitudes again. We denote the magnitude in any
arbitrary bandpass by a lower-case $m$, and thus the GCLF is essentially the
probability distribution function $dN/dm$. It is {\it not}
equivalent to the distribution of true GC luminosities, since of
course $m\equiv C-2.5\,\log\,L$ for some constant $C$,
so $dN/dL = (dN/dm)|\partial m/\partial L|$ has a functional
form different from that of $dN/dm$. 

In this paper, when we speak of GC masses, we denote them by an
upper-case $M$ and we almost always make the assumption that they are
related by a multiplicative constant to GC luminosities, such that
$m=C^{\prime} - 2.5\,\log\,M$, with $C^{\prime}$ another constant including
the logarithm of a mass-to-light ratio (generally taken to be the
same for all GCs in any one system, as is the case in the Milky Way;
McLaughlin 2000). We refer to the number of GCs per
unit mass, $dN/dM$, as a ``mass function'' or a ``mass distribution.'' In
the literature, it is sometimes also called a ``mass spectrum.'' Its
relation to the GCLF is
\begin{equation}
\frac{dN}{dM} \propto \frac{1}{M} \frac{dN}{d\,\log\,M} 
              \propto 10^{0.4 m} \, \frac{dN}{dm} \ .
\label{eq:defs}
\end{equation}

As we have already mentioned, most observed GCLFs show a ``peak'' or
``turnover'' at a cluster magnitude that is generally rather similar from
galaxy to galaxy. One important consequence of equation (\ref{eq:defs}) is
that any such feature in the GCLF does {\it not} correspond to a local maximum
in the GC mass 
distribution: if the first derivative of $dN/dm$ with respect to $m$ vanishes
at some magnitude $m_{\rm TO}$, then the derivative of $dN/dM$ with respect to
$M$ at the corresponding mass scale $M_{\rm TO}$ is strictly negative, i.e.,
the mass function still rises towards GC masses below the point where the GCLF
turns over.  (More specifically, the logarithmic slope of $dN/dM$ at the
GCLF turnover point $M_{\rm TO}$ is always exactly $-1$; see McLaughlin 1994
for further discussion.) 

\subsection{The Standard Model}
\label{ssec:gaussmod}

The function most commonly taken to describe
GCLFs is a Gaussian, which is  the easiest way to
represent the peaked appearance of most luminosity functions in terms of
number of clusters per unit magnitude. It is thus our first choice
to fit to each of the observed GCLFs in this paper. Denoting the mean
GC magnitude $\mu\equiv\langle m \rangle$ and the dispersion
$\sigma_m = \langle (m-\mu)^2 \rangle^{1/2}$, we have the usual
\begin{equation}
\frac{dN}{dm} = \frac{1}{\sqrt{2\pi}\,\sigma_m}
                       \exp\left[-\frac{(m-\mu)^2}{2\sigma_m^2}\right]\ .
\label{eq:gauss}
\end{equation}
In terms of GC masses, $M$, this standard distribution corresponds to a
mass function $dN/dM = (dN/dm)|\partial m/\partial M|$ or, since
$m={\rm constant} - 2.5\,\log\,M$ (assuming a single mass-to-light ratio for
all clusters in a sample),
\begin{equation}
\frac{dN}{dM} =\frac{1}{(\ln 10) M} 
\frac{1}{\sqrt{2\pi}\,\sigma_M} 
\exp\left[-\frac{(\log\,M-\langle\log\,M\rangle)^2}{2\sigma_M^2}\right]\ ,
\label{eq:lognorm}
\end{equation}
with $\sigma_M\equiv\sigma_m/2.5$.

As will be evident in what follows, the GCLFs in a large
sample such as ours show a variety of detail that is unlikely to be conveyed
in full by a few-parameter family of distributions. But it is also clear
that a Gaussian captures some of the most basic information
we are interested in investigating---the mean and the standard deviation
of the GC magnitudes in a galaxy---with a minimal number of
parameters. It is also the historical function of choice for GCLF fitting, and
in many cases the fit is indeed remarkably good.

Nevertheless, the Gaussian does have some practical limitations. Secker (1992)
showed that the tails of the GCLF in the Milky Way and M31 are heavier than
a Gaussian allows, and he argued that a Student's $t$ distribution
(with shape parameter $\nu\simeq 5$) gives a better match to the data.
More importantly, the observed GCLFs in our Galaxy and in M31 are {\it
asymmetric} about their peak magnitude, a fact which has been emphasized most
recently by Fall \& Zhang (2001). This was implicit in the work of McLaughlin
(1994), who advocated using piecewise power laws to fit the number of GCs per
unit linear luminosity---or piecewise exponentials to describe the usual
number of GCs per unit magnitude. Baum et al.~(1997) used an asymmetric
hyperbolic function to fit the strong peak and asymmetry in the combined
Galactic and M31 GCLF.

However, all of these alternative fitting functions still share another
shortcoming of the Gaussian, which is that there is no theoretical
underpinning to it. Moreover, with the exception of the power laws in
McLaughlin (1994), there is no obvious connection with the mass distributions
of the young massive clusters that form in mergers and starbursts in the local
universe. We therefore introduce an alternative fitting function---based
on existing, more detailed studies of initial cluster mass functions and their
long-term dynamical evolution---to address these issues.

\subsection{An Evolved Schechter Function}
\label{ssec:esmod}

\subsubsection{Initial GC Mass Function}
\label{sssec:esinit}

Observations of young star clusters indicate that the number of clusters per
unit mass is well described by a power law---$dN/dM \propto M^{-\beta}$
with $\beta \approx 2$---or alternatively by
a Schechter (1976) function with an index of about 2 in its power-law part and
an exponential cut-off above some large mass scale that might vary from
galaxy to galaxy (e.g., Gieles et al.~2006a).
Perhaps the best-observed mass distribution for a young cluster system is that
in the Antennae galaxies, NGC 4038/4039 (Zhang \& Fall 1999). In this specific
case, a pure power-law form suffices to describe the cluster $dN/dM$ as it is
currently known; but a Schechter function with an appropriately high cut-off
mass also fits perfectly well. Thus, assuming
$dN/dM \propto M^{-\beta}\,\exp(-M/M_c)$, we find from the data plotted by
Zhang \& Fall (1999) that $\beta=2.00\pm0.04$ and
$\log\,(M_c/M_\odot)=6.3^{+0.7}_{-0.3}$ for their sample of clusters with ages
2.5--6.3 Myr; and $\beta=1.92\pm0.14$ and $\log\,(M/M_c)=5.9^{+0.45}_{-0.25}$
for ages 25--160 Myr.

The mass functions
of old globular clusters in the Milky Way and M31 can also be described by
power laws with $\beta\simeq 2$ for clusters more massive than the GCLF peak
(McLaughlin 1994; McLaughlin \& Pudritz 1996; Elmegreen \& Efremov 1997).
And the GC mass distributions in large ellipticals follow power laws over
restricted high-mass 
ranges, although here the slopes are somewhat shallower (McLaughlin
1994; Harris \& Pudritz 1994) and there is clear evidence of curvature in
$dN/dM$ (McLaughlin \& Pudritz 1996) that is better described by the
exponential cut-off at very high cluster masses in a Schechter function
(e.g., Burkert \& Smith 2000). Theoretical models for GC formation,
which aim expressly to explain these high-mass features of GCLFs and relate
them to the distributions of younger clusters and molecular clouds, have been
developed by McLaughlin \& Pudritz (1996) and Elmegreen \& Efremov (1997).

The important difference between the mass functions of old GCs and
young massive clusters, then, is not the power-law or Schechter-function form
of the latter {\it per se}; it is the fact that the frequency of young
clusters continues to rise toward the low-mass limits of observations, while
the numbers of GCs fall well below any extrapolated power-law behavior for
$M\la 2\times10^5\,M_\odot$, i.e., for clusters fainter than the classic peak
magnitude of the GCLF. We therefore assume a Schechter function,
\begin{equation}
\frac{dN}{dM_0} \propto M_0^{-2}\,\exp(-M_0/M_c) \ ,
\label{eq:esinit}
\end{equation}
as a description of the initial mass distribution of globular clusters
generally. We emphasize again that the fixed power law 
of $M_0^{-2}$ at low masses is chosen for compatibility with current data on
systems of young massive clusters. The variable cut-off mass $M_c$ is required 
to match the well observed curvature present at $M\ga 10^6\,M_\odot$
in the mass distributions of old
GC systems in large galaxies. This feature is certainly allowed by the young
cluster data, even if it may not be explicitly required by them.

A strong possibility to explain the difference between such
an initial distribution and the present-day $dN/dM$ is the preferential
destruction of low-mass globular clusters by a variety of dynamical processes
acting on Gyr timescales (see Fall \& Zhang 2001; Vesperini 2000, 2001;
and references therein).

\subsubsection{Evolution of the Mass Function}
\label{sssec:esevolve}

Fall \& Zhang (2001) give a particularly clear recent overview of the
dynamical processes that act to destroy globular clusters on Gyr timescales as
they orbit in a fixed galactic potential. The main destruction mechanisms
are dynamical friction; shock-heating caused by passages through
galaxy bulges and/or disks; and evaporation as a result of two-body
relaxation. Only the latter two are important to the development
of the low-mass end of the GCLF, since dynamical-friction timescales grow
rapidly towards low $M$, as $\tau_{\rm df}\propto M^{-1}$.
(Cluster disruption due to stellar-evolution mass loss does not change
the shape of the GC mass function if the stellar IMF is universal, unless
a primordial correlation between cluster concentration and mass
is invoked; cf.~Fall \& Zhang 2001 and Vesperini \& Zepf 2003).

Tidal shocks drive mass loss on timescales
$\tau_{\rm sh}\propto \rho_h P_{\rm cr}$, where 
$\rho_h\propto M/R_h^3$ is the mean density of a cluster inside its half-mass
radius, and $P_{\rm cr}$ is the typical time between disk or bulge crossings.
Evaporation scales rather differently, roughly as
$\tau_{\rm ev}\propto M/\rho_h^{1/2}$. A completely general assessment
of the relative importance of the two processes can therefore be
complicated. However, tidal shocks are 
rapidly self-limiting in most realistic situations (Gnedin, Lee,
\& Ostriker 1999): clusters with high enough 
$\rho_h$ and on orbits that expose them only to ``slow'' and well-separated 
shocks (i.e., with both the duration of individual shocks and the interval
$P_{\rm cr}$ longer than an internal dynamical time, $t_{\rm dyn} \propto
\rho_h^{-1/2}$) will experience an early, sharp
increase in $\rho_h$ in response to the first few shocks. Thereafter
$\tau_{\rm ev}\ll \tau_{\rm sh}$, and in the long term shock-heating presents
a second-order correction to the dominant mass loss caused by
evaporation. Most GCs today, at least in our Galaxy, appear to be in this
evaporation-driven evolutionary phase (Gnedin et al.~1999; see also
Figure 1 of Fall \& Zhang 2001, and Prieto \& Gnedin 2006).

Fall \& Zhang (2001) therefore develop a model for the evolution of
the Galactic GCLF that depends largely on evaporation to erode
an initially steep $dN/dM_0$ (in fact, they adopt a Schechter function,
as in eq.~[\ref{eq:esinit}], for one of their fiducial cases).
They assume---as is fairly standard; e.g., see Vesperini
(2000, 2001)---that any cluster roughly conserves its mean half-mass
density $\rho_h$ as it loses mass, at least after any rapid initial
adjustments due to stellar mass loss or the first tidal shocks, and
when the evolution is dominated by evaporation. The mass-loss
rate\footnotemark 
\footnotetext{Note that we use $\mu_{\rm ev}$ to denote the evaporation
mass-loss rate in equation (\ref{eq:muev}) in order to be consistent with the
notation of Fall \& Zhang (2001). This should not be confused
with our use of the symbol $\mu$---with different subscripts---to represent
the mean magnitude in a Gaussian description of the GCLF.}
is then
\begin{equation}
\mu_{\rm ev} \equiv - dM/dt
    \propto M/t_{\rm rh} \propto (M/R_h^3)^{1/2} \sim {\rm constant} \ ,
\label{eq:muev}
\end{equation}
where
$t_{\rm rh}\sim \rho^{-1} \langle v^2\rangle^{3/2}
   \propto M^{1/2} R_h^{3/2}$
is the relaxation time at the half-mass radius $R_h$.
Under this assumption, the mass of a GC at any age $t$ is just
$M(t)=M_0 - \mu_{\rm ev} t$. For any collection of clusters with the same
density (for example, those on similar orbits, if $\rho_h$ is set by tides at
a well defined perigalacticon), $\mu_{\rm ev}$ is independent of cluster mass
as well as time, and if the GCs are coeval in addition, then
$\mu_{\rm ev}t$ is a 
strict constant. The mass function of such a cluster ensemble with any age $t$
is then related to the initial one as in equation (11) of Fall \& Zhang
(2001): 
\begin{equation}
\frac{dN}{dM(t)}
   = \frac{dN}{dM_0}\left|\frac{\partial M_0}{\partial M}\right|
   = \frac{dN}{dM_0} \ .
\label{eq:dndmtrans}
\end{equation}
Thus, simply making the substitution $M_0 = M+\mu_{\rm ev}t$ in the
functional form of the original GC mass function gives the evolved
distribution. An initially steep $dN/dM_0$ therefore always evolves to a flat
mass function, $dN/dM(t) \sim {\rm constant}$, at sufficiently low masses $M\ll
\mu_{\rm ev}t$. As Fall \& Zhang show for the Milky Way, and as we shall
also see for the early-type ACSVCS galaxies, this gives a  good fit
to observed GCLFs if the cumulative mass-loss term 
$\mu_{\rm ev}t > 10^5\,M_\odot$ by $t\simeq 13$~Gyr.

The key physical element of this argument, as far as the GCLF is concerned, is
the linear decrease of cluster mass with time. While the quickest way to
arrive at such a conclusion is to follow the logic of Fall \& 
Zhang, as just outlined, there are some caveats to be kept in mind.

Tidal shocks can be much more important than evaporation
for some globulars. In particular, clusters with low densities
and low concentrations (such that shocks
significantly disturb the cores as well as the halos), and/or those on
eccentric orbits with very short 
intervals between successive bulge or disk crossings, may never recover fully
from even one shock. Rather than re-adjusting quickly to a situation in which 
$\tau_{\rm ev}\ll \tau_{\rm sh}$, such clusters may be kept out of dynamical
equilibrium for most of their lives, significantly overflowing their nominal
tidal radii. Their entire evolution could then be
strongly shock-driven. This appears to be the case, for example, with the well
known Galactic globular, Palomar 5; see Dehnen et al.~(2004).
The extremely low mass and concentration of Palomar 5 make it
highly unusual in comparison to the vast majority of known GCs in any
galaxy currently, but many more clusters like its progenitor may well have
existed in the past. This then raises the question of whether considering
evaporation-dominated evolution alone gives a complete view of the dynamical
re-shaping of the GC mass function. Here, however, it is important that
the $N$-body simulations of Dehnen et al.~(2004) show that the late-time
evolution of even the most strongly shock-dominated clusters is still
characterized by a closely linear decrease of mass with time: rather than
$\rho_h$ being conserved in this case, the half-light radius $R_h$ is nearly
constant in time, and for a given orbit the mass-loss rate is
$M/\tau_{\rm sh}\propto M/\rho_h\propto R_h^3$. While the physical reasoning
changes, the end result for the GCLF of clusters in this physical regime
is the same as equation (\ref{eq:dndmtrans}).\footnotemark 
\footnotetext{Another process that may fall in this regime is impulsive
shocking due to encounters between GCs and massive concentrated objects like
giant molecular clouds; see Lamers \& Gieles (2006) for a recent
discussion of this. However, this is presumably most relevant to clusters
orbiting in disks, where the shocks can occur in fairly rapid
succession. It is not important for the large majority of GCs in galaxy
halos.}

The importance of Fall \& Zhang's (2001) assumption of a
constant $\rho_h$ for evaporation-driven cluster evolution is its
implication that the mass-loss rate is constant in time. This has some
direct support from N-body simulations (e.g., 
Vesperini \& Heggie 1997; Baumgardt \& Makino 2003). (Note the distinction in
Baumgardt \& Makino between the total cluster mass loss and that due only to
evaporation; see their Figure 6 and related discussion.) But more than this,
if evaporation is to be primarily responsible for the
strong depletion of a GC mass function at scales $M<\mu_{\rm ev}t$, then after
a Hubble time $\mu_{\rm ev}t$ must be roughly of order the current GCLF
turnover mass $M_{\rm TO}\sim 2\times 10^5\,M_\odot$. Since
$\mu_{\rm ev}\propto \rho_h^{1/2}$, this argument ultimately constrains the
average density required of the globulars, which can of course be
checked against data. In addition,
satisfying observational limits on the (small) variation of GCLFs
between different subsets of GCs in any one system---for example, as a
function of galactocentric position---puts constraints on the allowed
distribution of initial (and final) GC densities.

Fall \& Zhang calculate in detail the evolution of the Milky Way
GCLF over a Hubble time under the combined influence of stellar evolution
(which, as mentioned above, does not change the {\it shape} of $dN/dM$
except in special circumstances), evaporation, and tidal shocks
(which, again, contribute second-order corrections to the results of
evaporation in their treatment). They relate $\mu_{\rm ev}\propto
\rho_h^{1/2}$ to GC orbits, by assuming that $\rho_h$ is set by tides
at the pericenter of a cluster orbit in a logarithmic
potential with a circular speed of 220 km~s$^{-1}$. They then find the
GC orbital distribution that allows both for an average cluster
density high enough to give a good fit to the GCLF of the Milky Way as
a whole, and for a narrow enough spread in $\rho_h$ to reproduce the
observed weak variation in $M_{\rm TO}$ with Galactocentric radius
(e.g., Harris 2001).

Ultimately, the GC distribution function found by Fall \& Zhang in this
way is too strongly biased towards radial orbits with small pericenters
to be compatible with the observed kinematics and $\rho_h$ 
distributions of globulars in the Milky Way and other galaxies---as
both they and others (e.g., Vesperini et al.~2003) have pointed
out. However, Fall \& Zhang also note that the difficulties at this level
of detail do not necessarily disprove the basic idea that long-term
dynamical evolution is primarily responsible for the present-day shape
of the GCLF at low masses. The problem may lie instead in the specific
relation adopted to link the densities, and thus the disruption rates,
of GCs to their orbital pericenters. In particular, Fall \&
Zhang---along with almost all other studies along these lines---assume
a spherical and time-independent Galactic potential. Both assumptions
obviously break down in a realistic, hierarchical cosmology. Once
time-variable galaxy potentials are taken properly into account in
more sophisticated simulations, it could still be found that cluster
disruption on Gyr timescales can both explain the low-mass side of GC
mass functions and be consistent with related data on the
present-day cluster orbital properties, $\rho_h$ distributions, and so
on. Recent work in this vein by Prieto \& Gnedin (2006) appears
promising, though it is not yet decisive. 

We will return to these issues in \S\ref{ssec:evaporation}.
First, however, we describe an analytical form for $dN/dM$, which
combines the main idea in Fall \& Zhang (2001)---that evaporation
causes cluster masses to decrease linearly with time---with a
plausible, Schechter-function form for the initial $dN/dM_0$. We fit the
evolved function to the GCLF of the Milky Way, to show that it 
provides a good approximation to the fuller, numerical models of Fall
\& Zhang; and then we fit it to our ACSVCS data, to produce new
empirical constraints for detailed modeling of the formation and
evolution of GC mass functions under conditions not specific only to
our Galaxy.

\subsubsection{Fitting Functions for $dN/dM$ and the GCLF}
\label{sssec:eschar}

To summarize the discussion above,
we assume that the mass-loss rate of any globular cluster is constant
in time. Following Fall \& Zhang (2001), we expect that this will occur
naturally if the disruption process most relevant to the GCLF in
the long term is evaporation, which plausibly conserves the average
densities $\rho_h$ of individual clusters inside their half-mass radii.
Thus, we continue to denote the
mass-loss rate by $\mu_{\rm ev}$. However, it should be recognized
that tidal shocks can contribute second-order corrections 
to $\mu_{\rm ev}$ and may even, in some extreme cases,
dominate evaporation (though the net result arguably could still be
a constant total $dM/dt$).

For any set of clusters with similar
ages $t$ and similar $\rho_h$ (and on similar
orbits, if these significantly affect $\rho_h$ or add
tidal-shock contributions to $\mu_{\rm ev}$), the cumulative mass loss
$\Delta\equiv \mu_{\rm ev} t$ is a 
constant, so that each cluster has $M(t)=M_0 - \Delta$. 
Combining equation (\ref{eq:esinit}) for the initial mass distribution
with equation (\ref{eq:dndmtrans}) for its evolution then
yields an ``evolved Schechter function''
\begin{equation}
\frac{dN}{dM}
       \propto \frac{1}{(M+\Delta)^2}\,
       \exp\left(-\frac{M+\Delta}{M_c}\right) \ ,
\label{eq:esmass}
\end{equation}
with $M_c$ allowed to vary between galaxies.
Once again, $\Delta$ in this expression may vary between different
sets of GCs, with different densities or orbits, in the same
galaxy. The detailed modeling of Fall \& Zhang (2001) takes this
explicitly into account. But in what follows, we
fit equation (\ref{eq:esmass}) to GC data taken from large
areas over galaxies, which effectively returns an estimate of the
average mass loss per cluster over a Hubble time.
Since $\mu_{\rm ev}\propto \rho_h^{1/2}$ when
evaporation dominates shocks, this implicit averaging is essentially done
over the distribution of GC mean half-mass densities.

To relate this evolved mass function to the standard
observational definition of a GCLF---the number of GCs per unit
magnitude---we write 
$m\equiv C-2.5\,\log\,M$, $\delta\equiv C-2.5\,\log\,\Delta$, and $m_c\equiv
C-2.5\,\log\,M_c$, where $C$ is related to the solar absolute magnitude and
the typical cluster mass-to-light ratio in an appropriate bandpass. The model
then reads
\begin{equation}
\frac{dN}{dm} \propto
\frac{10^{-0.4(m-m_c)}}
     {\left[10^{-0.4(m-m_c)}+10^{-0.4(\delta-m_c)}\right]^{2}} \,
                   \exp\left[- 10^{-0.4(m-m_c)}\right] \ .
\label{eq:esmag}
\end{equation}
In both of equations (\ref{eq:esmass}) and (\ref{eq:esmag}), the
constants of proportionality required to normalize the distributions
can be evaluated numerically.

\begin{figure}
\epsscale{1.4}
\plotone{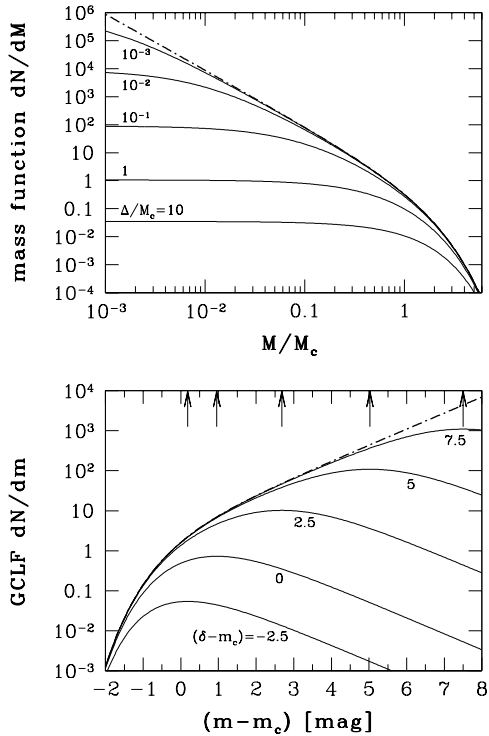}
\caption[]{{\it Top:} Evolved Schechter mass functions $dN/dM$
(eq.~[\ref{eq:esmass}]), for various values of the ratio $\Delta/M_c$,
which fixes the shape of the distribution. Curves are arbitrarily normalized.
The uppermost, broken curve corresponds to $\Delta=0$, i.e., a regular
Schechter (1976) function with a power-law exponent of $-2$. 
For non-zero $\Delta$, $dN/dM$ is flat at low masses.
{\it Bottom:} GCLFs $dN/dm$ corresponding to the mass functions in the
upper panel (see eq.~[\ref{eq:esmag}]). Curves are again arbitrarily
normalized, and the parameter controlling the shape is the magnitude
difference $(\delta-m_c)=-2.5\,\log(\Delta/M_c)$. For any finite
$(\delta-m_c)$, the GCLF peaks and turns over at the magnitude
$m_{\rm TO}$ given by equation (\ref{eq:es_magto}) (corresponding to the mass
in eq.~[\ref{eq:es_massto}]), and the faint side of the GCLF always
approaches the limiting shape $dN/dm \propto 10^{-0.4 m}$. Arrows mark
the turnover points of the models shown here. In the limit
$(\delta-m_c)\rightarrow +\infty$ (i.e. $\Delta\ll M_c$), we have that
$m_{\rm TO}\rightarrow \delta$, while in the limit
$(\delta-m_c)\rightarrow -\infty$ (large $\Delta\gg M_c$),
the turnover $m_{\rm TO}\rightarrow m_c$.
For $\Delta/M_c \ga 10$ or $(\delta-m_c)\la -2.5$, the GCLF has an essentially
fixed shape.
}
\label{fig:schematic}
\end{figure}

Figure \ref{fig:schematic} illustrates the form of the evolved
Schecter function, in terms of both the mass distribution $dN/dM$ and
the GCLF $dN/dm$. (Note that mass $M$ increases to the right along the
$x$-axis in the upper panel, but---as usual---larger $M$ corresponds to
brighter magnitudes $m$, at the left of the axis in the lower panel.)
From the equations above, it is clear that the mass
$M_c$ or the magnitude $m_c$ sets the scale of the function, while the
ratio $\Delta/M_c$ or the magnitude difference $(\delta-m_c)$ controls
its overall shape. For very small $\Delta\ll M_c$
(faint $\delta\gg m_c$), the function approaches an
unmodified Schechter (1976) function. This is drawn in Figure
\ref{fig:schematic} as the bold, broken curves that rise unabated toward
low cluster masses or faint magnitudes. The magnitude $m_{\rm TO}$ at
which the GCLF peaks in general can be found by setting to zero the
derivative of equation (\ref{eq:esmag}) with respect to $m$. This
yields
\begin{equation}
10^{-0.8 (m_{\rm TO}-m_c)}
  + 10^{-0.4 (m_{\rm TO}-m_c)}\left[1 + 10^{-0.4 (\delta-m_c)}\right]
  - 10^{-0.4 (\delta-m_c)}  =  0  \ ,
\label{eq:es_magto}
\end{equation}
the solution to which corresponds to a mass of
\begin{equation}
M_{\rm TO} =
     \frac{-(M_c + \Delta) +
                    \sqrt{(M_c + \Delta)^2 + 4\Delta M_c}}{2} \ .
\label{eq:es_massto}
\end{equation}

\placefigure{fig:schematic}

From either of equations (\ref{eq:es_magto}) or (\ref{eq:es_massto}),
or from the sequence of curves in Figure \ref{fig:schematic}, it can
be seen that when $\Delta\ll M_c$, the GCLF peaks at a 
magnitude $m_{\rm TO}\simeq \delta$, i.e., the turnover reasonably
approximates the average cluster mass loss in the model (although
$m_{\rm TO}$ is formally always fainter than $\delta$).
As the ratio $\Delta/M_c$ increases, the GCLF turnover initially tracks
$\Delta$ but eventually approaches an upper limit set by the exponential
cut-off scale in the mass function: $m_{\rm TO} \rightarrow m_c$ as
$(\delta-m_c) \rightarrow -\infty$ ($\Delta\gg M_c$).

For any fixed value of $\Delta/M_c$,
Figure \ref{fig:schematic} shows that in the limit of low masses,
$M \ll \Delta$, the mass function in equation (\ref{eq:esmass}) is
essentially flat. As Fall \& Zhang (2001) first pointed out, this is a
direct consequence of the assumption of a mass-loss rate
that is constant in time. It follows generically
from equation (\ref{eq:dndmtrans}) above, independently of the 
specific initial GC mass function.
At the other extreme, for very  high masses $M\gg \Delta$ the evolved $dN/dM$
just approaches the assumed underlying initial function with $\Delta=0$. In
terms of the GCLF, this means that $dN/dm$ tends (always) to an
exponential, $dN/dm \propto 10^{-0.4\, m}$, at magnitudes much fainter
than the turnover; and (for initial Schechter function assumed here) to the
steeper $dN/dm \propto 10^{0.4\, m} \exp\,[-10^{-0.4\, (m-m_c)}]$ for very
bright magnitudes. The faint half of the GCLF in this model is therefore
significantly broader than the bright half.

Finally, it is worth considering the widths of the GCLFs in the lower
panel of Figure \ref{fig:schematic} in more detail. For $\Delta=0$,
the full width at half-maximum (FWHM) of $dN/dm$ is undefined, since
there is no turnover. As the ratio $\Delta/M_c$ increases and a
well-defined peak appears in the GCLF, the distribution clearly
becomes narrower and narrower. As we have already discussed, even
though formally $\Delta/M_c$ can increase without limit, the turnover
magnitude ultimately has a maximum brightness
$m_{\rm TO}\rightarrow m_c$. Similarly, the FWHM of the GCLF approaches a firm
{\it lower limit} of  ${\rm FWHM}\simeq 2.66$~mag. This includes
a limiting half width at half-maximum of
${\rm HWHM}\simeq 1.59$~mag on the faint side of the GCLF, and a
smaller ${\rm HWHM}\simeq 1.07$~mag on the bright side. All of these
numbers can be obtained from analysis of equation (\ref{eq:esmag}) by
letting $(\delta-m_c)\rightarrow -\infty$, i.e.,
$\Delta/M_c\rightarrow + \infty$.
In this limit, the GCLF approaches a fixed shape and is free
only to shift left or right depending on the value of
$m_c\simeq m_{\rm TO}$. This limiting shape is already
essentially achieved with $\Delta/M_c=10$ or $(\delta-m_c)=-2.5$, which
is plotted in Figure \ref{fig:schematic} (even though the
turnover is still about 0.18 mag fainter than $m_c$ in this case).

As we will see in \S\ref{ssec:esfits} and \S\ref{sssec:masstrend}, the GCLFs
observed in the ACSVCS are all best fit with $\Delta/M_c\ga 0.1$, or
$(\delta-m_c)\la 2.5$ mag. This is the case also in the Milky Way.

\subsection{Comparison with the Milky Way GCLF}
\label{ssec:modcomp}

\begin{figure}
\epsscale{1.40}
\plotone{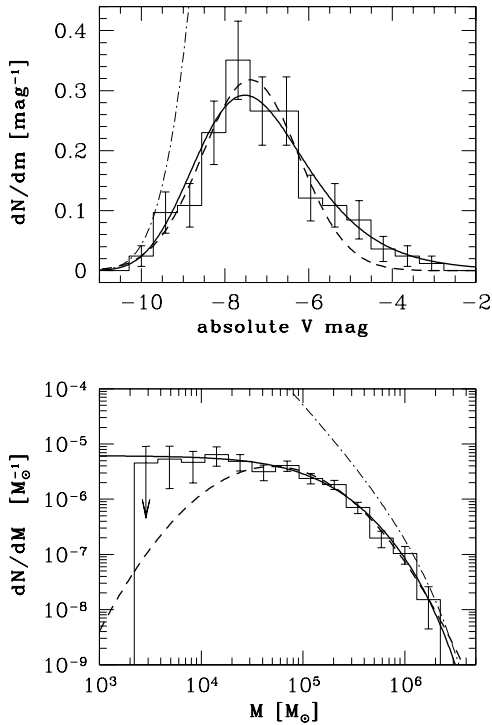}
\caption[]{{\it Top:} Fits of a Gaussian (dashed curve) and 
an evolved Schechter function (solid curve) to
the Milky Way GCLF, expressed as the (normalized) number of clusters per unit
of absolute $V$ magnitude. The dot-dashed curve is a Schechter function with
the same value for $M_c$ as the solid curve but with mass-loss parameter
$\Delta$ set to zero. 
{\it Bottom:} Corresponding observed GC mass
function $dN/dM$, and model fits, derived from the GCLF assuming a $V$-band
mass-to-light ratio of $2\ M_\odot\ L_{V,\odot}^{-1}$ for all clusters
(McLaughlin \& van der Marel 2005).
}
\label{fig:MWGCLF}
\end{figure}

Figure \ref{fig:MWGCLF} plots the GCLF and the
corresponding GC mass function in the Milky Way.
The upper
panel of this figure shows the GCLF $dN/dm$, in terms of clusters per unit
absolute $V$ magnitude, for 143 GCs in the online catalogue of
Harris (1996).\footnotemark
\footnotetext{http://physwww.mcmaster.ca/$\sim$harris/mwgc.dat}
(Note again that cluster luminosity and mass increase to the left in this
standard magnitude distribution.)
The bold, dashed line is the usual Gaussian representation
(eq.~[\ref{eq:gauss}]) with parameters given by Harris (2001):
\begin{equation}
\mu_V = -7.4 \pm 0.1\ {\rm mag} \ ; \quad  \sigma_V = 1.15 \pm 0.10\ {\rm mag}
\ .
\label{eq:MWgmag}
\end{equation}
The bold solid curve is our fit of the evolved Schechter function in
equation (\ref{eq:esmag}), with
\begin{equation}
\delta_V = -8.0 \pm 0.3\ {\rm mag} \ ; \quad  m_{c,V} = -9.3 \pm 0.3\ {\rm mag}
\ .
\label{eq:MWesmag}
\end{equation}
The lighter, broken line rising steeply towards faint magnitudes is a normal
Schechter function with $m_c$ as in equation (\ref{eq:MWesmag}) but no
mass-loss parameter, i.e., $\delta \rightarrow \infty$ in equation
(\ref{eq:esmag}). The shape of this curve is therefore
typical of the distribution of {\it logarithmic} mass for young massive
clusters in nearby galaxies.

\placefigure{fig:MWGCLF}

The lower panel of Figure \ref{fig:MWGCLF} contains a log-log representation
of the Galactic GC mass function, $dN/dM$. To construct this distribution, we
converted the absolute
$V$ magnitude of each GC into an equivalent mass by assuming a mass-to-light
ratio of $\Upsilon_V=2\ M_\odot\,L_\odot^{-1}$ for all clusters
(as implied by population-synthesis models; see McLaughlin \& van der
Marel 2005). The curves
here are the mass equivalents of those in the upper panel. Thus the bold,
dashed curve traces equation (\ref{eq:lognorm}) with
\begin{equation}
\langle\log(M/M_\odot)\rangle = 5.2\pm0.04 \ ; \quad \sigma_M=0.46 \pm 0.04
\label{eq:MWgmass}
\end{equation}
while the solid curve is equation (\ref{eq:esmass}) with
\begin{equation}
\log\,(\Delta/M_\odot)=5.4\pm0.1 \ ; \quad \log\,(M_c/M_\odot) = 5.9 \pm 0.1
\label{eq:MWesmass}
\end{equation}
and the lighter broken curve is equation (\ref{eq:esmass}) with
$\log\,(M_c/M_\odot)=5.9$ and $\Delta=0$---again, representative of
young cluster mass functions.

Although both model fits to the GCLF are
acceptable in a statistical sense, the evolved Schechter
function yields a significantly lower $\chi^2$ value. This is because
of the clear asymmetry in the observed GCLF, which appears as a 
faintward skew in the top panel and as a failure of the
mass function $dN/dM$ to decline toward low masses in the bottom
panel. This behavior is described well by the evolved Schechter
function but is necessarily missed by the Gaussian, which systematically
underestimates the number of clusters with $M\la 3\times10^4\,M_\odot$.

As a result of this, the
best-fit evolved Schechter function yields a GCLF peak which is slightly
brighter than the Gaussian. From the parameters given just above and either of
equations (\ref{eq:es_magto}) or (\ref{eq:es_massto}), we find a turnover
magnitude of $m_{\rm TO}=-7.5\pm0.1$ in the $V$ band, some 0.1~mag
brighter than the Gaussian turnover in equation (\ref{eq:MWgmag}). The
turnover mass implied by the evolved Schechter function is thus
$M_{\rm TO}\simeq (1.75\pm0.15)\times10^5\,M_\odot$, just over 10\% more
massive than the Gaussian fit returns.
The intrinsically symmetric Gaussian model is
forced to a fainter or lower-mass turnover in order to better fit the
relatively stronger low-mass tail of the observed GCLF. We find
similar offsets in general 
between the GCLF turnovers from the two model fits to our ACSVCS data
(see \S\ref{sec:results} and \S\ref{sec:trends} below).

We reiterate that the parameter $\Delta$ in the evolved Schechter
function represents the average total mass loss per cluster (presumably
due mostly to evaporation) that is required to transform an initial
mass function like that of young clusters in the local universe, into
a typical old GCLF.
Both qualitatively and quantitatively, our model fits in Figure
\ref{fig:MWGCLF} correspond to the various similar plots in Fall \&
Zhang (2001). In fact, the value $\Delta\simeq
(2.5\pm0.5)\times10^5\,M_\odot$ obtained here for the 
Milky Way agrees well with the mass losses
required by Fall \& Zhang for their successful models with the
second-order effects of tidal shocks included. The simple function in
equation (\ref{eq:esmass}) is thus a good approximation to their much
fuller treatment of the GCLF.

It is also worth emphasizing just how close $\Delta$ is
to the GCLF turnover mass scale. This implies that essentially {\it all}
globulars currently found in the faint ``half'' of the GCLF are remnants of
substantially larger initial entities. Equivalently, any clusters initially
less massive than $\simeq\!2$--$3\times10^5\,M_\odot$ are inferred
to have disappeared completely from the GC system.

Despite any difficulties in detail (\S\ref{sssec:esevolve} and
\S\ref{ssec:evaporation}) that might remain to be resolved
in this evaporation-dominated view of the GCLF, and of GC systems in
general, it is important just to have at hand a fitting formula like the
evolved Schechter function. In purely phenomenological terms, it fits
the GCLF of the Milky Way---which is, after all, still the best
defined over the largest range of cluster masses---at least as well as
any other function yet tried in the literature. In particular, it
captures the basic asymmetry of the distribution without sacrificing
the small number of parameters and the simplicity of form that have
always been the primary strengths of a Gaussian description. 
But at the same time, it is grounded in a detailed physical model
with well specified input assumptions (Fall \& Zhang 2001). Fitting it
to large datasets, such as that afforded by the ACSVCS, thus offers the
chance to directly, quantitatively, and economically assess the
viability of these ideas, in much more general terms than has been
possible to date.

\section{Fitting Methodology and Technical Considerations}
\label{sec:method}

\subsection{Maximum-Likelihood Fitting}
\label{ssec:ML}

Given either of the models just discussed---or, of course, any other---we
wish to estimate a set of
parameters for the intrinsic GCLF of a cluster sample using 
the method of maximum likelihood, following
an approach similar to that of Secker \& Harris (1993).
To do so, we make use of all the observational material described in
\S\ref{ssec:gclfhists}.

First, we denote the set of GC magnitudes and uncertainties in any galaxy, in
either the $z$ or the $g$ band, by $\{m_i,\epsilon_{m,i}\}$. Second, we
write the three-dimensional completeness function discussed above as
$f(m,R_h,I_b)$, which again depends not only on GC apparent magnitude but also
on a cluster's half-light radius and the background (``sky'' and 
galaxy) light intensity
at the position of the cluster. Third, from our 17 control fields we are able
to estimate the luminosity function of contaminants in the field of any ACSVCS
galaxy. We call this function $b(m)$, and we determine it by constructing a
normal-kernel density estimate, with bandwidth chosen using cross-validation
(see Silverman 1986, \S\S~2.4, 3.4). Finally, this further
allows us to estimate the net fractional contamination
in the GC sample of each galaxy: $\widehat{\cal B}=N_C/N$, where
$N_C \equiv (1/17) \sum_{i=1}^{17}N_{C,i}$
with $N_{C,i}$ the total number of contaminants present in the $i$-th
customized control field, and $N$ is the total number of all GC candidates in
the sample.  

Now, given this observational input, we assume that an intrinsic GCLF
is described by
some function $G(m|\Theta)$, where $\Theta$ is the set of model
parameters to be fitted. The choices for $G$ that we explore in this paper
were discussed in detail in \S\ref{sec:models}. Thus, for example, for the
Gaussian model of equation (\ref{eq:gauss}), $\Theta\equiv\{\mu, \sigma_m\}$,
while for the evolved Schechter function of equation (\ref{eq:esmag}),
$\Theta\equiv\{\delta, m_c\}$. We further assume that
magnitude measurement errors are Gaussian distributed, so that---in the
absence of contamination---the probability of finding an apparent
magnitude $m$ for a GC with given effective radius $R_h$, galaxy
background $I_b$, and magnitude uncertainty $\epsilon_m$ would be
\begin{equation}
G_T(m|\Theta,R_h,I_b,\epsilon_m) =
    {\cal A}\ [ h(m|\epsilon_m) \otimes G(m|\Theta)] f(m,R_h,I_b),
\end{equation}
where
$h(m|\epsilon_m)=(2\pi\epsilon_m^2)^{-1/2}\exp (-m^2/2\epsilon_m^2)$;
$\otimes$ denotes convolution; and the normalization
${\cal A}$---a function of the GCLF parameters $\Theta$ and the GC
properties $R_h$, $I_b$, and $\epsilon_m$---is fixed by requiring that
the integral of $G_T$ over the entire magnitude range covered by the
observations be unity.\footnotemark
\footnotetext{In principle there should be another factor 
multiplying $G_T$ proportional to the marginalization over $R_h$  of the joint
GC distribution in $m$ and $R_h$ times an indicator function which is 
1 over the area that satisfies $p_{\rm GC} \ge 0.5$. We neglect this factor here, which
is justified {\it a posteriori} by the agreement of results using
GC samples constructed using different selection functions.}

If a fraction ${\cal B}$ of sources in a galaxy are
contaminants, then the probability of having a bona fide GC with
magnitude $m$ (and given $R_h$, etc.) is reduced to
$(1-{\cal B})G_T$, and thus the likelihood that a set of GCLF model
parameters $\Theta$ can account for $N$ total objects with observed
magnitudes $\{m_i\}$ and properties
$\{R_{h, i}, I_{b, i}, \epsilon_{m, i}\}$ is
\begin{equation}
{\cal L}(\Theta, {\cal B}) = \prod_{i=1}^N \left[
   (1-{\cal B})G_T(m_i|\Theta,R_{h,i},I_{b,i},\epsilon_{m,i})
                                          +{\cal B}b(m_i)\right] 
\label{eq:lik}
\end{equation}
in which it is assumed that the luminosity function $b(m)$ of
contaminants is also normalized.

For any chosen functional form $G(m|\Theta)$ of the intrinsic GCLF, we
specify some initial parameter values $\Theta$, compute $G_T$ and $b$ for each
observed object in a galaxy, and maximize on $\Theta$ the product
in equation (\ref{eq:lik}). 
In principle, it is possible simultaneously to determine the
contamination fraction in this way, but in practice we found this to
be a rather unstable procedure (even small inadequacies in the chosen
model for $G$ can lead to a maximum-likelihood solution that converges
to quite unreasonable values for ${\cal B}$). Thus, we instead made
direct use of our prior information from the 17 control fields, and 
fixed this fraction to the rather precise average
$\widehat{\cal B}$ that we have measured for each galaxy.

The uncertainties in the fitted parameters $\Theta$ are 
estimated by using the covariance matrix calculated at the point of maximum
likelihood (e.g., Lupton 1993).
These uncertainties include the effects of possible 
correlation between the parameters, but they do not
include the additional, unavoidable uncertainty arising from cosmic
variance in the form of $b(m)$ and the expected number
$\widehat{\cal B}$ of contaminants in any field. As such, they 
constitute 
lower limits to the total uncertainty. This is not a significant issue
for GCLF fits to cluster samples combined from several galaxies (see below),
but it can be important for fits to individual galaxies.

To deal with this, when we fit any individual
GC system, we re-run our maximum-likelihood algorithm 17 times, each
time using the background contamination fraction ${\cal B}$ as estimated from
a different one of our 17 control fields (versus using
$\widehat{\cal B}$ from an average of all control fields to obtain the
nominal best fit). We
record the different sets of best-fit GCLF parameters obtained in
these trials and use the variance in them to evaluate the additional
uncertainty arising from cosmic variance of the background
contamination.

\subsection{Bias Tests}
\label{ssec:bias}

Maximum-likelihood estimators are biased in general. It is thus
important when deriving conclusions to test the bias properties of the 
estimator used, under circumstances similar to the ones under study. We
have done this specifically for the benchmark case of Gaussian fits to
the GCLF. After obtaining mean magnitudes and dispersions from our
maximum-likelihood routine for the 89 ACSVCS galaxies, we analyzed 20
simulated datasets per galaxy, using the following procedure. First,
we subtracted the number of contaminants $\widehat{\cal B}N$ in the galaxy
from the total number $N$ of GC candidates there, to estimate the expected
population $N_{\rm GC}$ of bona fide GCs. We then randomly drew a
sample of $N_{\rm GC}$ magnitudes from a
Gaussian distribution with a mean $\mu$ taken to be
the fitted maximum-likelihood estimate for that galaxy, and a dispersion 
chosen from $\sigma_m=0.4,0.7,1$ or 1.3~mag. (We did 5 simulations for
each of these dispersions, giving the total of 20 simulations per galaxy.)
The randomly generated objects replaced the $N_{\rm GC}$ objects
in that galaxy's sample with the highest
$p_{GC}$ values. 
The values of $R_h$ and $I_h$ of the latter objects plus the simulated
magnitude are used to determine the completeness value $f$ for each
source. A uniform random deviate is then computed and if that is larger
than the value of $f$ the source is discarded, a new magnitude drawn
from the Gaussian and the process repeated until the condition is
met. In this way the effects of completeness are taken into account.
The maximum-likelihood procedure was finally run 
on each simulated sample and the output parameters compared to the
input ones.

\begin{figure}
\epsscale{0.8}
\plotone{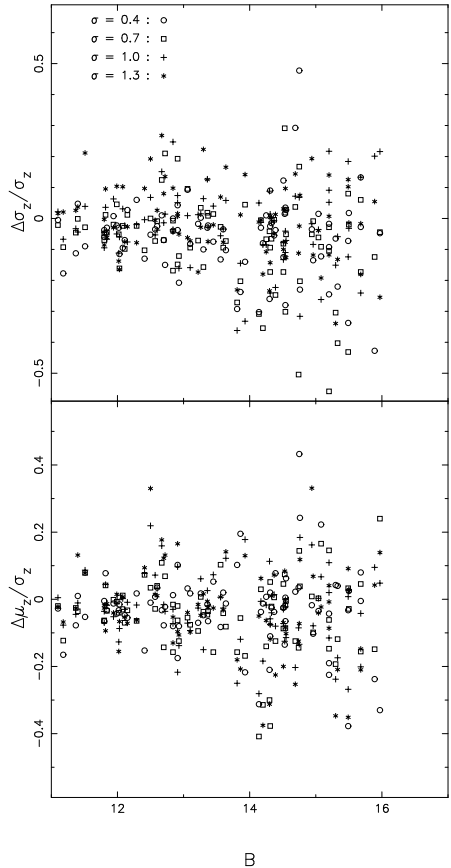}
\caption[]{{\it Top:} Fractional difference $\Delta \sigma_z/\sigma_z$
between input and recovered Gaussian dispersion for simulated GCSs with four
different Gaussian dispersions assumed ($\sigma=0.4,0.7,1.0,1.3$).
{\it Bottom:} Difference $\Delta \mu_z/\sigma_z$ between input and recovered
Gaussian mean $\mu$ for the same simulated GCSs as in the upper panel.
}
\label{fig:MLbias}
\end{figure}

The results of these simulations in the $z$ band
are summarized in Figure~\ref{fig:MLbias}.
There may be slight biases in the recovered parameters, with
$\langle \Delta\sigma_m/\sigma_m \rangle \approx\! -0.03$ and 
$\langle \Delta\mu/\sigma_m \rangle \approx\! -0.03$, although there are no
significant trends in these average offsets with galaxy luminosity (i.e.,
sample size). Moreover, the statistical significance of these biases is not
high ($< 3 \sigma$), and so we choose not to correct for them. As a result, it
is possible that our output best-fit parameters are biased at the level of 3\%
of the GCLF dispersion; but with the possible exception only of the most
populous GC system (that of M87=VCC 1316), this turns out always to be smaller
than the formal uncertainties on the GCLF parameters
(see \S\ref{ssec:gaussian}).
Note that the {\it scatter} of the retrieved parameters compared with the
input ones increases towards fainter galaxy magnitudes because
the candidate-GC sample size is decreasing, and the variance in the estimates
of both $\sigma$ and $\mu$ scales as $\sim\! 1/N$.

\placefigure{\ref{fig:MLbias}}

\subsection{Effects of Selection Procedure}
\label{ssec:selec}

As we mentioned in \S\ref{sec:obs}, the procedure we used to construct
a sample of GC candidates for each galaxy involved assigning a
probability $p_{\rm GC}$ to each source and allowing into the sample
only those objects with $p_{\rm GC}\ge 0.5$. This  may influence
the resulting observed luminosity function and consequently
affect the derived parameters of any fitted model. In order
to check that we do not unduly bias our GCLF fitting by this selection
technique, we also constructed alternate candidate-GC
samples that do not use the selection on $p_{\rm GC}$ but only apply
a magnitude cut and an upper limit of $R_h\la 5$~pc (cf.~the second half of
Table \ref{tab:gclf_hists} in \S\ref{ssec:gclfhists}).
The magnitude distributions of such samples are free of any selection
effects arising from using the $p_{\rm GC}$ values
and are useful for testing the robustness of any result. Thus, when
we fit GCLFs to any of our data, we have verified that consistent conclusions
are obtained using either of our sample definitions.

\subsection{Binned Samples}
\label{ssec:binned}

While we always perform GCLF fits to individual
galaxies, some of the fainter systems suffer from small-number
statistics and/or excessive contamination. We thus constructed
still more GC samples by combining all candidate clusters from as
many galaxies as required to reach a total sample size above some
minimum. Going down the list of our target galaxies sorted by apparent $B$
magnitude, we accumulate galaxies
until the expected number of bona fide GCs (i.e., the total number of
candidates minus the number of contaminants estimated from our
customized control fields) is $\ga 200$. Although many of the brighter
galaxies satisfy this condition by themselves, we refer to the samples
defined in this way as ``binned'' samples.\footnotemark
\footnotetext{We excluded 5 galaxies when constructing the
binned samples, namely VCC~798, VCC~1192, VCC~1199, VCC~1297  and
VCC~1327. The first was excluded due to the presence of a strong excess of
diffuse clusters (Peng et al. 2006b) and the rest because of their proximity
to either M87 or M49, making their GC systems dominated by those of their
giant neighbour; see \S\ref{ssec:gradius}. We additionally excluded all
galaxies without available SBF  distances.}
There are 24 of them 
in all, and they are used in \S\ref{sec:trends} particularly, to
assess trends in GCLF parameters as a function of galaxy luminosity
without the significant scatter caused by the small numbers of GCs in
faint systems.

Our SBF analysis has shown that some of the ACSVCS
galaxies 
have distance moduli significantly different from the mean 
$(m-M)_0=31.09$ mag for Virgo (Mei et al.~2007),
and thus simply combining the apparent magnitudes of GCs from different
galaxies with no
correction could artificially inflate the dispersion of any composite
GCLF. To avoid this, we do the binning by first using the SBF
distances to transform all candidate GC luminosities to the value they
would have at a distance of 31.1 mag ($D=16.5$~Mpc). 

\section{Model Fits}
\label{sec:results}

In this section we present the results of our maximum-likelihood
fitting of Gaussians and evolved Schechter functions to the GCLFs in the Virgo
Cluster Survey. Recall that any alternative model may be
fit to the GCLF histograms in Table \ref{tab:gclf_hists}, which can be
downloaded from the electronic edition of the {\it Astrophysical Journal}.

\subsection{Gaussian Fits}
\label{ssec:gaussian}

The parameter estimates for an intrinsic Gaussian fitted to our 89
individual GCLFs are given in Table~\ref{tab:gclfpars}. There we list each
galaxy's ID number in the VCC and its total apparent magnitude $B_{\rm gal}$,
both taken from Binggeli, Sandage \& Tammann (1985). Following this are
the maximum-likelihood values of the mean GC magnitude and dispersion
and their uncertainties in the $g$-band ($\mu_g, \sigma_g$), 
the same quantities in the $z$-band ($\mu_z, \sigma_z$),
the fraction $\widehat{\cal B}$ of the sample that is expected to be 
contamination, and the total number $N$ of all objects (including contaminants
and uncorrected for incompleteness) in the galaxy's
candidate-GC sample. The last column of Table \ref{tab:gclfpars} gives 
comments on a few galaxies with noteworthy aspects.
Note that the uncertainties in the Gaussian parameters include
contributions from cosmic variance in the shape and normalization
of the contamination luminosity function $b(m)$ (see \S\ref{ssec:ML}).

\placetable{\ref{tab:gclfpars}}
 
\begin{figure*}
\epsscale{1.14}
\plottwo{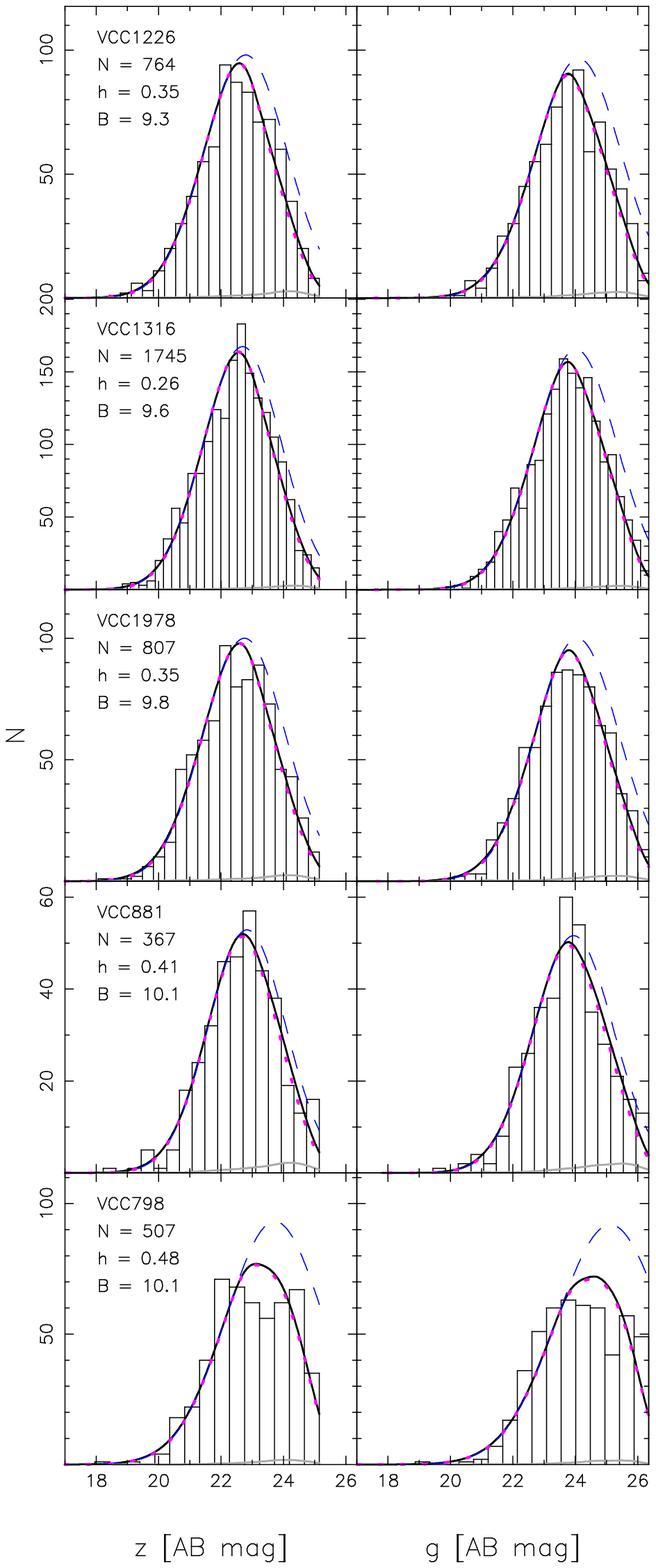}{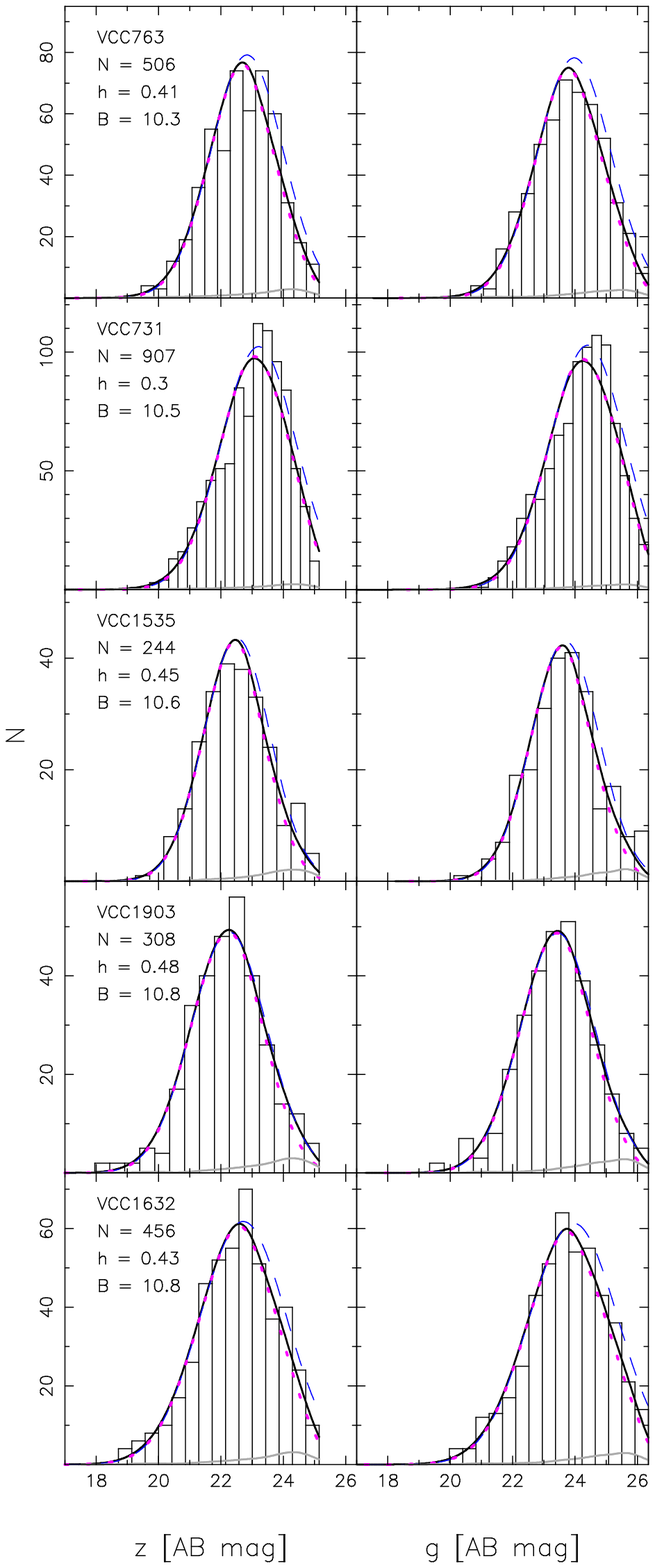}
\caption[]{Histograms of the GCLFs for our sample galaxies. For each galaxy we
present the $z$-band and $g$-band GCLFs side by side. The VCC name 
and $B$ magnitude  of the
galaxy are indicated in the upper left corner of the left panel, where we also 
indicate the total number of sources in each histogram and the bin-width $h$
used to construct the histogram. Additionally we show 
the best-fit model (solid black curve), the intrinsic Gaussian
component (dashed curve), the Gaussian component
multiplied by the expected completeness (dotted curve)
and a kernel-density estimate of the expected contamination in the sample
(solid gray curve). The solid black curve is the sum of the solid gray and dotted curves.
The galaxies are ordered by decreasing apparent $B$-band total luminosity,
reading down from the upper left-hand corner.
The parameters of the fits are given in Table \ref{tab:gclfpars}.
}
\label{fig:gaussfits}
\end{figure*}

In Figure \ref{fig:gaussfits} we present
histograms of the observed GCLFs along with 
the best fitting maximum-likelihood models. The galaxies are arranged in order
of decreasing apparent $B_{\rm gal}$ magnitude (i.e., the same order as in
Table \ref{tab:gclfpars}), and there are two panels per galaxy: one presenting
the $z$-band data and model fits, and one for the $g$ band. The bin width
chosen for display purposes here is not the same for all galaxies, but follows
the rule $h=2 (IQR) N^{-1/3}$, where $(IQR)$ is the interquartile range of the
magnitude distribution and $N$ is the total number of objects in each GC
sample (Izenman 1991).

\placefigure{\ref{fig:gaussfits}}

%\clearpage
\begin{figure*}
{\epsscale{1.14} \plottwo{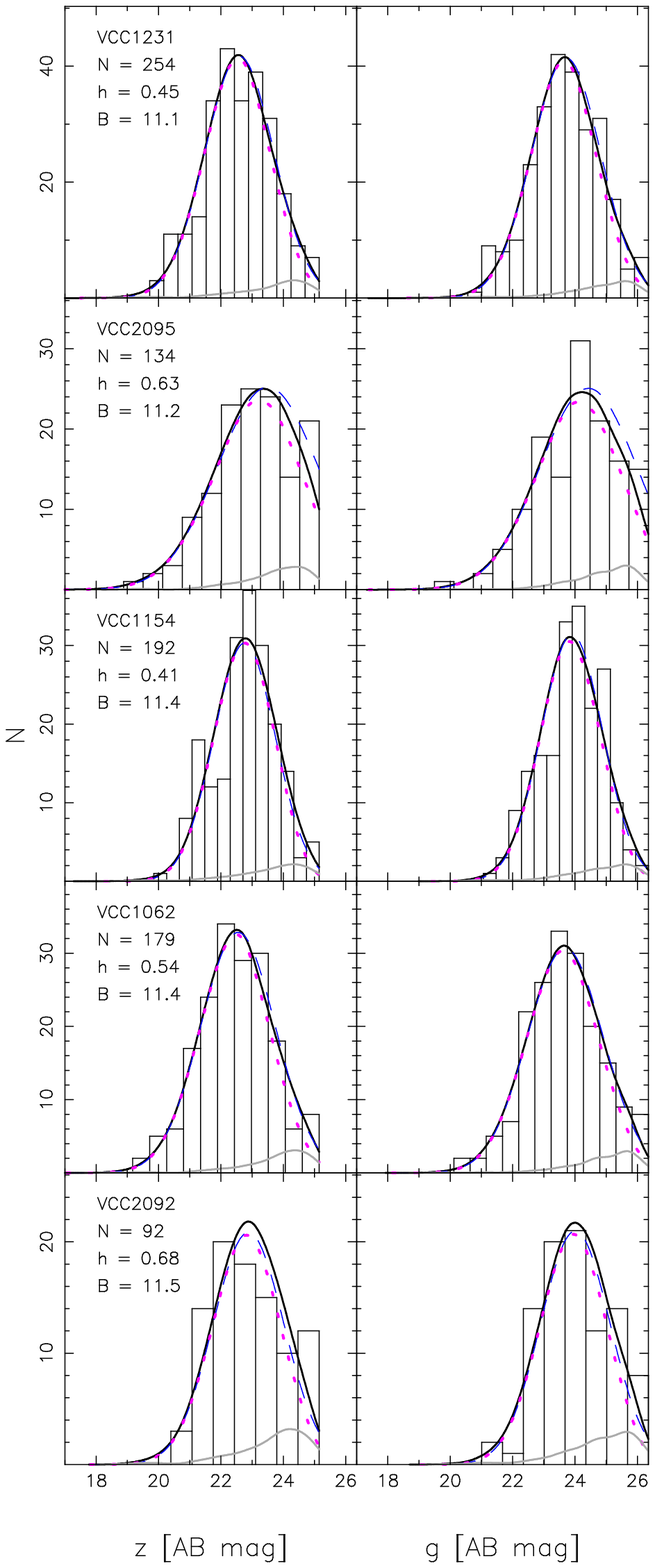}{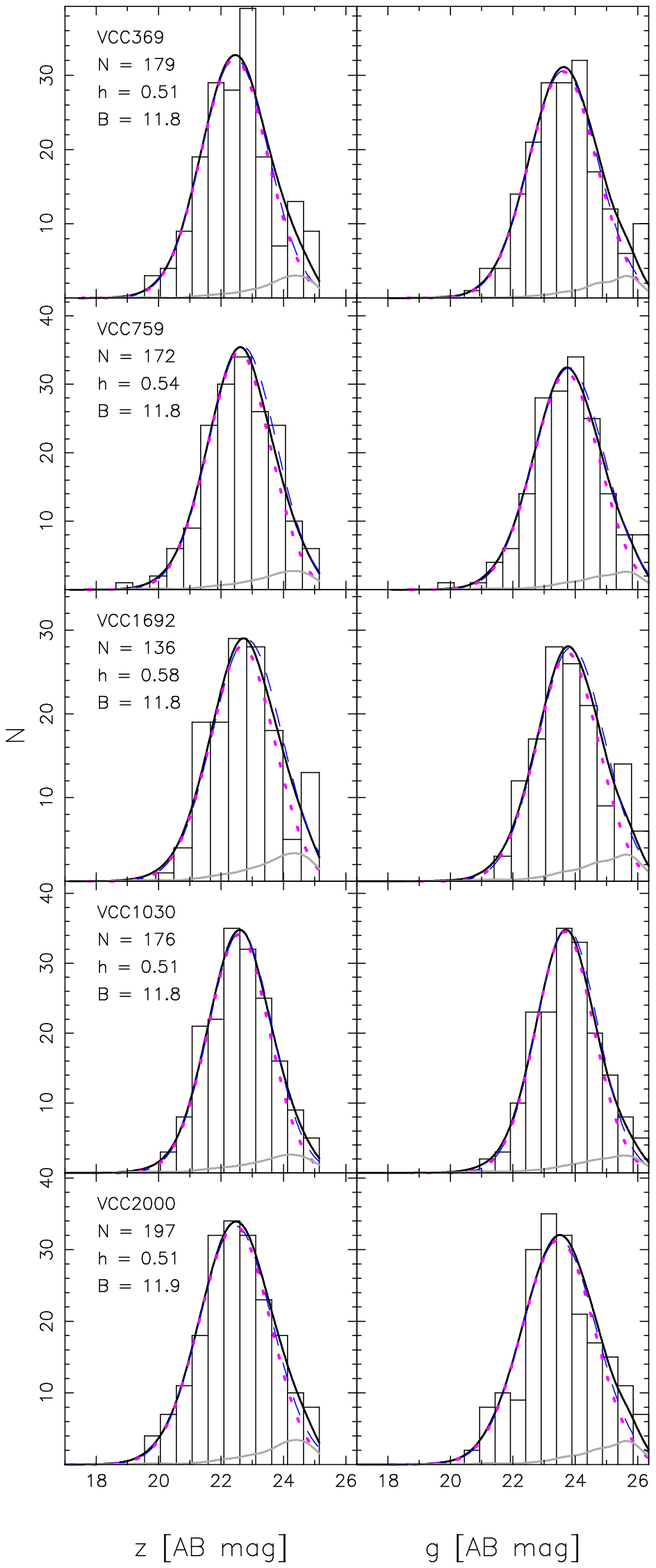}}\\[5mm]
\centerline{Fig. 4. --- {\it Continued}}
\end{figure*}

\begin{figure*}
{\epsscale{1.14} \plottwo{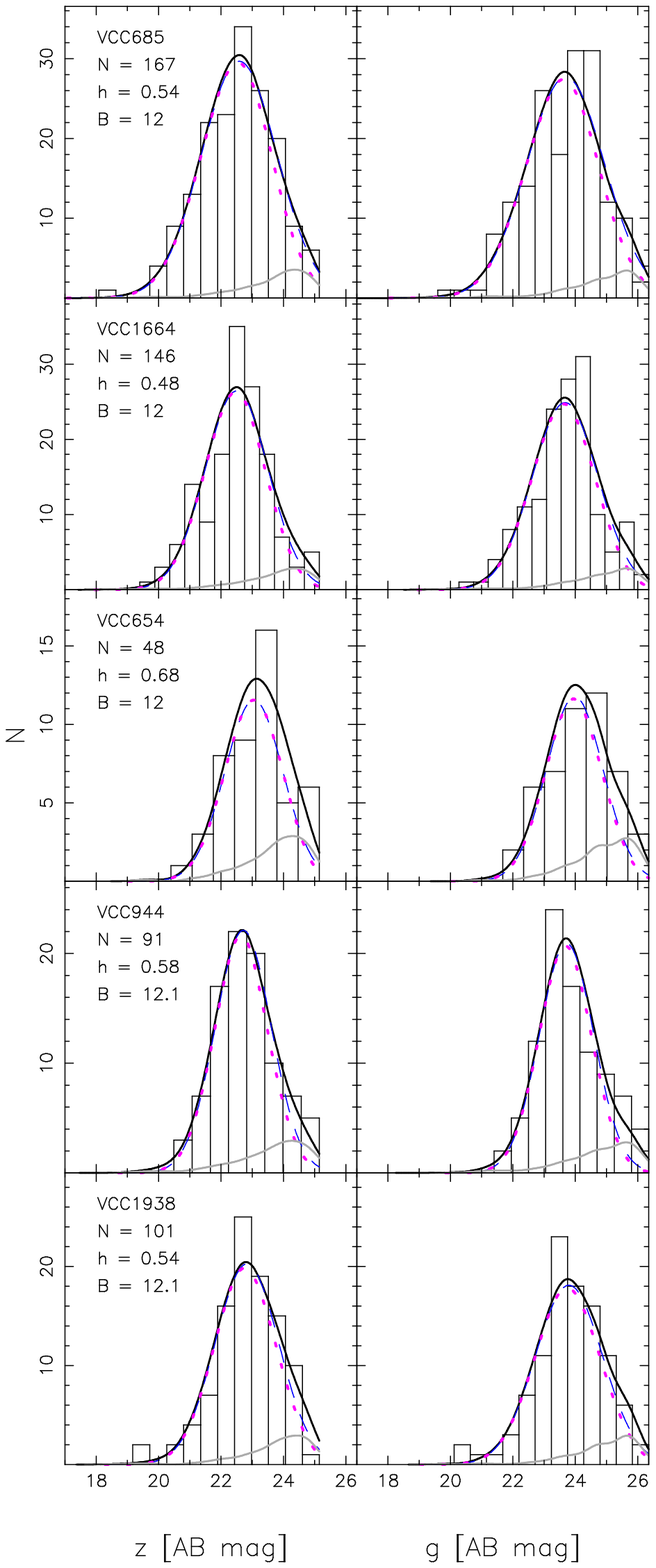}{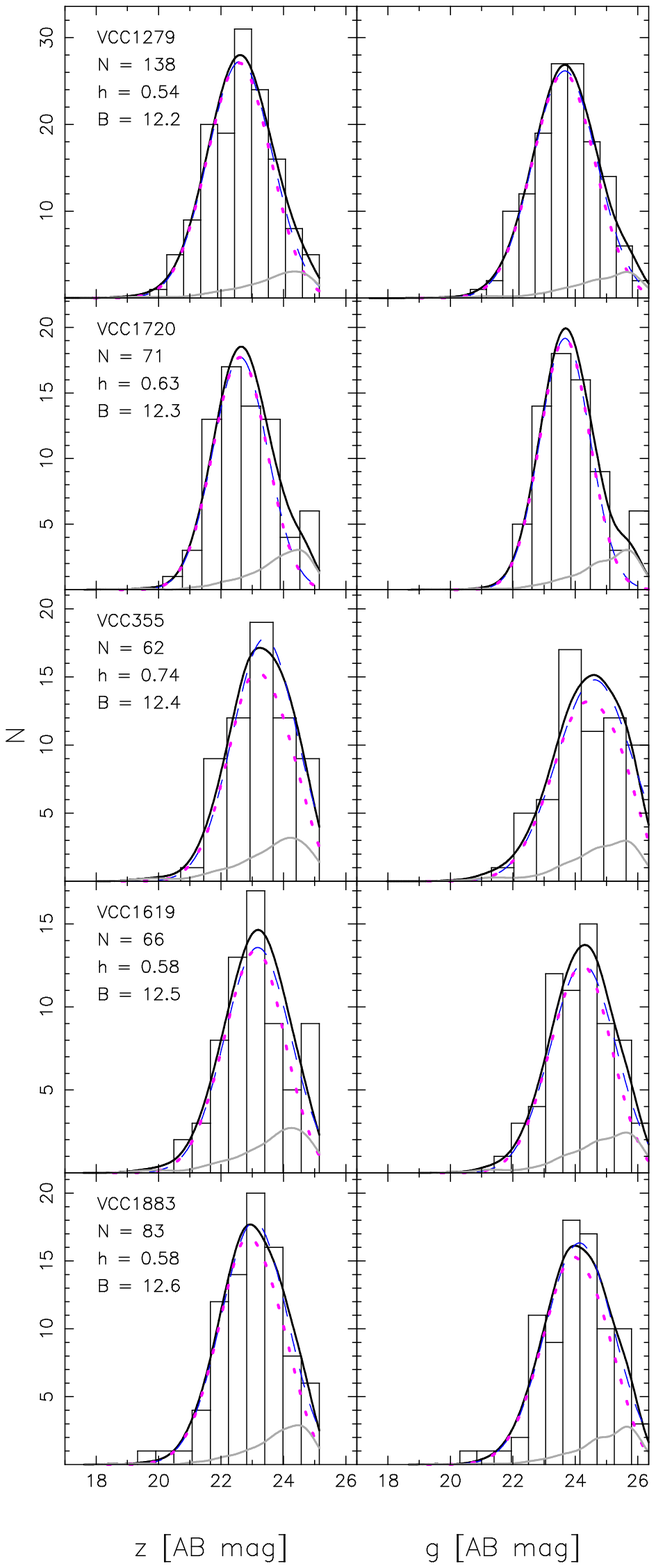}}\\[5mm]
\centerline{Fig. 4. --- {\it Continued}}
\end{figure*}

\begin{figure*}
{\epsscale{1.14} \plottwo{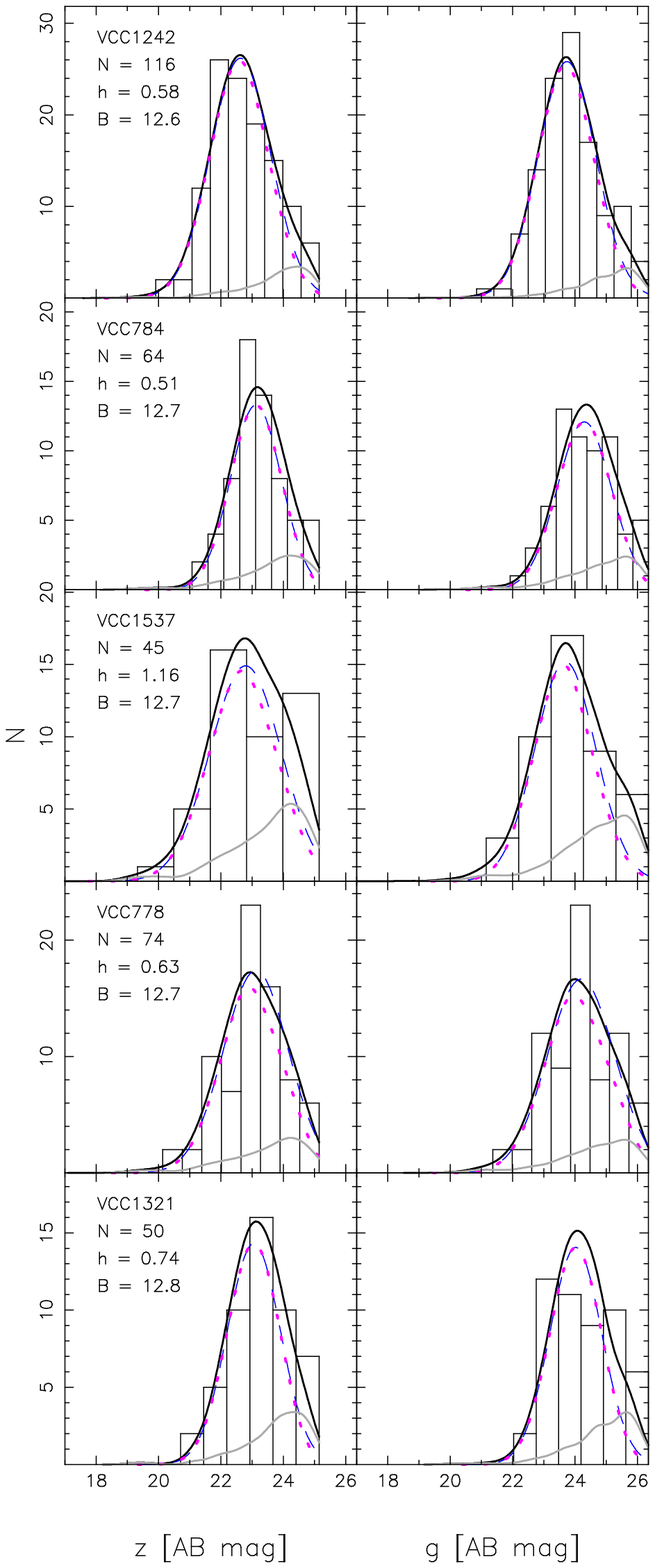}{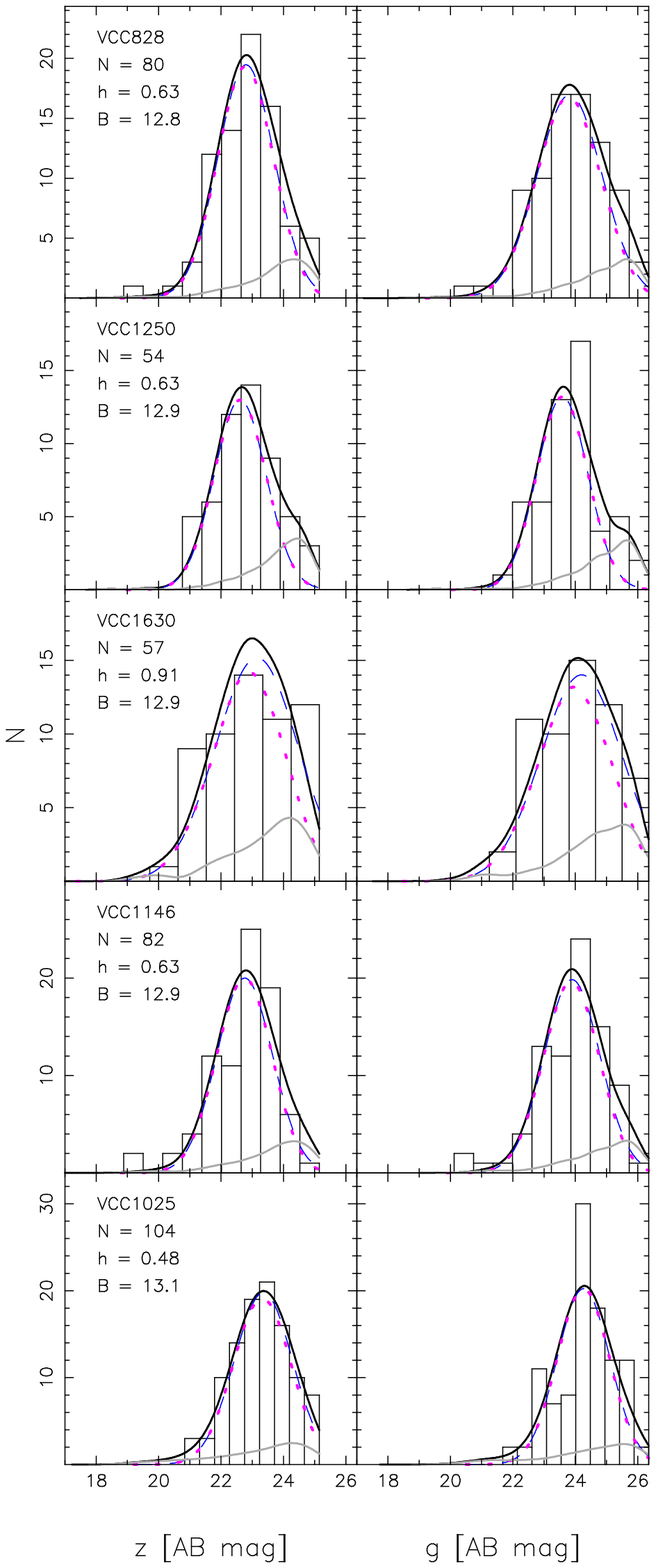}}\\[5mm]
\centerline{Fig. 4. --- {\it Continued}}
\end{figure*}

\begin{figure*}
{\epsscale{1.14} \plottwo{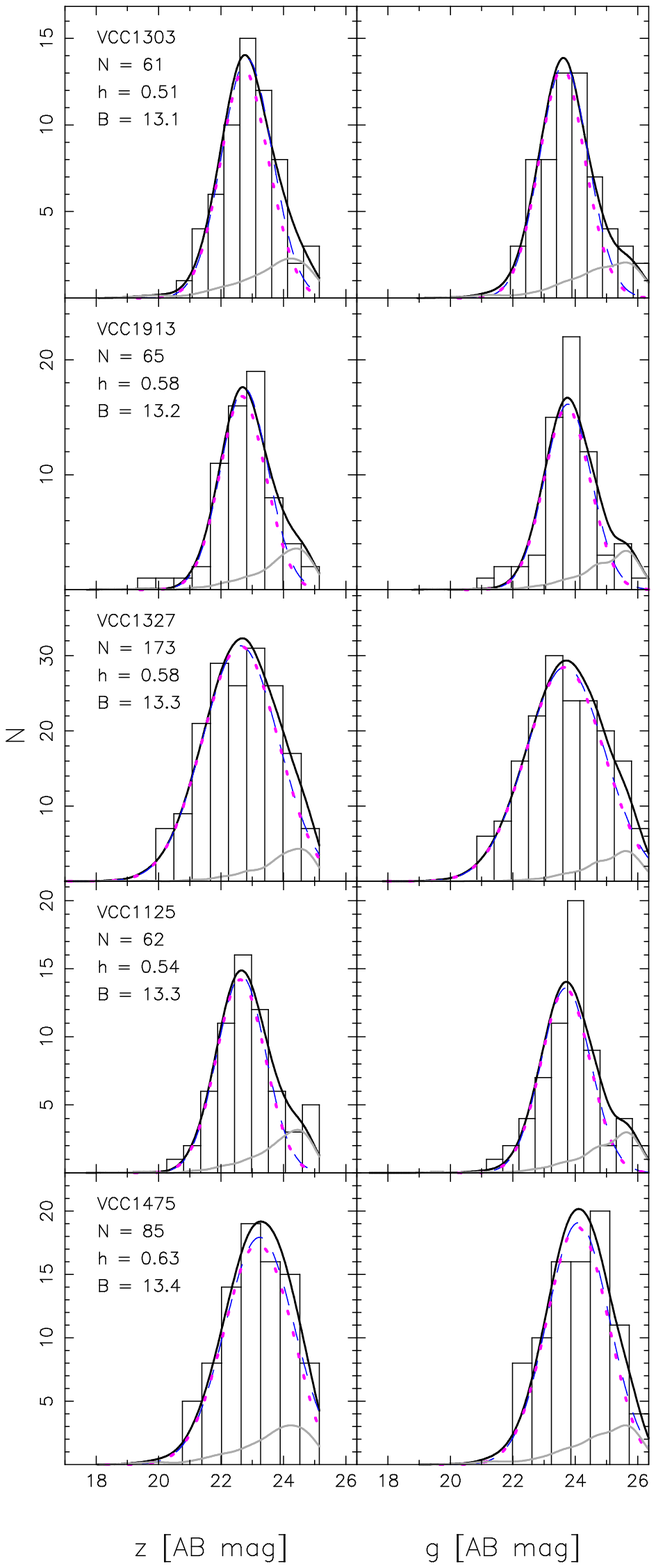}{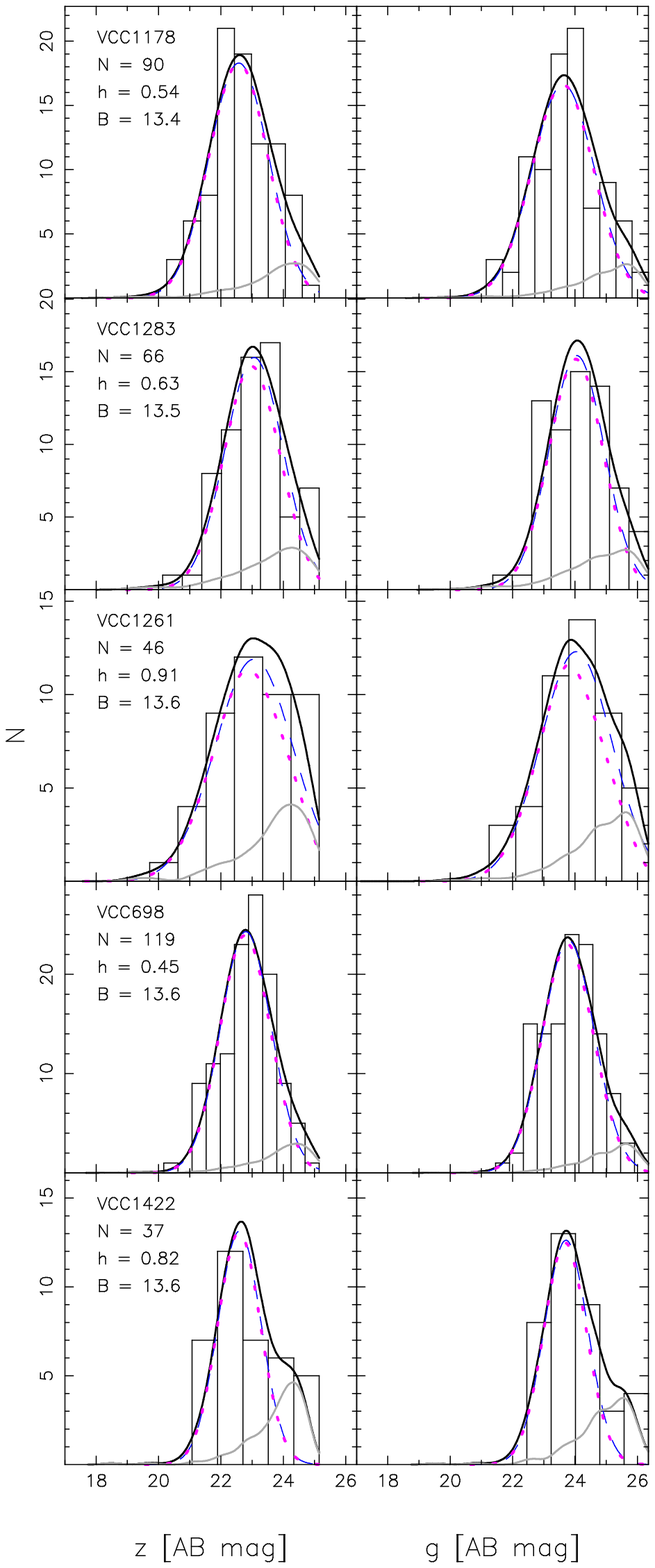}}\\[5mm]
\centerline{Fig. 4. --- {\it Continued}}
\end{figure*}

\begin{figure*}
{\epsscale{1.14} \plottwo{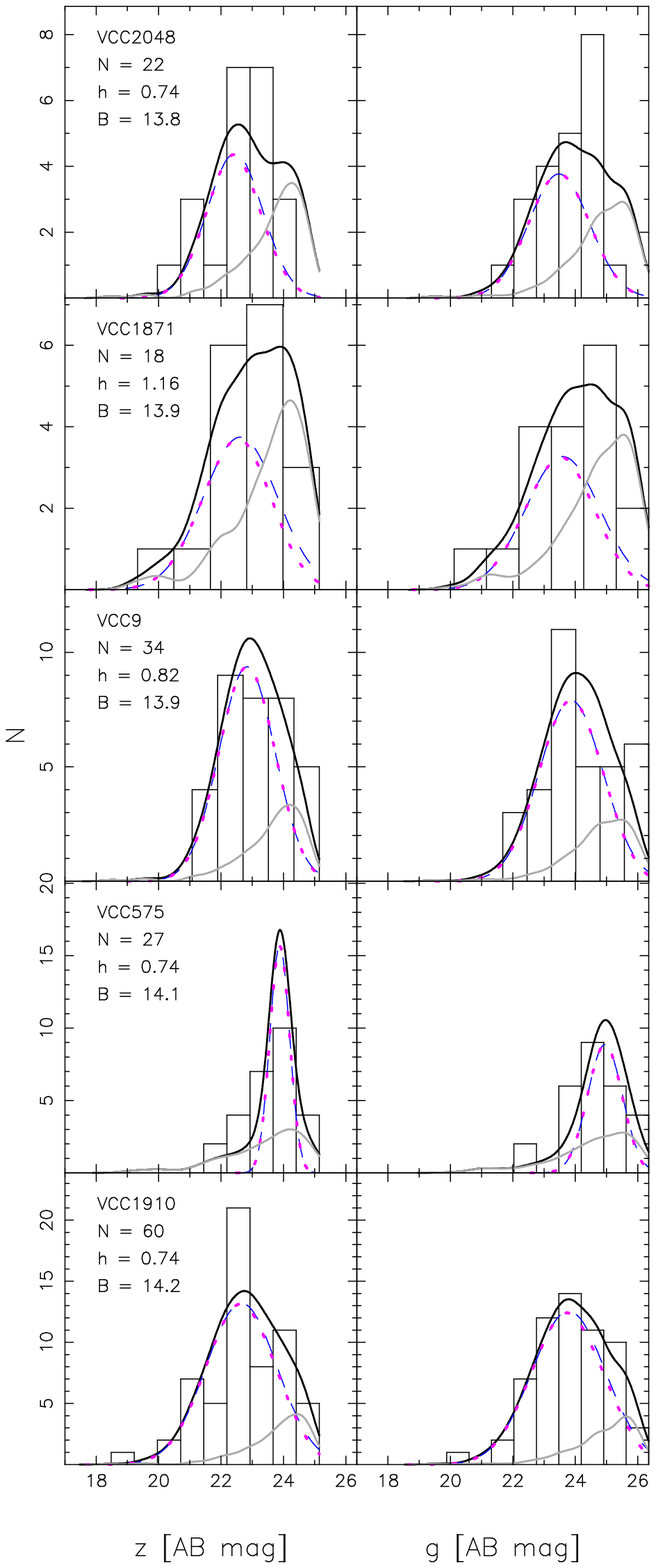}{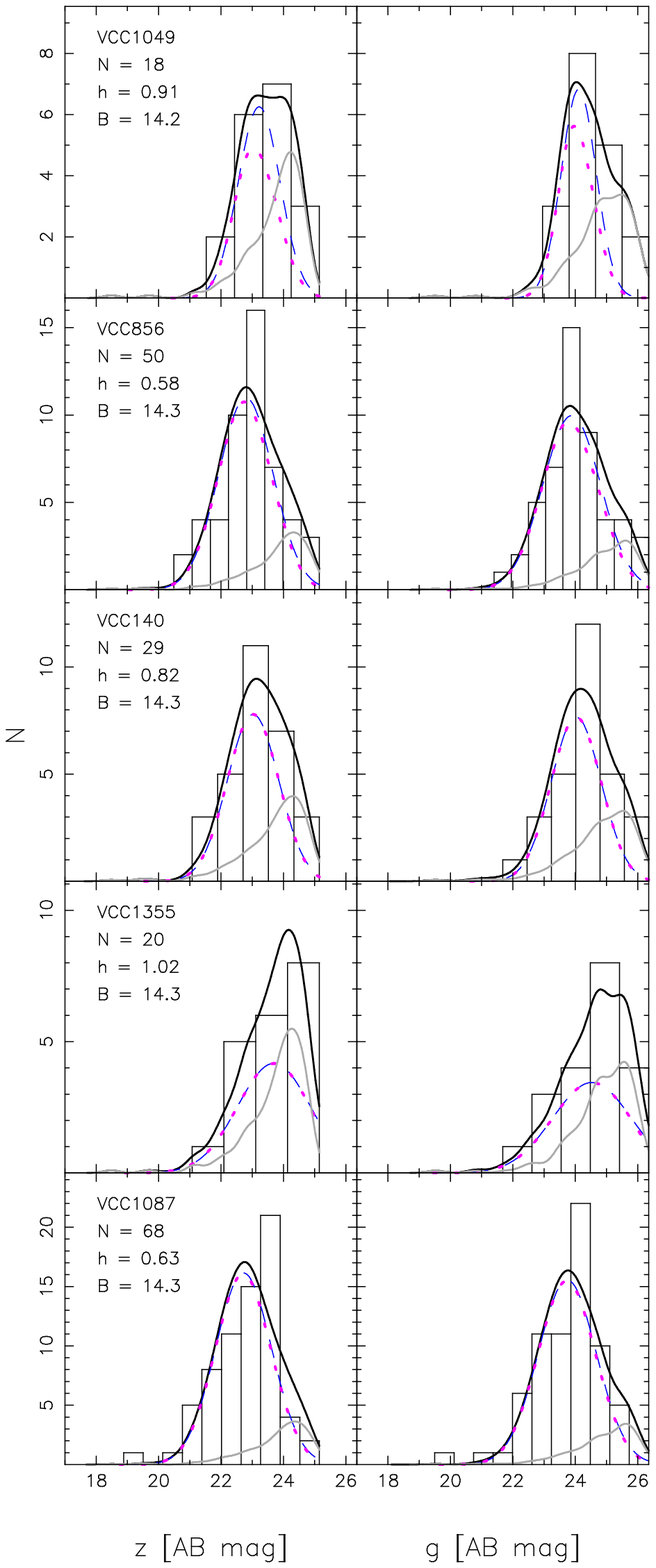}}\\[5mm]
\centerline{Fig. 4. --- {\it Continued}}
\end{figure*}

\begin{figure*}
{\epsscale{1.14} \plottwo{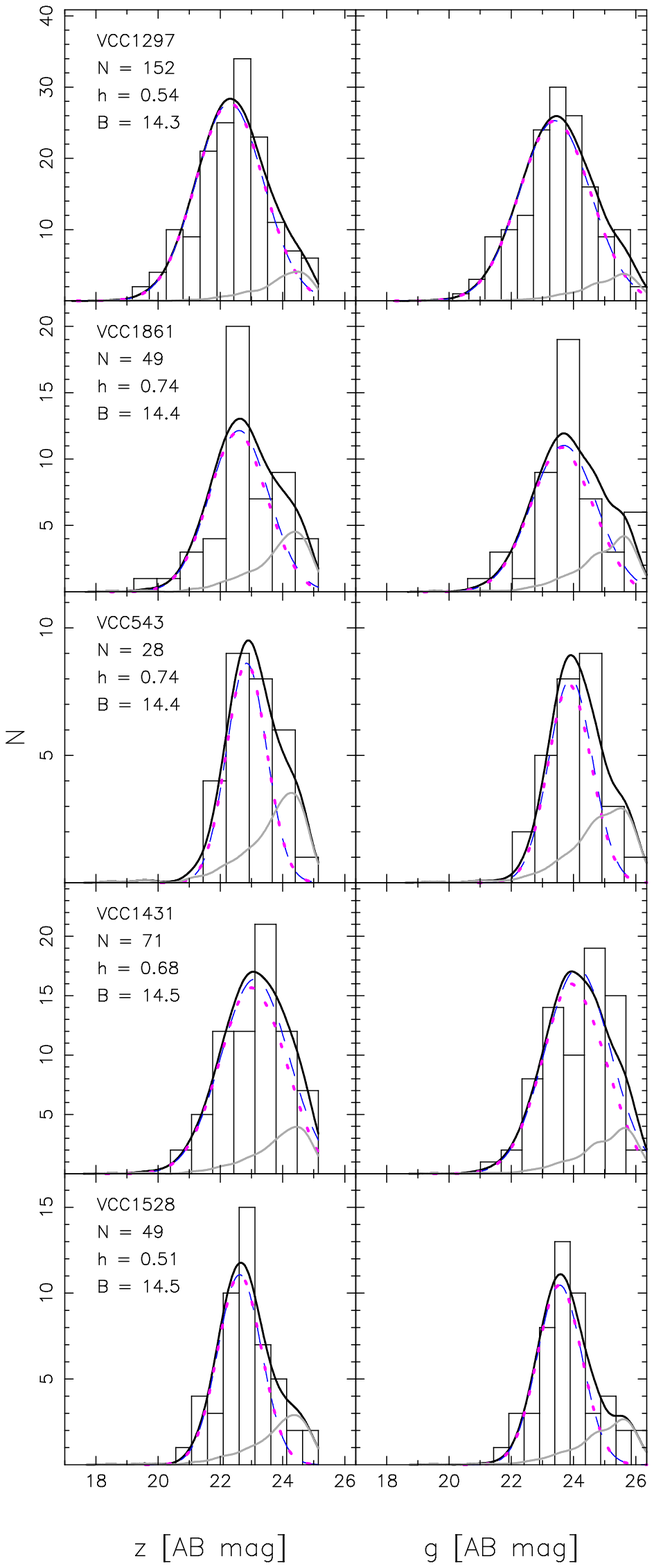}{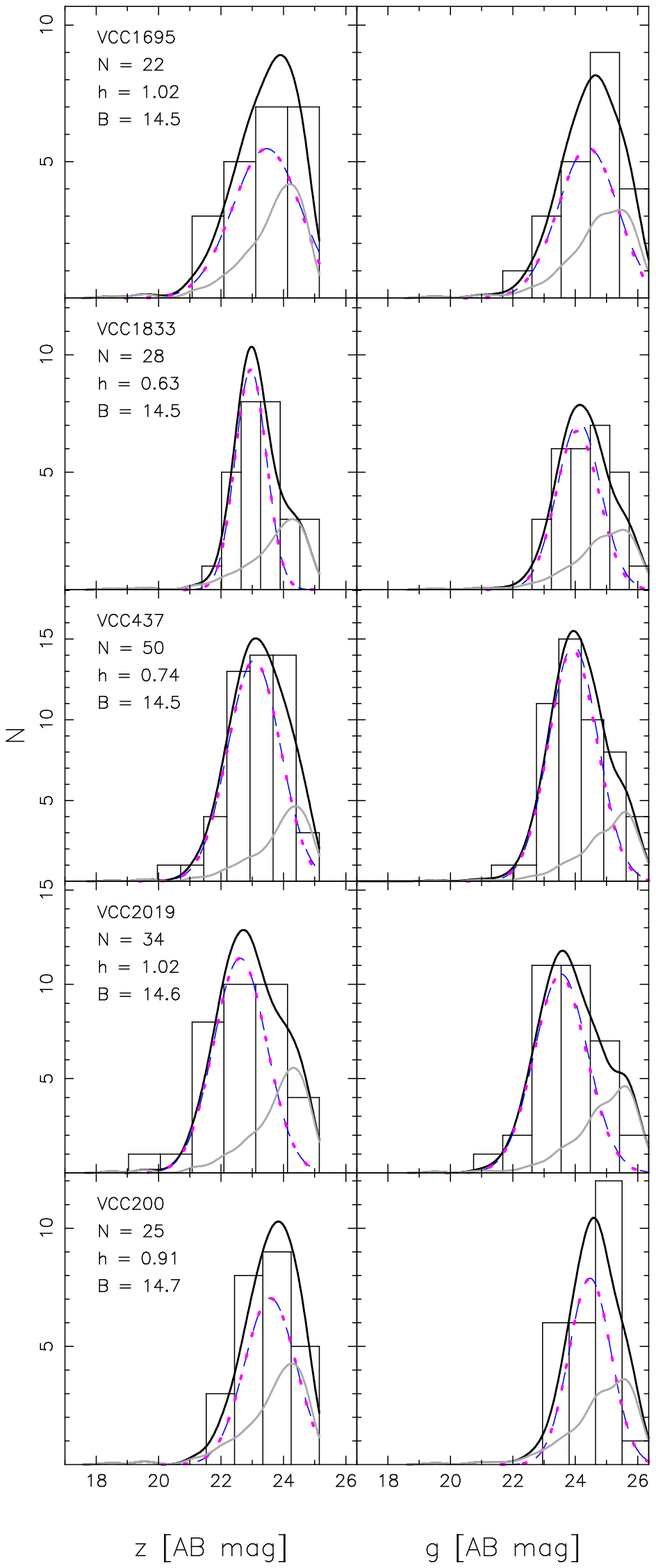}}\\[5mm]
\centerline{Fig. 4. --- {\it Continued}}
\end{figure*}

\begin{figure*}
{\epsscale{1.14} \plottwo{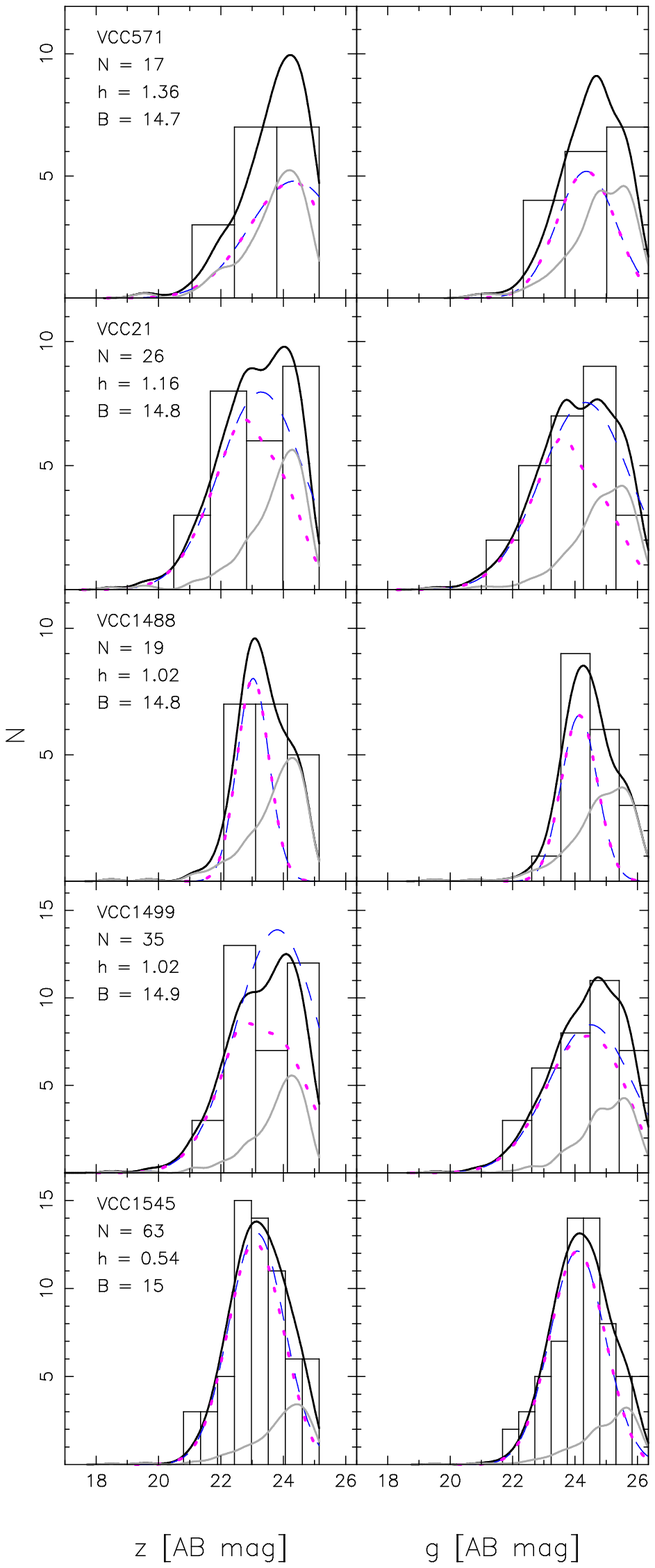}{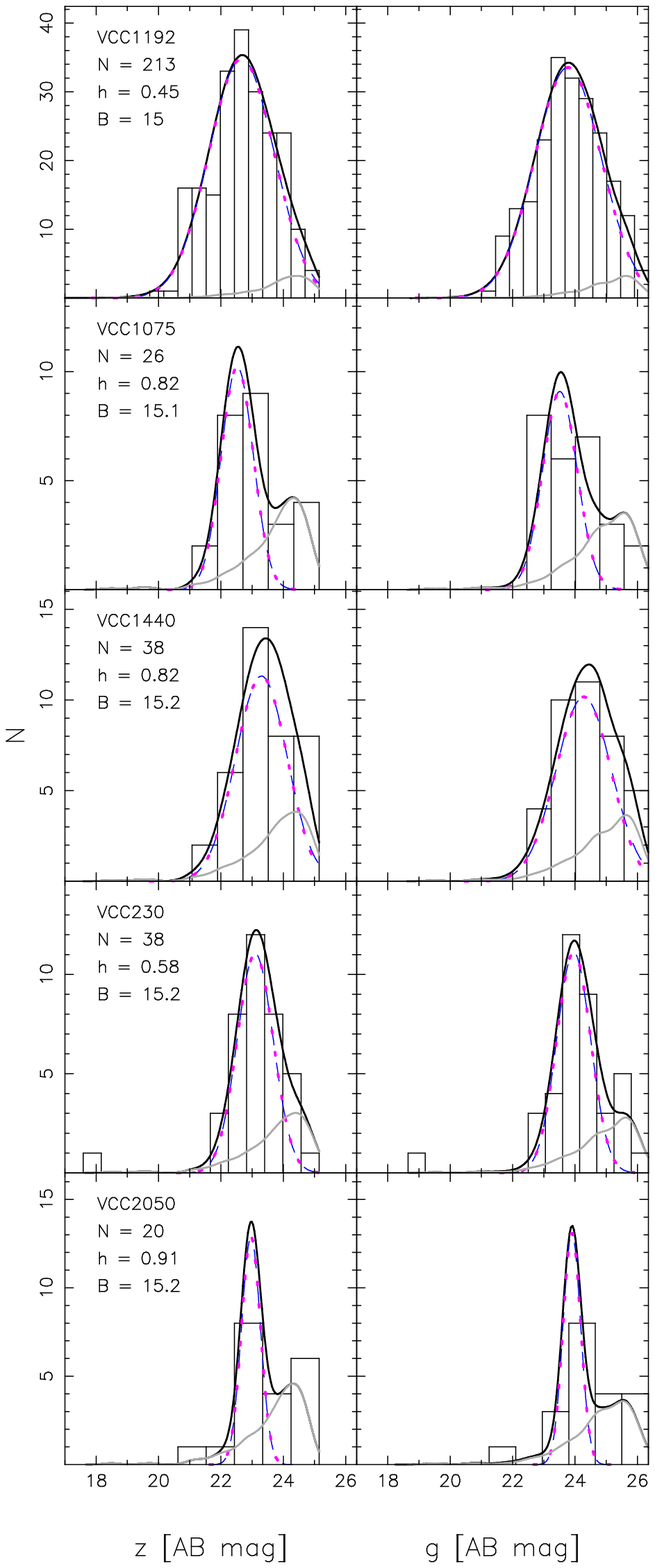}}\\[5mm]
\centerline{Fig. 4. --- {\it Continued}}
\end{figure*}

\begin{figure*}
{\epsscale{1.14} \plottwo{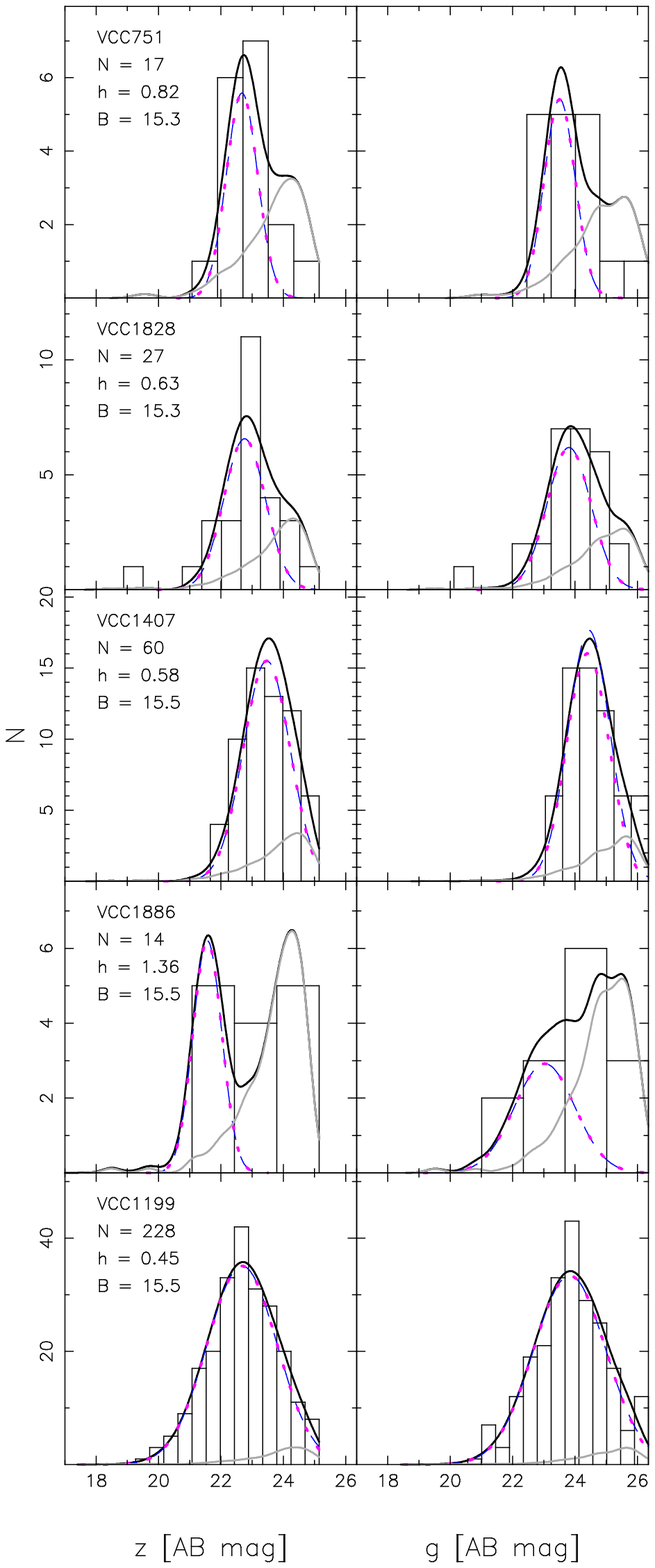}{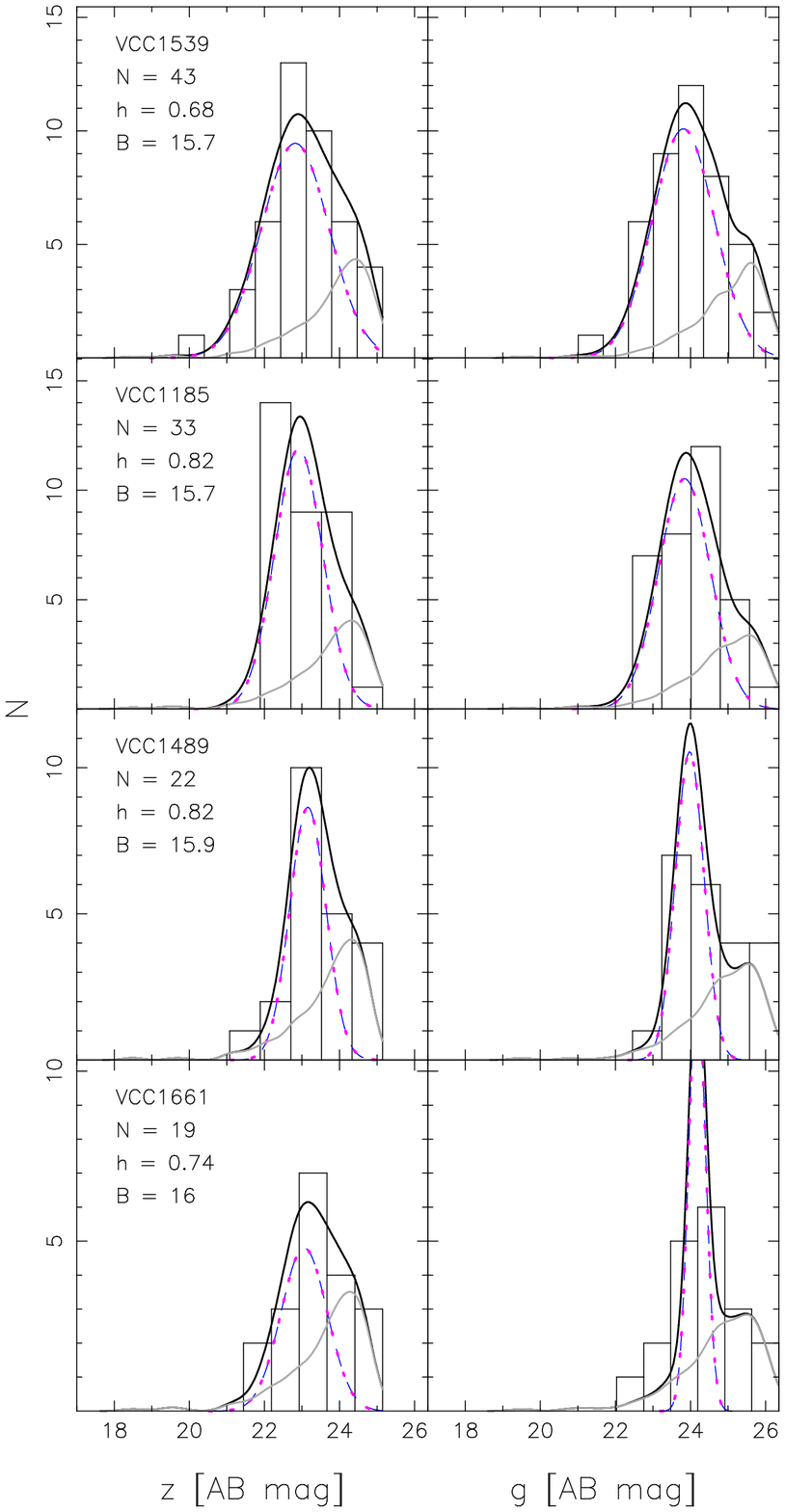}}\\[5mm]
\centerline{Fig. 4. --- {\it Continued}}
\end{figure*}

There are four curves drawn in every panel
of Figure \ref{fig:gaussfits}. The long-dashed  curve is
the best-fit intrinsic Gaussian GCLF, given by equation (\ref{eq:gauss}) with
the parameters listed in
Table \ref{tab:gclfpars}. The dotted  curve is this intrinsic
model multiplied by the completeness function, $f(m, R_h, I_b)$, after
marginalizing the latter over the distribution of
$R_h$ and $I_b$ for the observed sources in each galaxy.\footnotemark
\footnotetext{In order to
marginalize  $f(m, R_h, I_b)$ one needs to know the distributions of $R_h$ and
$I_b$---information which is not available {\it a priori}. Using the full
observed distributions of $R_h$ and $I_b$ is not possible, because they are
affected by completeness (e.g., faint GCs with large $R_h$ are less
likely to be detected). We therefore marginalize $f$ assuming that the 
underlying distributions in of $R_h$ and $I_b$ are given by the observed
distributions for objects satisfying $z < 22.5$ and $g < 23.7$, which gives
samples of objects that can be considered complete with high confidence,
anywhere in any of our galaxies.}
The solid gray curve is our
kernel-density estimate of the expected contaminant luminosity
function. Finally, the solid black curve is the sum of the solid gray and dotted
curves; it is the net distribution for which the likelihood in equation
(\ref{eq:lik}) above is maximized.

\begin{figure}%[t!]
\epsscale{1}
\plottwo{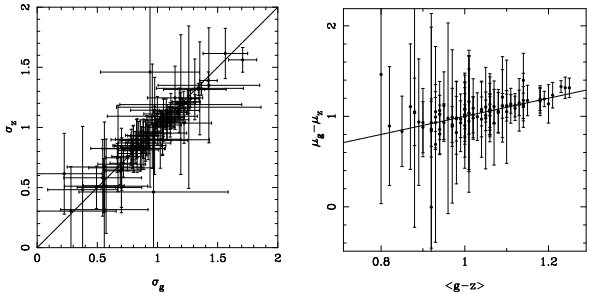}{f5b.eps}
\caption[]{({\it Left}) Estimate of Gaussian dispersion in the $z$-band, 
$\sigma_z$, versus the same quantity in the $g$-band, $\sigma_g$, for the
GCLFs of our
sample galaxies. Uncertainties are 1$\sigma$. The line marks the one-to-one
correspondence between these two quantities.
({\it Right}) Difference between estimates of Gaussian means in the 
$g$- and $z$-bands, $\mu_g - \mu_z$, versus the mean color
$\langle g-z \rangle$ of the GC systems of our sample galaxies. Uncertainties
are 1$\sigma$. The line marks the one-to-one
correspondence between these two quantities.
}
\label{fig:consistency}
\end{figure}

Since we have two realizations of the GCLF for every
galaxy---one in the $z$ band and one in the $g$ band---we are able to check
the internal consistency of our model parameter estimates.
Thus, in Figure~\ref{fig:consistency} we compare the measured Gaussian
means and dispersions in the two bands. The left-hand panel of this plot shows
the scatter of $\sigma_z$ vs.~$\sigma_g$ about a line of equality, while the
right-hand panel shows the difference in fitted means $(\mu_g-\mu_z)$ vs.~the
average GC $(g-z)$ color in each galaxy (from Peng et al.~2006a), again
compared to a line of equality. Both cases show excellent agreement between
the maximum-likelihood results for the two bandpasses. We conclude
that the measurements are internally consistent and that our
uncertainty estimates are reasonable.

\placefigure{\ref{fig:consistency}}

\begin{figure*}
\epsscale{1}
\plottwo{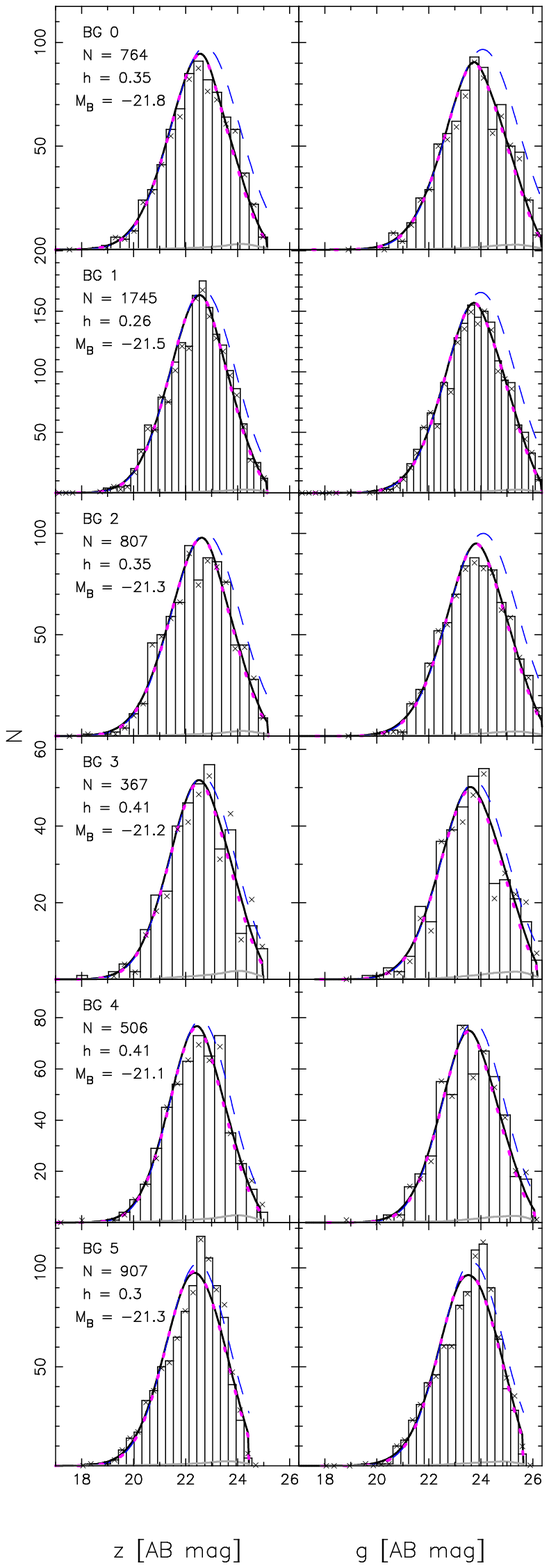}{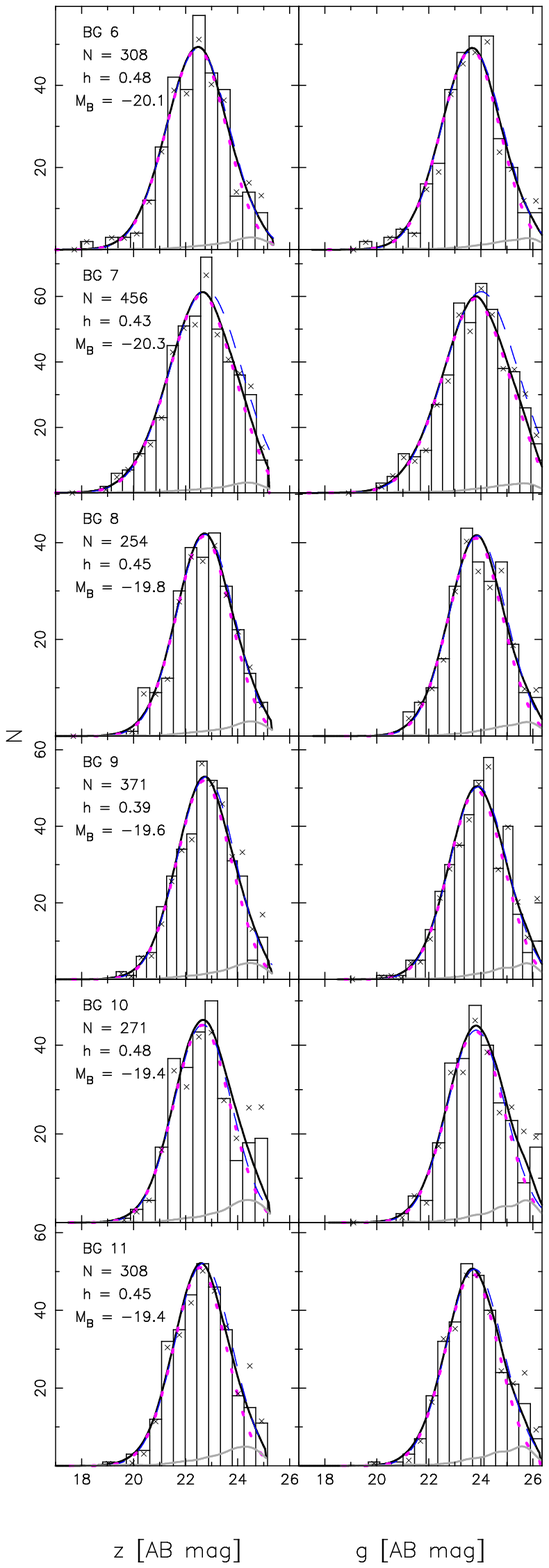}
\caption[]{Histograms and Gaussian fits to the GCLFs for our binned-galaxy
samples. For each sample, named BG$n$ with $n=0,\ldots,23$,  we
present the $z$-band and $g$-band GCLFs side by side. The identifier of the
galaxy bin is indicated in the upper left corner of the left panel, where we
also indicate the number $N$ of all sources in the histogram (as chosen by
requiring $p_{\rm GC}\ge 0.5$) and the bin-width $h$
used when constructing the histograms. In each panel we show 
the best fitting model (solid black curve), the intrinsic Gaussian
component (dashed curve), the Gaussian component
multiplied by the completeness fraction (dotted  curve)
and a kernel-density estimate of the expected contamination in the sample
(solid gray curve). The solid black curve is the sum of the solid gray and dotted curves.The
galaxy bins are ordered by decreasing mean apparent $B$-band luminosity
of the galaxies that went into the sample construction. Crosses in all panels
show the histograms that result when GC candidates are selected on the basis
of cuts in magnitude and half-light radius; see \S\ref{ssec:selec}.
The parameters of all fits are given in Table \ref{tab:galbins}.}
\label{fig:gaussbin}
\end{figure*}

Finally, we also fit Gaussians to our 24 ``binned'' GC samples, constructed by
combining the candidates in as many galaxies as necessary to reach net sample
sizes of at least 200 (see \S\ref{ssec:binned}). The IDs and total magnitudes
of the galaxies going into each of these bins are summarized in Table
\ref{tab:galbins}, along with the best-fit $z$- and $g$-band Gaussian
parameters for each binned GCLF and the best-fit parameters for the evolved
Schechter function discussed in \S\ref{sec:models} (see just below). In Figure
\ref{fig:gaussbin} we display the binned GCLFs in histogram format, along with
a number of curves representing the maximum-likelihood Gaussian fits. The
curves in every panel have exactly the same meaning as in the individual GCLF
fits of Figure \ref{fig:gaussfits}. We additionally show in this Figure
(as the crosses in each magnitude bin of each histogram) alternative GCLFs
for the binned-galaxy samples, obtained by defining GCs on the basis of
absolute magnitude and an upper limit on the half-light radius $R_h$
(\S\ref{ssec:selec}).\footnotemark 
\footnotetext{Note that these alternative GCLFs do not have exactly the same 
numbers of objects as the bar histograms corresponding to GC samples defined by
$p_{\rm GC}\ge 0.5$.}

\placetable{\ref{tab:galbins}}

\placefigure{\ref{fig:gaussbin}}

\begin{figure*}
{\epsscale{1} \plottwo{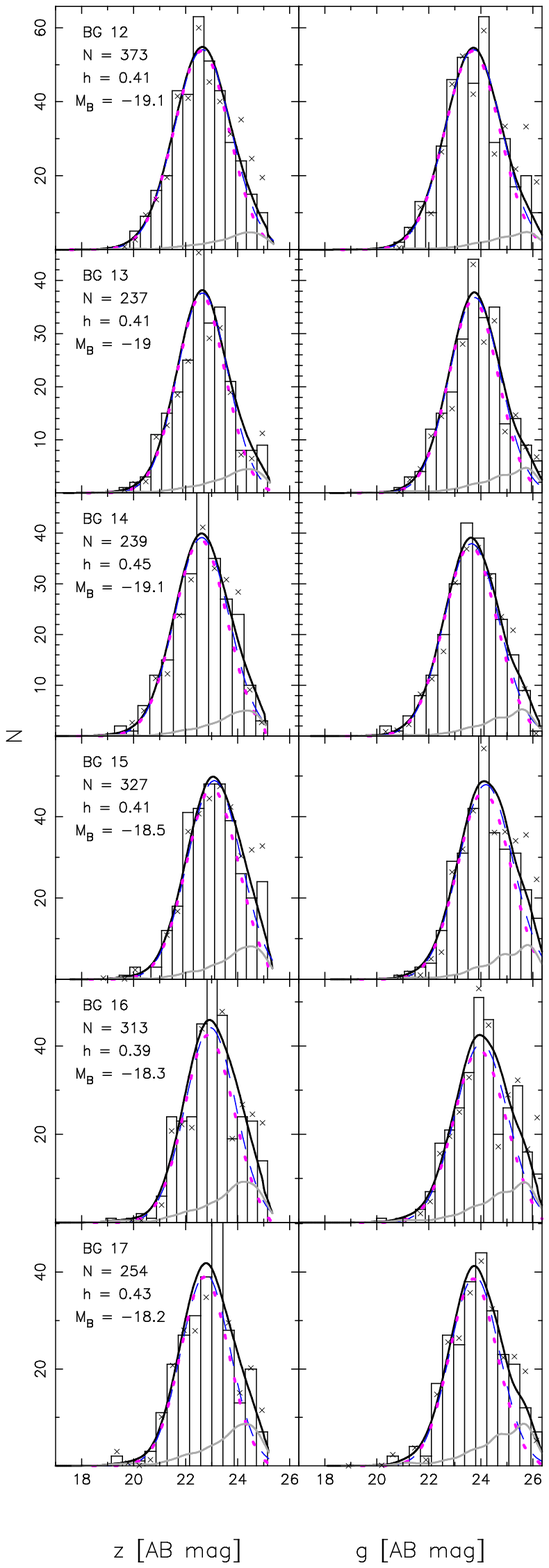}{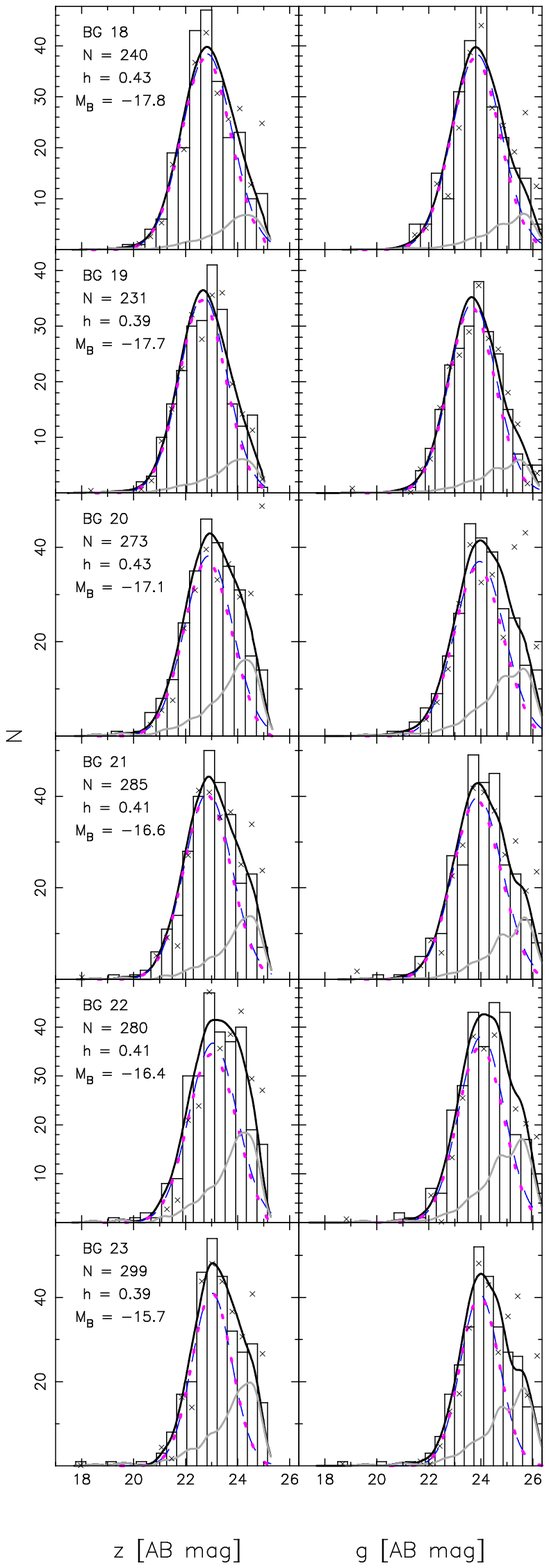}}\\[5mm]
\centerline{Fig. 6. --- {\it Continued}}
\end{figure*}

\begin{figure*}
\epsscale{1}
\plottwo{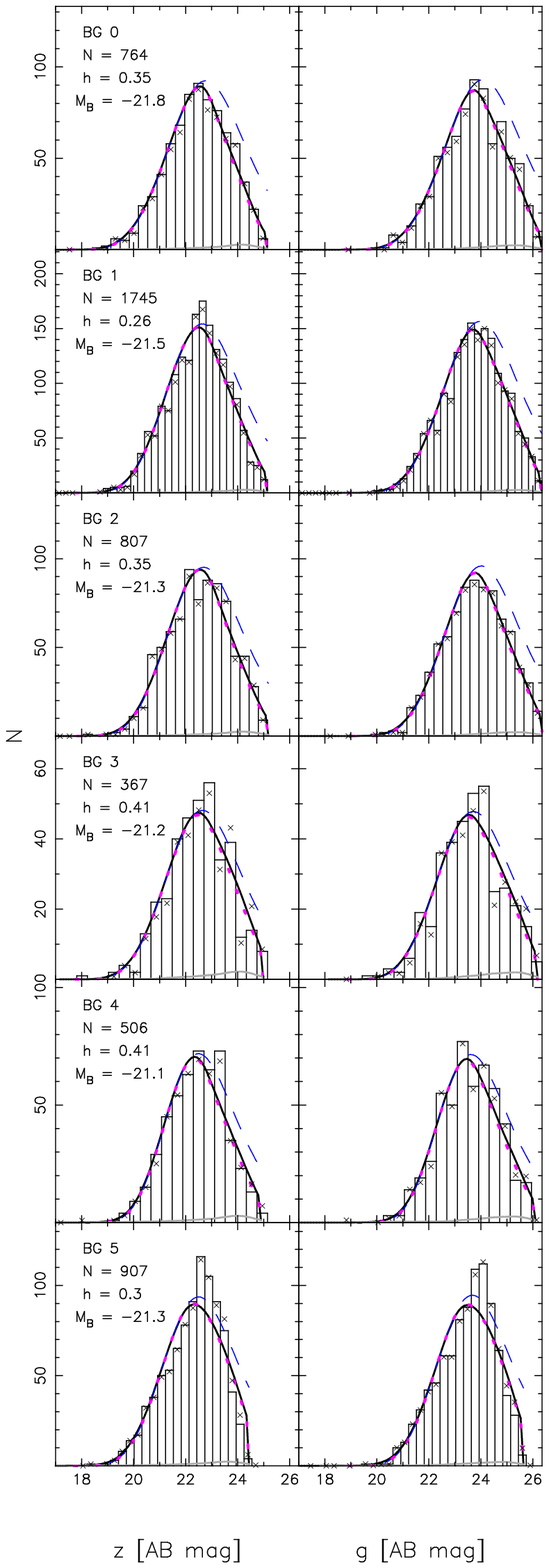}{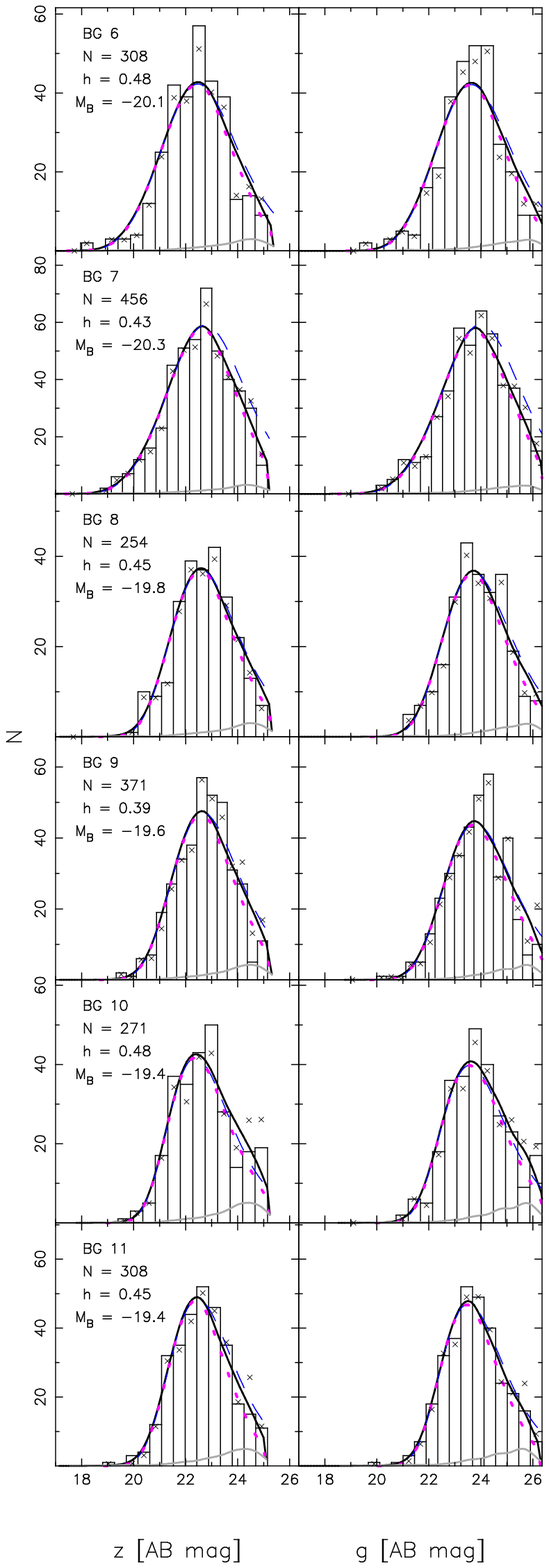}
\caption[]{Histograms and evolved Schechter function fits to the GCLFs for our
binned-galaxy samples. For each sample, named BG$n$ with $n=0,\ldots,23$,  we
present the $z$-band and $g$-band GCLFs side by side. The identifier of the
galaxy bin is indicated in the upper left corner of the left panel, where we
also indicate the number $N$ of all sources in the histogram (as chosen by
requiring $p_{\rm GC}\ge 0.5$) and the bin-width $h$ used
when constructing the histograms. In each panel we show 
the best fitting model (solid black curve), the intrinsic evolved Schechter
component (dashed curve), the evolved Schechter component
multiplied by the completeness fraction (dotted  curve)
and a kernel-density estimate of the expected contamination in the sample
(solid gray curve). The solid black curve is the sum of the solid gray and dotted curves. The
galaxy bins are ordered by decreasing mean apparent $B$-band luminosity
of the galaxies that went into the sample construction. Crosses in all panels
show the histograms that result when GC candidates are selected on the basis
of cuts in magnitude and half-light radius; see \S\ref{ssec:selec}.
The parameters of all fits are given in Table \ref{tab:galbins}.}
\label{fig:ESbin}
\end{figure*}

\begin{figure*}
{\epsscale{1}\plottwo{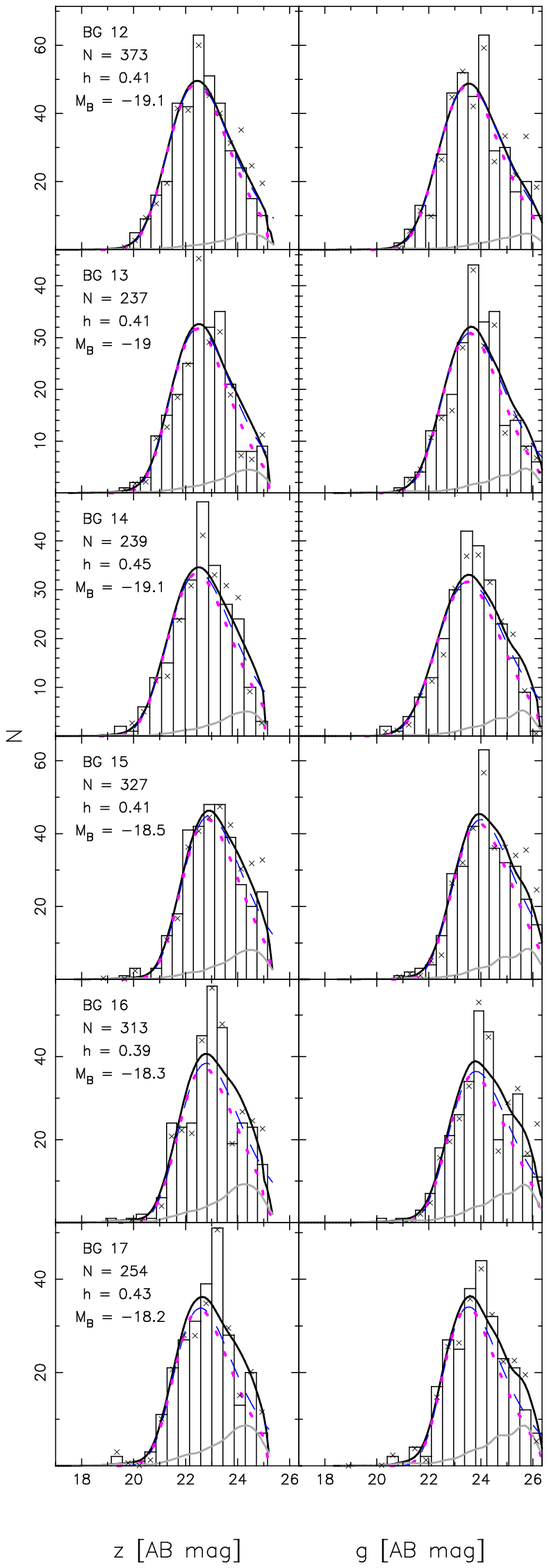}{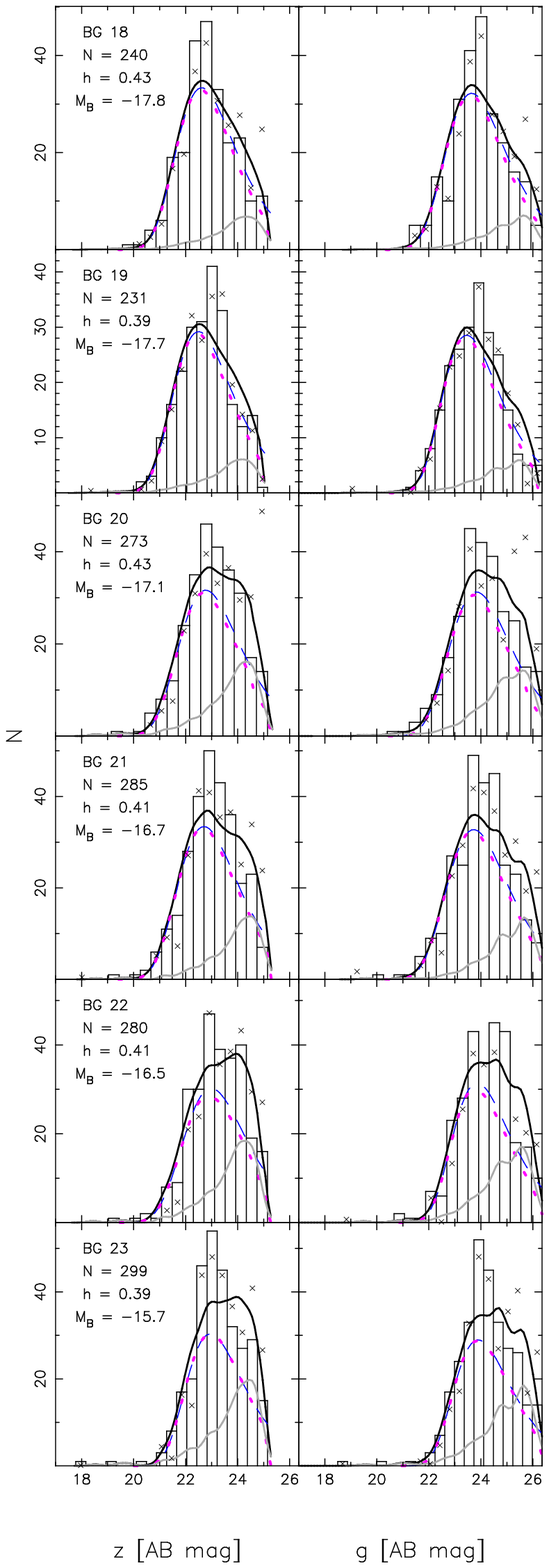}}\\[5mm]
\centerline{Fig. 7. --- {\it Continued}}
\end{figure*}

In \S\ref{sec:trends} below, we will compare GCLF systematics as a
function of galaxy properties for these binned samples vs.~the fits to
individual galaxies. We also note here, without showing further details,
that repeating the exercises of this Section using the samples of GC
candidates selected only by magnitude and $R_h$, rather than by a $p_{\rm GC}$
criterion, leads to results that are consistent in all ways with those we
present below.

\subsection{Fits of Evolved Schechter Functions}
\label{ssec:esfits}

We have performed fits of the evolved Schechter function in 
equation (\ref{eq:esmag})---or equivalently, the more transparent equation
(\ref{eq:esmass})---to the GCLFs of our individual galaxies and binned
samples. Here we discuss only the results of fitting the 24 binned GC
samples, as the results from fitting to all 89 galaxies separately lead to
similar conclusions.

In all these fits, we enforced the constraint that the fitted
(average) mass loss $\Delta$ be less than ten times 
the exponential cut-off mass scale $M_c$: $\Delta/M_c < 10$, or 
$(\delta-m_c)>-2.5$ in magnitude terms.
This was done because, as was discussed in 
\S\ref{ssec:esmod} (see Figure \ref{fig:schematic}), for such large ratios of
$\Delta$ to $M_c$ the evolved Schechter function has essentially attained a
universal limiting shape. The likelihood surface then becomes very flat for
any greater $\Delta/M_c$, and the fitting procedure has difficulty
converging if this parameter is allowed to vary to arbitrarily high values.
The majority of our evolved Schechter function fits do converge to
$\Delta/M_c$ values that satisfy our imposed constraint; in only one case does
the ``best-fit'' model have the limiting $\Delta/M_c=10$.

We show in Figure~\ref{fig:ESbin} the binned-sample GCLF histograms, along with
model curves analogous to those in Figure \ref{fig:gaussbin}. 
Again, then, the intrinsic evolved-Schechter model GCLF is
the long-dashed curve; this model multiplied by the marginalized
completeness function is the dotted curve; a kernel-density
estimate of the contaminant luminosty function is shown as the solid gray
curve; and the net best-fitting model (sum of dotted and solid gray curves) is drawn
as a solid black curve. Also as in Figure \ref{fig:gaussbin}, we use
crosses in Figure \ref{fig:ESbin} to show the GCLFs inferred in every galaxy
bin when we define GC samples by simple magnitude cuts and $R_h$ limits, 
rather than by using our $p_{\rm GC}$ probabilities.

Comparing Figure \ref{fig:ESbin} with Figure \ref{fig:gaussbin}, it is
apparent that an evolved Schechter function describes the GCLFs of bright
galaxies about as well as a Gaussian does. In some of the fainter galaxies
there is possibly a tendency for the Schechter function to overestimate the
relative number of faint GCs, but it is difficult to assess how serious this
might be. The worst
disagreements between the model fits and the data tend to occur in the very
faintest extents of the histograms for the handful of the faintest galaxy bins
at the end of Figure \ref{fig:ESbin}. Indeed, the largest discrepancies appear
at magnitudes where contaminants account for $\ga50\%$ of the total
observed population. Any impression of success or failure for {\it
any} model in these extreme regimes of the GCLFs must be tempered by the
realization that the fitting itself is something of a challenge under such
conditions. 

This is further illustrated by contrasting, in both Figures \ref{fig:ESbin}
and \ref{fig:gaussbin}, the GCLFs for cluster samples selected by magnitude
and $R_h$ only (crosses in the figures), to those for samples selected on the
basis of $p_{\rm GC}$ probabilities (bars). The former samples generally tend
to put more objects in the faintest GCLF bins, an effect particularly apparent
in the faintest galaxies. The low-mass end of the GCLF for faint galaxies
is thus not tightly constrained by our observations; there is a fundamental
uncertainty, due to contamination, that cannot be overcome by any selection
procedure. (Note that some of the more extreme discrepancies between
the different GCLF definitions---such as in the faintest magnitude bin of
BG~20---are due to the presence in some galaxies of a strong excess of
diffuse clusters that are classified as contaminants when using $p_{\rm GC}$
to construct the sample; see Peng et~al. 2006b). 
But it is still worth recalling, in this context, that the ``overabundance''
of low-mass clusters in the evolved Schechter function, vs.~a Gaussian, is in
fact a demonstrably better description of the Milky Way GCLF; see Figure
\ref{fig:MWGCLF}.

The fitted magnitude-equivalents $\delta$ and $m_c$ of the mass
scales $\Delta$ and $M_c$, in each of the $z$ and $g$ bands, are recorded for
each of our binned GCLFs in Table \ref{tab:galbins}.
In \S\ref{sec:trends} we discuss in detail the conversion of these to masses
and also consider dependences of $\Delta$ and $M_c$ on galaxy luminosity.

\begin{figure}
\epsscale{0.8}
\plotone{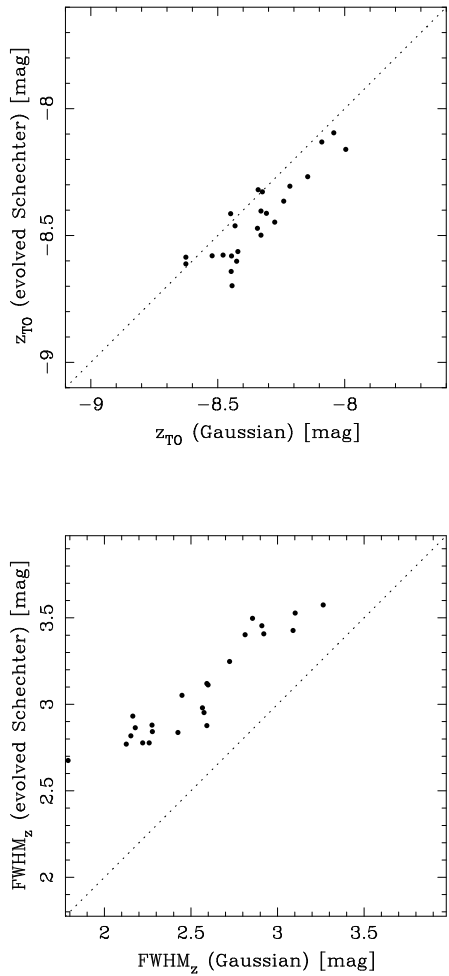}
\caption[]{{\it Upper panel:} Comparison of the absolute $z$-band magnitude of
the GCLF turnover, as inferred from the maximum-likelihood fitting of
intrinsic evolved Schechter functions versus that inferred 
from Gaussian fits.
{\it Lower panel:} Comparison of the FWHM of the intrinsic $z$-band GCLFs,
as returned by the Gaussian and evolved Schechter fits. Similar plots for
the fits to our $g$-band GCLFs look the same as these $z$-band results.}
\label{fig:gaussVSes}
\end{figure}

Just before looking at these issues, Figure \ref{fig:gaussVSes}
compares the turnover magnitudes and full-widths at 
half maximum (FWHM) for the binned $z$-band 
GCLFs as returned by the fits of evolved Schechter functions (see
eq.~[\ref{eq:es_magto}]), against the same quantities implied by our Gaussian
fits. For the turnovers, there is a slight offset, in that the fitted
Schechter functions tend to
peak at slightly brighter magnitudes (typical difference $\la\! 0.1$ mag,
corresponding to a turnover mass scale that is $<\! 10\%$ larger than
implied by the Gaussian fits). This is very similar to 
the offset in the two fitted turnover magnitudes for the Milky Way GCLF
in \S\ref{ssec:modcomp}. As we discussed there, the discrepancy is a
result of the intrinsic symmetry assumed in the Gaussian model, vs.~the
faint-end asymmetry built into the evolved Schechter function.

\placefigure{\ref{fig:gaussVSes}}

The FWHMs differ more substantially between the two functional forms,
with the evolved-Schechter fits being typically $\simeq\! 0.5$~mag
broader (or about 0.2~dex in terms of mass) than the Gaussian fits.
But this is again only to be expected from the asymmetry of the former
function vs.~the symmetry of the Gaussian. As was 
noted at the end of \S\ref{ssec:esmod}, the shape of the evolved
Schechter function is universally flat in terms of $dN/dM$ for low
GC masses, or universally $\propto\! 10^{-0.4 m}$ in terms of $dN/dm$
for magnitudes much fainter than the peak of the GCLF. As a result,
the faint side of the GCLF is always broader than any Gaussian, and so
if the two models give comparable descriptions of the bright
halves of all GCLFs, the FWHM of the evolved Schechter functions must
always be larger than those of the Gaussian fits. Moreover, for
very narrow observed GCLFs, fit by small Gaussian
$\sigma_m$ (primarily to reproduce the steepness of the bright side of the
GCLF, as discussed below), the evolved Schechter function fits are limited by
a minimum FWHM of $\simeq\! 2.66$~mag (\S\ref{ssec:esmod}), explaining the
tendency towards a plateau at the left side of the lower panel of Figure
\ref{fig:gaussVSes}.

\section{Trends Between and Within Galaxies}
\label{sec:trends}

Having fitted two different GCLF models to each of our individual
galaxies and binned samples, we now outline some systematic
variations in the properties of GC mass distributions indicated by this
work. First, we examine the dependence of GCLF parameters on host
galaxy luminosity; then---even though the ACSVCS data are not ideal
for this purpose---we look for any evidence of GCLF trends with radius
inside the two brightest Virgo galaxies, M49 (VCC 1226) and M87 (VCC 1316).

\subsection{Variations with Galaxy Luminosity}
\label{ssec:galmb}

\subsubsection{Gaussian Parameters}
\label{sssec:gausstrend}

\begin{figure}
%\epsscale{0.50}
\plotone{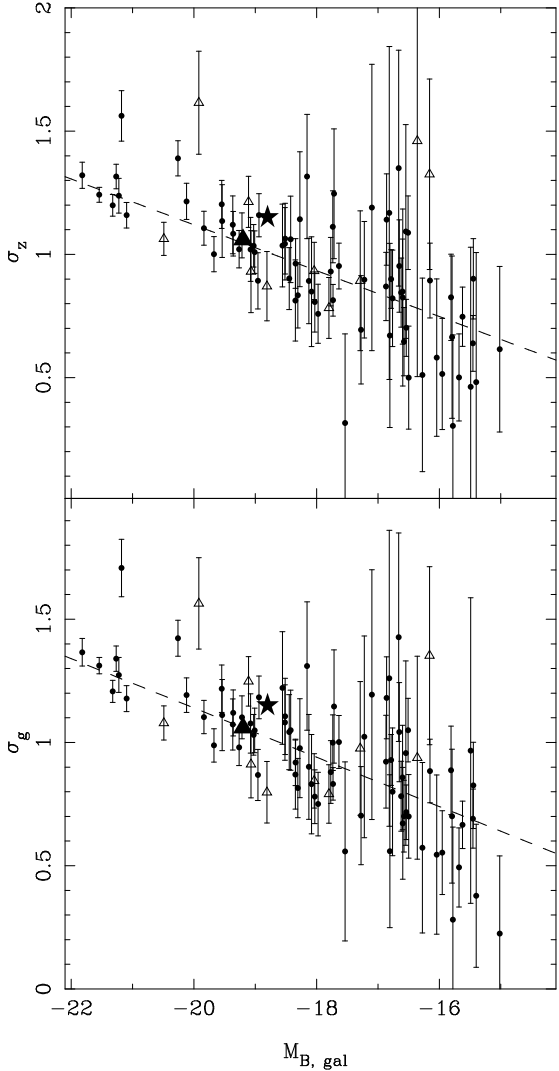}
\caption[]{{\it Top:} GCLF dispersion $\sigma_z$, inferred from Gaussian
fits to the $z$-band data, versus galaxy $M_{B, {\rm gal}}$. Filled symbols
are galaxies for which we have 
available SBF distances while open triangles represent galaxies for which we
do not and for which we have assumed a distance modulus of $(m-M)_0=31.1$. The
dashed line is the linear relation between $\sigma_z$ and $M_{B, {\rm gal}}$
in equation~(\ref{eq:sigmaz}).
{\it Bottom:} Same comparison, but for the intrinsic Gaussian dispersion
of the $g$-band GCLFs, $\sigma_g$. Dashed line is equation (\ref{eq:sigmag}).
In both panels the star shows values for the MW and the triangle represents
M31. The outliers at $M_{B, \rm gal} \simeq -21.2$ and  $M_{B, \rm gal} \simeq -19.9$ 
in both panels are VCC~798 and VCC~2095, galaxies
which have an excess of faint, diffuse star clusters
(Peng et~al. 2006b).}
\label{fig:sigma_B}
\end{figure}

Figure~\ref{fig:sigma_B} shows one of the main results of this paper:
GCLFs are narrower in lower-luminosity galaxies
(see also Jord\'an et al. 2006).

\placefigure{\ref{fig:sigma_B}}

The upper panel of
this figure plots the Gaussian dispersion
that best fits the $z$-band GCLF, as a function of
absolute galaxy magnitude $M_{B,{\rm gal}}$ for our 89 
individual galaxies. Filled circles represent
galaxies with measured (SBF) distance moduli, while open triangles
correspond to galaxies for which no distance modulus is available and
for which we assume  $(m-M)_0=31.1$ (consistent with the average  
Virgo distance modulus of Mei et~al. 2007) to
compute $M_{B,{\rm gal}}$.
The lower panel shows the analogous result for our $g$-band GC
data. The straight lines drawn in the panels are  convenient
linear characterizations of the $\sigma_m$--$M_{B,{\rm gal}}$ trends:
\begin{equation}
\sigma_z = (1.12\pm0.01) - (0.093 \pm 0.006)(M_{B, {\rm gal}}+20)
\label{eq:sigmaz} 
\end{equation}
and
\begin{equation}
\sigma_g = (1.14\pm0.01) - (0.100 \pm 0.007)(M_{B, {\rm gal}}+20) 
.
\label{eq:sigmag}
\end{equation}

While it has been reported before that there is a tendency
for the GCLFs in lower luminosity galaxies to show 
somewhat lower dispersions (e.g., Kundu \& Whitmore 2001a),
the homogeneity of our sample and analysis make this the most convincing
demonstration to date of the existence of a continous trend over a
factor of $\approx\! 400$ in galaxy luminosity.
It is particularly noteworthy that the fainter galaxies in our
sample---all of which are early type---have very modest $\sigma_m \la 1$,
values more usually associated with the GCLFs of late-type
galaxies. In fact we have also plotted on Figure \ref{fig:sigma_B} the
$V$-band GCLF dispersions (Harris 2001)
and absolute bulge luminosities of the
Milky Way (large filled star at
$M_{B, {\rm gal}}=-18.8$; de Vaucouleurs \& Pence 1978)
and M31 (large filled triangle at
$M_{B, {\rm gal}}=-19.2$; from Kent 1989, but assuming
a distance of 810 kpc).
Clearly these fall well in the midst of our new data, and thus the
correlation of $\sigma$ with $M_{B,{\rm gal}}$ would appear to be more
fundamental than the older view, that GCLF dispersions depend on
galaxy Hubble type (Harris 1991).

At this point it should be noted that the GCs in brighter galaxies are known
to have broader color distributions, and hence larger dispersions in
metallicity, than those in fainter galaxies (e.g., Peng et
al.~2006a). But cluster mass-to-light ratios, $\Upsilon$, are functions of
[Fe/H] in general, so there will be some galaxy-dependent spread in their
values. Since the
variance in an observed luminosity distribution is related to
that in the mass distribution, by the usual
$\sigma^2(\log\,L)=\sigma^2(\log\,M) + \sigma^2(\log\,\Upsilon)$, this then
suggests
the possibility that the trend we see in the GCLF $\sigma_z$ and $\sigma_g$
vs.~galaxy luminosity might result from systematics in
$\sigma(\log\,\Upsilon)$ vs.~$M_{B,{\rm gal}}$ on top of a more nearly constant
$\sigma(\log\,M)$.
In fact, this idea was recently invoked by Waters et al.~(2006) as a 
potential explanation for the fact that the $I$-band GCLF of M87 is 
broader than that of the Milky Way; and by Strader et al.~(2006) as a 
possible reason for the narrower composite GCLF of a subsample of ACSVCS 
dwarfs versus the GCLFs of Virgo giants. However, neither of those 
works checked these claims quantitatively. We have done so here
(see also Jord\'an et al.~2006), and we find that the explanation is not 
tenable.

As we will
discuss further in \S\ref{sssec:masstrend}, GC mass-to-light ratios in the
longer-wavelength $z$ band vary by less than $\pm10\%$ over the entire range
$-2\le {\rm [Fe/H]}\le 0$, which includes the large majority of clusters.
Thus $\sigma(\log\,\Upsilon_z) < 0.04$ no matter 
what the details of the GC metallicity distribution are---making for
an utterly negligible ``correction'' to the observed
$\sigma(\log\,L_z)=\sigma_z/2.5$ for all of our GCLFs.
In the shorter-wavelength $g$
band, mass-to-light ratios are more sensitive to cluster colors. But here the
close agreement of our $g$- and $z$-band GCLF dispersions shows immediately
that the former must be reflecting the properties of the GC mass
functions just as closely as the latter are. Indeed, more detailed
calculations, which include the observed specifics of the color distributions
in our galaxies (Peng et al.~2006a), confirm that the spread in expected GC
$\Upsilon_g$ values contributes $\sim\! 0.02$~mag to the total
observed GCLF dispersion---an amount well within the observational
uncertainties on $\sigma_g$ in the first place\footnote{We note that 
the median value of $(\sigma_g - \sigma_z)$ for our sample galaxies
is 0.02 mag.}. Thus, we proceed knowing that
the correlations between GCLF dispersion and galaxy luminosity that we are
discussing here are very accurate reflections of equivalent trends in the more
fundamental GC mass distributions.

Because of the symmetry assumed in the model, the trend of decreasing
Gaussian $\sigma_m$ in Figure \ref{fig:sigma_B} might appear to imply a
steepening of the GCLF
on both sides of the turnover mass. However, as we have already
discussed, if we take the more physically
based, evolved-Schechter function of equation (\ref{eq:esmag}) or
(\ref{eq:esmass}) to describe the distribution of GC masses, then
{\it all} GCLFs must have the same basic shape (and thus half-width)
for clusters fainter than about the turnover magnitude---in which case the
trends in Figure \ref{fig:sigma_B} can only be driven by systematics in
the bright side of the GCLF. Indeed, as was mentioned in 
\S\ref{sec:models} above (and discussed at length by, e.g., McLaughlin
\& Pudritz 1996), it has long been clear that power-law
representations of the GC mass function above the turnover mass
in the Milky Way and
M31 are significantly steeper than those in M87, M49, and other bright
ellipticals; there is no ``universal'' power-law slope for present-day
GC mass functions. 

\begin{figure}
\epsscale{0.8}
\plotone{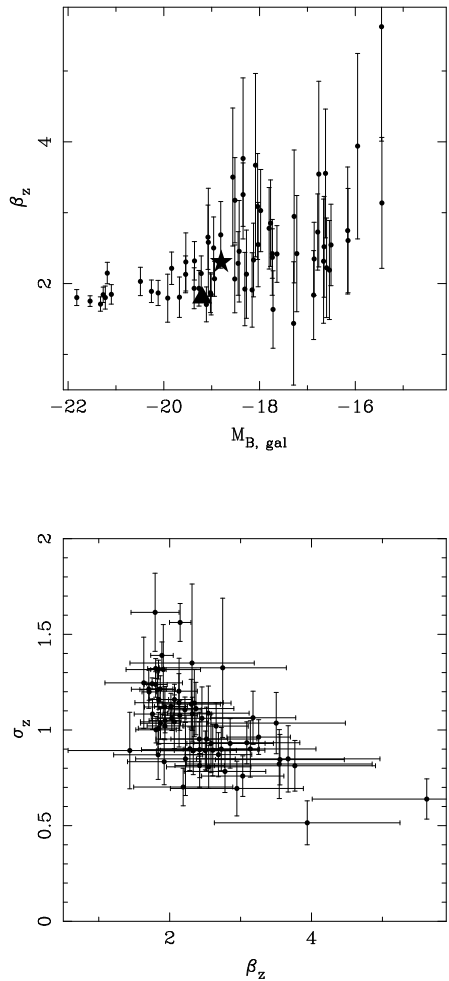}
\caption{{\it Upper panel} shows the slope $\beta_z$ of the power law that
best fits the $z$-band GCLF data for GC masses
$ 3\times10^5 \la (M/M_{\odot}) \la 2\times 10^6$,
against host galaxy absolute $B$ magnitude. 
The star and triangle show $\beta$ values for the Milky Way and M31 
respectively, 
measured in the same mass regime using the data from Harris (1996) 
and Reed et~al. (1994) assuming a $V$-band mass-to-light ratio
$M/L_V = 2\ M_\odot\,L_{V,\odot}^{-1}$. 
{\it Lower panel} shows the
correlation between this power-law index and the dispersion $\sigma_z$ in a
Gaussian representation of the GCLF. These graphs illustrate that the
systematic ``narrowing'' of the GCLF for decreasing galaxy luminosity, as seen
in Figs.~\ref{fig:sigma_B} and \ref{fig:binVSind}, is a real phenomenon
rather than an artifact of the Gaussian model: it shows up clearly as a
steepening of the (largely complete and relatively contamination-free)
high-mass end of observed GCLFs.
Corresponding plots for the $g$-band GCLFs are very similar to these.}
\label{fig:power}
\end{figure}

Given these points, we have also performed maximum-likelihood fits of
pure power-law mass distributions ($dN/dM \propto M^{-\beta}$; or,
in terms of magnitude, $dN/dm \propto 10^{0.4 (\beta-1) m}$) to GCs
between $\simeq\! 0.5$--2.5 mag brighter than the turnover magnitude in
the cluster samples of our individual galaxies. (Such subsamples are both
highly complete and essentially uncontaminated in all of our galaxies). 
The best-fit $\beta$ for the 66 galaxies in which we were able to measure
it are presented in Table~\ref{tab:beta}.
The results from fitting to the $z$- and $g$-band data are similar, and thus we
show only the former here, in the
upper panel of Figure~\ref{fig:power}. This confirms that the high-mass
end of the GCLF steepens systematically for decreasing galaxy luminosity,
independently of how the low-mass GC distribution behaves. 
In Figure~\ref{fig:power} we also plot  a star and triangle showing 
$\beta$ values for the Milky Way and M31 respectively, 
measured in the same mass regime using the data from Harris (1996) 
and Reed et~al. (1994) assuming a $V$-band mass-to-light ratio $M/L_V = 2$.
The lower panel of Figure \ref{fig:power} then plots the fitted
power-law exponent for high GC masses against the Gaussian GCLF
dispersion from Figure \ref{fig:sigma_B}, showing that there is indeed
a clear correlation between these two parameters in the sense that a narrower
Gaussian $\sigma$ reflects a steeper high-mass power-law $\beta$.

\placefigure{\ref{fig:power}}

The regularity and the high significance of the narrowing of the
GCLF as a function of galaxy luminosity---or the steepening of the
mass distribution above the classic turnover point---places a new and
stringent constraint on theories of the 
formation and evolution of the mass function of GCs. In one sense,
this is then on a par with the modest amount of variation seen in the
turnover mass. An important difference may be that the GCLF turnover
could be imprinted to some large extent by long-term dynamical evolution
(Fall \& Zhang 2001; though see, e.g., Vesperini 2000, 2001, and Vesperini \&
Zepf 2003 for a differing view, and \S\ref{ssec:evaporation} below for a
discussion of caveats). By contrast, most analyses agree that the
shape of $dN/dm$ above the turnover is
%, at least in large galaxies, 
largely resistant to change by dynamical processes (\S\ref{ssec:dynfric})---in
which case it seems most likely that
the systematic variations in Figures \ref{fig:sigma_B} and
\ref{fig:power} are reflecting a fundamental tendency to form massive
star clusters in greater {\it relative} numbers in more massive galaxies.

\begin{figure}
%\epsscale{0.5}
\plotone{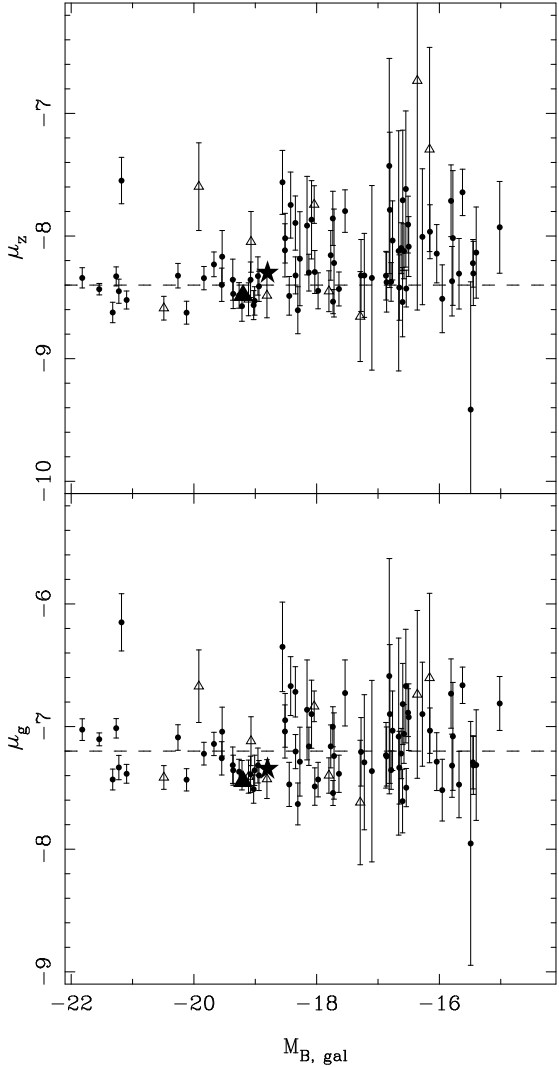}
\caption[]{{\it Top:} Absolute magnitude $\mu_z$ of the GCLF turnover, versus
$M_{B, {\rm gal}}$, inferred from
Gaussian fits to the $z$-band GCLFs. Filled symbols are galaxies for which
we have available SBF distances while open triangles represent galaxies for
which we do not and for which we have assumed a distance modulus of
$(m-M)_0=31.1$. The dashed line is at $\mu_z=-8.4$, the average for galaxies
brighter than $M_{B, {\rm gal}}=-18$. Fainter galaxies have turnover
magnitudes that tend to scatter fainter than this.
{\it Bottom:} Same comparison, for the $g$-band magnitude of the GCLF
turnover. Horizontal line is $\mu_g=-7.2$. In both panels the star shows
values for the MW and the triangle represents M31. 
The outliers at $M_{B, \rm gal} \simeq -21.2$ and  $M_{B, \rm gal} \simeq -19.9$ 
in both panels are VCC~798 and VCC~2095, galaxies
which have an excess of faint, diffuse star clusters
(Peng et~al. 2006b).
}
\label{fig:mu_B}
\end{figure}

Moving now to the GCLF turnover magnitude, in Figure~\ref{fig:mu_B} we
show the absolute $\mu_z$ and $\mu_g$ as functions of host galaxy
absolute magnitude $M_{B,{\rm gal}}$. In both panels of this figure,
horizontal lines are drawn at the levels of the typical turnovers in large
ellipticals: excluding VCC~798, which has an anomalously large excess of
faint, diffuse star clusters (Peng et al.~2006b), the average Gaussian
turnovers for ACSVCS galaxies with $M_{B,{\rm gal}}<-18$ are
\begin{equation}
\begin{array}{ccl}
\langle \mu_z \rangle & = & -8.4 \pm 0.2  \\
\langle \mu_g \rangle & = & -7.2 \pm 0.2  \ .
\label{eq:brightmu}
\end{array}
\qquad  (M_{B,{\rm gal}} < -18)
\end{equation}
The turnover in the Milky Way
is shown as a large filled star and that in M31 is represented by a large
filled triangle, as in Figure \ref{fig:sigma_B}. We estimated these turnovers
from the $V$-band values given in Table 13 of Harris (2001), by applying
$(g-V)$ and  $(V-z)$ colors calculated for 13-Gyr old clusters with
${\rm [Fe/H]}=-1.4$ for the Milky Way (Harris 2001) and  ${\rm [Fe/H]}=-1.2$
for M31 (Barmby et~al. 2000)
using the PEGASE population-synthesis model (Fioc \& Rocca-Volmerange 1997).

\placefigure{\ref{fig:mu_B}}

The $z$-band turnovers in the upper panel of Figure \ref{fig:mu_B}
show a tendency to scatter systematically above (fainter than) the
bright-galaxy value for systems with $M_{B,{\rm gal}}\ga -18$, but there are
no such systematics in the $g$-band turnovers in the lower panel. Interpreting
these results is most easily done in terms of equivalent turnover {\it
mass} scales, and thus we defer further discussion to \S\ref{sssec:masstrend},
where we use the PEGASE model to convert all of our GCLF
parameters to their mass equivalents.
We note here, however, that the near constancy of
$\mu_g$ in Figure \ref{fig:mu_B} is equivalent to the well known
``universality'' of the GCLF turnover in the more commonly used $V$ band
(since our $g$ is the HST F475W filter, which is close to
standard $V$).

\begin{figure}
\epsscale{1}
\plotone{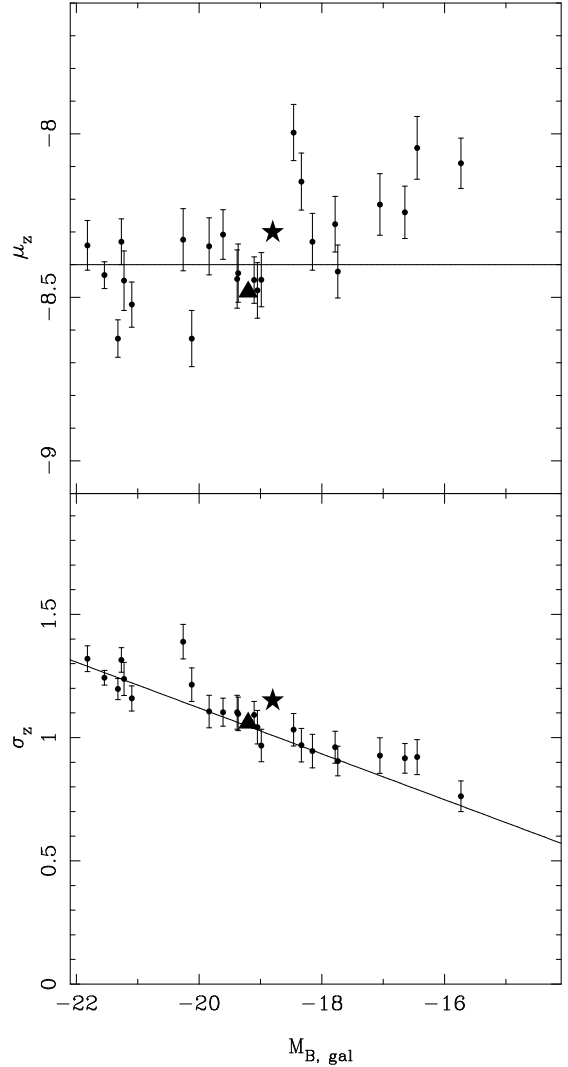}
\caption[]{$z$-band GCLF turnover magnitude ({\it upper panel}) and dispersion
({\it lower panel}) inferred from the Gaussian fits to the binned-galaxy
samples in Fig.~\ref{fig:gaussbin}. The horizontal line in the upper panel is
the same $\mu_z=-8.4$ that characterizes the bright galaxies in the upper
panel of Fig.~\ref{fig:mu_B}. The line in the lower panel is the fit of
equation (\ref{eq:sigmaz}) to the Gaussian dispersions obtained from fitting
all 89 of our galaxies individually (cf.~Fig.~\ref{fig:sigma_B}).
In both panels the star shows values for the MW and the triangle represents
M31. Note that VCC~798, the bright outlier galaxy in Figs.~\ref{fig:sigma_B}
and \ref{fig:mu_B}, has been excluded from our ``binned'' samples due to its
excess of faint, diffuse star clusters.}
\label{fig:binVSind}
\end{figure}

Before discussing masses in detail, we plot in Figure
\ref{fig:binVSind} the Gaussian means and dispersions of the $z$-band GCLFs in
our 24 binned samples, vs.~the average absolute magnitude of the galaxies in
each bin (see Table \ref{tab:galbins}). The straight lines in each panel are
just those from the upper panels of Figs.~\ref{fig:mu_B} and
\ref{fig:sigma_B}, characterizing the fits to all 89 individual galaxies.
This comparison shows that the results 
from our single- and binned-galaxy GC samples are completely consistent, so
that our binning process has served---as intended---to decrease the
scatter in the observed behavior of $\mu$ and $\sigma$ at low galaxy
luminosities. It also confirms the results of our simulations in
\S\ref{ssec:bias} above, which showed that our maximum-likelihood model
fitting is not significantly biased by size-of-sample effects.
A plot like Figure \ref{fig:binVSind}, but using our Gaussian fits to
the individual and binned $g$-band GCLFs, leads to the same conclusions.

\placefigure{\ref{fig:binVSind}}

\subsubsection{Mass Scales}
\label{sssec:masstrend}

To better understand the GCLF trends discussed above, and to mesh the
Gaussian-based results with those from fits of the more physically motivated
evolved-Schechter function, it is advantageous to work in terms of GC
mass, rather than $z$ and $g$ magnitudes. To make this switch, we rely on
population-synthesis model calculations of $(g-z)$ colors and $g$- and
$z$-band mass-to-light ratios as functions of metallicity for ``simple''
(single-burst) stellar populations.

\begin{figure}
%\epsscale{0.6}
\plotone{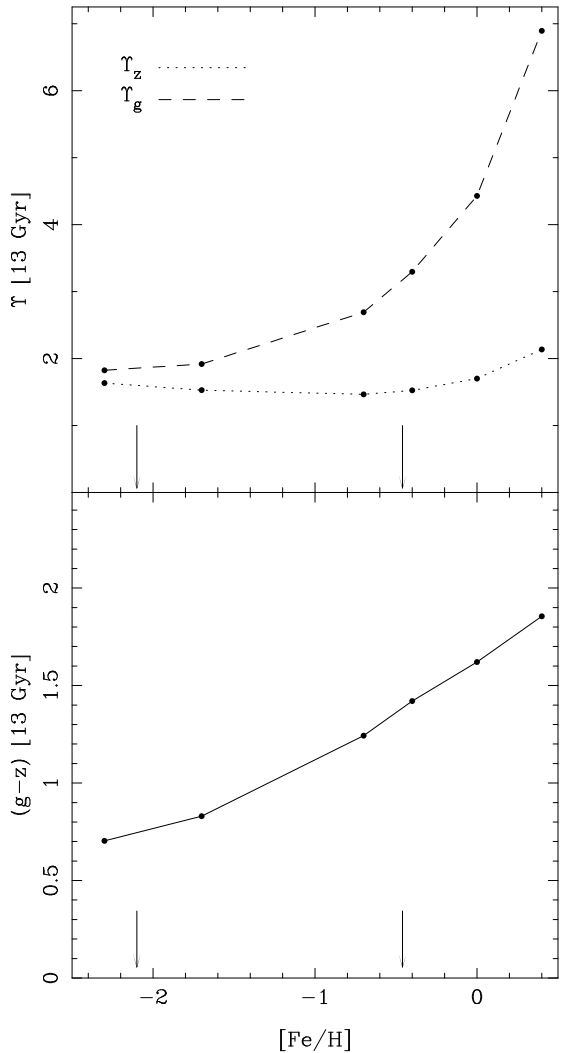}
\caption[]{Predicted mass-to-light ratio $\Upsilon$ ({\it upper Panel}) in solar units in the
$z$ (dotted line) and $g$ (dashed line) bands, and $(g-z)$ color ({\it lower panel}),
all as functions of metallicity for a 13-Gyr old simple stellar population
according to the PEGASE population-synthesis model (Fioc \& Rocca-Volmerange
1997) assuming a Kennicutt (1983) stellar IMF. The arrows in both panels indicate the minimum
and maximum average GC [Fe/H] in the ACSVCS galaxies, as inferred from their
mean $(g-z)$ colors (see Table \ref{tab:Ups}).} 
\label{fig:upsilon}
\end{figure}

The model we use is version 2.0 of the
PEGASE code (Fioc \& Rocca-Volmerange 1997), which we have run by inputting
the stellar initial mass function of Kennicutt (1983) to compute cluster masses
and $g$ and $z$ luminosities as functions of age for several fixed
values of [Fe/H]. The results, at an assumed uniform GC age of 13 Gyr, are
illustrated in Figure \ref{fig:upsilon}, which plots the mass-to-light ratios
$\Upsilon_g$ and $\Upsilon_z$ in solar units and the $(g-z)$ color against
[Fe/H]. Given the average $(g-z)$ of the GCs in any of our
galaxies (from Peng et al.~2006a), we interpolate on these PEGASE model
curves to estimate average $g$ and $z$ mass-to-light ratios. Table
\ref{tab:Ups} lists the mean GC color in each galaxy and the $M/L$ values we
have derived.

\placefigure{\ref{fig:upsilon}}

\placetable{\ref{tab:Ups}}

It is clear from Figure \ref{fig:upsilon} and Table \ref{tab:Ups} that the
$z$-band mass-to-light ratio varies by only a modest amount for
most GCs in our samples: we generally have
$0.8\la \langle (g-z) \rangle \la 1.2$
in these cluster systems, and thus
$1.45\la \Upsilon_z\la 1.55\ M_\odot\,L_\odot^{-1}$. A $z$-band luminosity is
therefore a very good proxy for total cluster mass. By contrast, over the 
same range of GC color or metallicity, the $g$-band mass-to-light ratio
increases monotonically from $\Upsilon_g\simeq1.9\ M_\odot\,L_\odot^{-1}$ to
$\Upsilon_g\simeq2.7\ M_\odot\,L_\odot^{-1}$. Note that if any of our GCs
were much younger than 13 Gyr, then the numerical values of these
mass-to-light ratios would all be lower (by $\sim\! 30\%$--40\% at an
age of 8 Gyr, for example), but the basic constancy of $\Upsilon_z$ and the
systematic increase of $\Upsilon_g$ for redder/more metal-rich GC systems
would remain.

This has immediate implications for our plots of the GCLF turnover magnitudes
in Figures \ref{fig:mu_B} and \ref{fig:binVSind} above. In particular, the
GCs are systematically bluer, on average, in lower-luminosity galaxies
(e.g., Peng et al.~2006a; see also Table \ref{tab:Ups}).
Assuming that this reflects a correlation between average cluster
metallicity and galaxy luminosity (rather than one between cluster age and
$M_{B,{\rm gal}}$), the typical $\Upsilon_g$ must be somewhat lower for GCs in
faint galaxies than in bright galaxies, while $\Upsilon_z$ is
essentially the same. The fact that the Gaussian GC $\mu_z$ scatters slightly
faintward towards fainter $M_{B,{\rm gal}}$ should then reflect a modest
downward scatter in the turnover {\it mass} scale. But in the $g$ band,
this would be balanced to at least some extent by the decrease in
mass-to-light ratio, and $\mu_g$ should stay more steady as a
function of $M_{B,{\rm gal}}$.

\begin{figure}
\epsscale{1}
\plotone{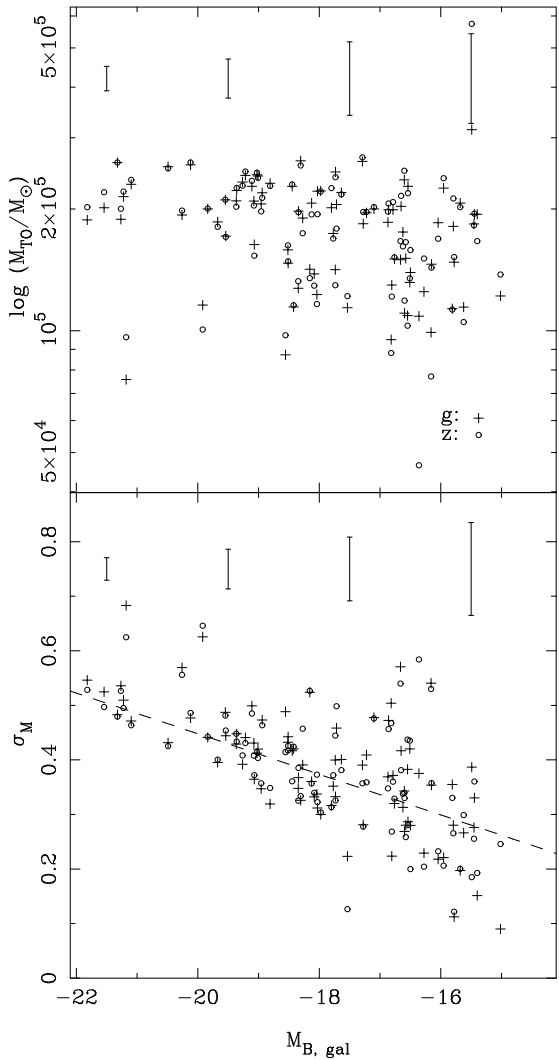}
\caption[]{Turnover {\it mass} $M_{\rm TO}$ ({\it upper panel}) and dispersion
of logarithmic mass ({\it lower panel}) implied by our Gaussian fits to the
$g$- and $z$-band GCLFs of individual galaxies. The turnover masses are
obtained from the magnitudes $\mu_g$ and $\mu_z$ by applying the
PEGASE model mass-to-light ratios summarized in Fig.~\ref{fig:upsilon}
and Table \ref{tab:Ups}. The dispersion in logarithmic mass is 
$\sigma_M=\sigma_g/2.5$ or $\sigma_z/2.5$. In both plots, results from the
$g$-band data are represented by circles, and results from the $z$-band by
crosses. In the upper part of both panels we show the typical behaviour of
error bars as a function of $M_{B, {\rm gal}}$.
The outlying points at $M_B \simeq -21.2$ in both panels correspond to
VCC~798, a galaxy which has a strong excess of faint, diffuse star clusters
(Peng et~al. 2006b).}
\label{fig:MTO_GZ}
\end{figure}

This interpretation of the situation is confirmed in Figure
\ref{fig:MTO_GZ}, where in the upper panel we plot the Gaussian turnover
masses, derived from the
$z$- and $g$-band GCLF fits as just described, vs.~parent galaxy absolute
magnitude. The average turnover magnitudes in equation (\ref{eq:brightmu}) and
the typical GC mass-to-light ratios in Table \ref{tab:Ups} together imply an
average turnover mass of
\begin{equation}
\langle M_{\rm TO} \rangle = (2.2 \pm 0.4) \times 10^5\ M_\odot
\label{eq:brightmTO}
\end{equation}
for the brightest ACSVCS galaxies with $M_{B,{\rm gal}} < -18$ (here we
have taken the absolute magnitude of the sun to be 4.51 in 
the $z$-band and 5.10 in $g$).
The consistency in most systems between the turnover masses estimated
from the two bandpasses shows that, indeed, for $M_{B,{\rm gal}}\ga -18$,
there is an overall tendency to find more GC systems with turnover masses
somewhat below the average for giant
ellipticals, by as much as a factor of 2 in some cases.
It also implies that the dependence of GC $\langle (g-z) \rangle$ on galaxy
luminosity does primarily reflect metallicity variations, since if GCs
had very similar metallicities but much younger ages in fainter galaxies,
the $z$- and $g$-band estimates of $M_{\rm TO}$ would differ by as much as the
fitted turnover magnitudes in \S\ref{sssec:gausstrend}.

\placefigure{\ref{fig:MTO_GZ}}

\begin{figure*}
\epsscale{1}
\plottwo{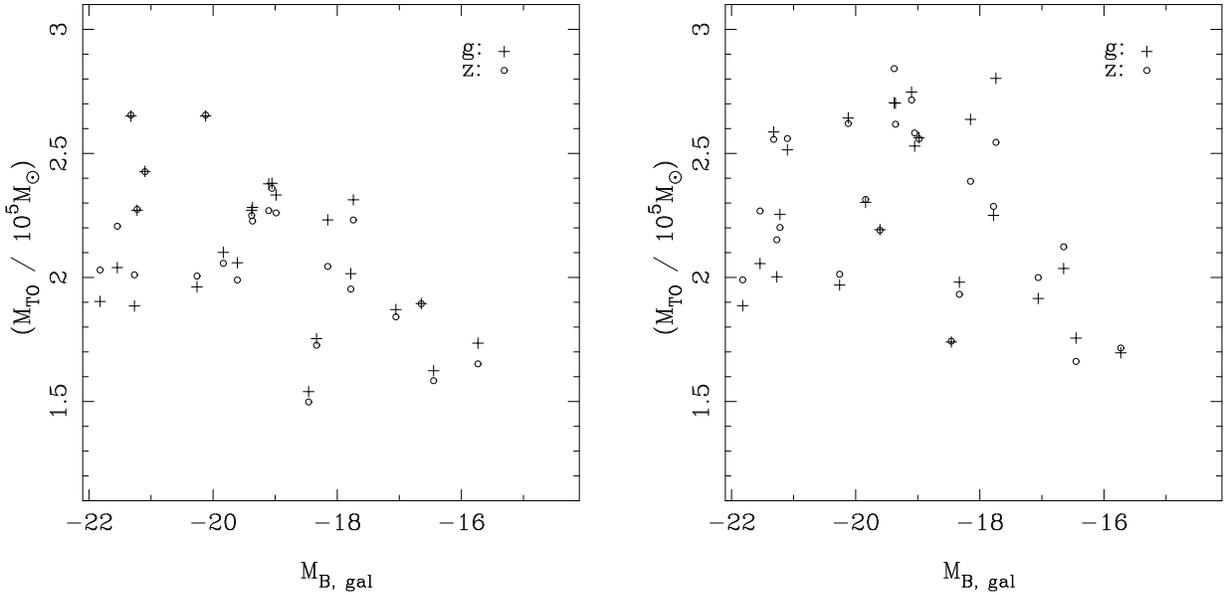}{f15b.eps}
\caption[]{{\it Left:} GCLF turnover mass $M_{\rm TO}$, inferred from the
maximum-likelihood estimate of Gaussian $\mu$, as a function of
$M_{B, {\rm gal}}$ for binned-galaxy samples.
$M_{\rm TO}$ has been inferred from $\mu$ using the PEGASE model, as
summarized in Fig.~\ref{fig:upsilon} and Table \ref{tab:Ups}. Crosses indicate
values of $M_{\rm TO}$ obtained from $\mu_g$, while circles are the values of
$M_{\rm TO}$ obtained from $\mu_z$.
({\it Right}) GCLF turnover mass $M_{\rm TO}$ versus galaxy absolute
magnitude, as inferred from the fits of evolved Schechter function to the
binned-galaxy samples (using eq.~[\ref{eq:es_magto}] and PEGASE model
mass-to-light ratios). The trend of decreasing $M_{\rm TO}$ with decreasing
galaxy mass is as in the left panel, showing again that the choice of
functional form does not affect our results. Note that VCC~798, responsible
for the outlying points at bright $M_{B,{\rm gal}}\simeq-21.2$ in
Fig.~\ref{fig:MTO_GZ}, has been excluded from our ``binned'' samples due to
its excess of faint, diffuse star clusters.}
\label{fig:mto_bd}
\end{figure*}

\placefigure{\ref{fig:mto_bd}}

For completeness, in the lower panel of Figure \ref{fig:MTO_GZ}, we show the
$g$- and $z$-band based estimates of the Gaussian dispersion of logarithmic GC
masses. Since $\sigma_M$ does not depend on the cluster mass-to-light
ratio, but is just the magnitude dispersion divided by 2.5, this plot is
completely equivalent to Figure \ref{fig:sigma_B}. Thus we have also drawn in
equation (\ref{eq:sigmaz}) above, multiplied by 0.4.

An interesting corollary to all of this is that the reliability of the GCLF
as a distance indicator would appear to be somewhat bandpass-dependent, at
least when applied to sub-$L_*$ galaxies with $M_{B,{\rm gal}}\ga -19$. We
have just argued that the near-universality of the turnover magnitude in the
$g$-band---and thus in the very closely related $V$ band---is at some
level the fortuitive consequence of quantitatively similar
decreases in both the turnover mass and the typical GC mass-to-light ratio in
smaller galaxies. At longer wavelengths, however, mass-to-light ratios are not
so sensitive to GC metallicity, and variations in the turnover mass carry over
more directly into variations in turnover magnitude. We will explore this
issue in more detail in future work. However,
any such pragmatic concerns about the precision of the
GCLF peak magnitude as a standard candle should not detract from the main
point of physical interest here: although the differences in
GCLF turnover mass that we find are real, they are nevertheless relatively
modest. While the galaxies in the ACSVCS range over a factor of $\simeq\!400$
in luminosity, $M_{\rm TO}$ never falls more than $\simeq\! 30\%$--40\% 
away from the (Gaussian) average of $2.2\times10^5\,M_\odot$ for the giant
ellipticals.

In the left panel of Figure~\ref{fig:mto_bd} we show the turnover masses
derived from the Gaussian GCLF means for our binned-galaxy GC samples. 
This again highlights the tendency to slightly less massive GCLF
peaks, on average, in lower-luminosity galaxies. In the right panel of this
figure we also show $M_{\rm TO}$  as derived from our fits of an
evolved Schechter function to the same GCLFs (see
eqs.~[\ref{eq:es_magto}] and [\ref{eq:es_massto}]). The close similarity of
the two graphs in Figure \ref{fig:mto_bd} is entirely in keeping with the
slight average offset between the Gaussian and extended-Schechter
turnover magnitudes in Figure \ref{fig:gaussVSes} above. It also illustrates
that our main results are not overly dependent on the particular choice of
model to fit the GCLFs.

\begin{figure}
\epsscale{1}
\plotone{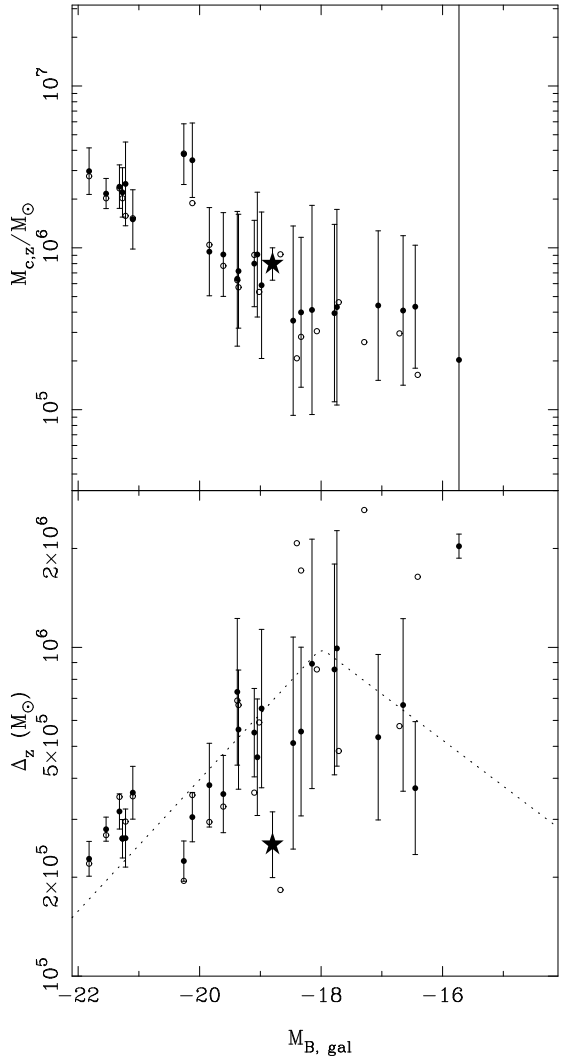}
\caption[]{{\it Top:} Cut-off mass scale $M_{c,z}$, inferred from our fits of
evolved Schechter functions to the $z$-band GCLFs of binned-galaxy cluster
samples, versus $M_{B, {\rm gal}}$.
{\it Bottom:} Average mass loss per globular cluster, $\Delta_{z}$,
versus $M_{B,{\rm gal}}$, from fits of evolved Schechter functions
to the $z$-band GCLFs of the binned-galaxy samples. 
In both panels, filled points are for fits to GC samples defined by the
criterion $p_{\rm GC}\ge 0.5$ while open symbols are for samples constructed
using only cuts in magnitude and half-light radius (see \S\ref{ssec:selec}).
It is clear that selection effects can be
safely ignored when investigating broad trends in $M_c$ and $\Delta$.
The stars in the two panels show the values of $M_c$ and $\Delta$ from fits
to the Milky Way GCLF (eq.~[\ref{eq:MWesmass}]). The dotted lines in the
lower panel show the rough expected scaling of $\Delta$ vs.~$M_{B,{\rm gal}}$
indicated by equations (\ref{eq:deltabright}) and  (\ref{eq:deltafaint})
in \S\ref{ssec:evaporation}. In order to show the scaling we have arbitrarily
assumed that $\Delta = 2.5\times 10^5 M_\odot$ at $M_{B, \rm gal} = -21$.
}
\label{fig:MB_ES}
\end{figure}

Last, in Figure \ref{fig:MB_ES} we show the GC mass scales $M_c$ (the
high-mass exponential cut-off) and $\Delta$ (interpreted as the average mass
lost per GC by evaporation) for our fits of evolved Schechter functions to the
binned-galaxy GCLFs, as inferred from their magnitude equivalents $m_c$ and
$\delta$ in Table \ref{tab:galbins}. The upper panel of the figure 
first plots $M_c$ vs.~$M_{B,{\rm gal}}$, using solid points to represent fits
to GC samples selected on the basis of our probabilities $p_{\rm GC}$ and open
symbols for fits to samples defined only by cuts on magnitude and GC effective
radius (see \S\ref{ssec:selec}). 
There is a clear, systematic decrease of $M_c$ with decreasing galaxy
luminosity. In terms of the structure of the mass function
(eq.~[\ref{eq:esmass}]), this corresponds to a steeper fall-off in the
frequency of GCs more massive than the turnover point.
It is therefore equivalent to our findings
in Figures \ref{fig:sigma_B} and \ref{fig:power} that the Gaussian $\sigma$
is narrower, and the high-mass power law $\beta$ steeper, for the GCLFs in
fainter galaxies. As we discuss in \S\ref{sec:disc}, features such
as this likely reflect the initial condition of the GC mass distribution.
Thus, if GC systems were indeed born with Schechter-like mass functions, it
would seem that the ``truncation'' mass scale $M_c$ was higher in
larger galaxies right from the point of cluster formation.

\placefigure{\ref{fig:MB_ES}}

The graph of $\Delta$ vs.~galaxy luminosity in the
lower panel of Figure \ref{fig:MB_ES} shows, first, that it is roughly
comparable to (though slightly larger than) the GCLF turnover mass in
general. This is certainly not unexpected, given the characteristics of the
model itself (see the discussion in \S\ref{ssec:esmod}). In physical terms,
though, if the model is taken at face value, the correspondence
reflects the fundamental role that evaporation is assumed to play in defining
any turnover point at all (see our discussion in \S\ref{sec:models},
and the more detailed exposition of Fall \& Zhang 2001). Beyond this, our fits
imply that there is a 
tendency for $\Delta$ to increase as galaxy luminosity decreases, but this is
not a particularly regular trend. All in all, there appears to be a fairly
narrow range of GC mass loss, $\Delta \approx 2-10 \times 10^5 M_{\odot}$,
required to account for our GCLF observations over a large range of galaxy
luminosity.

Note that several of the faintest galaxy bins in Figure
\ref{fig:MB_ES} have $\Delta/M_c \approx 2$, to be compared with
$\Delta/M_c \sim 0.1$ for the brightest systems. This reflects once again
the systematic narrowing of the GCLF, due to the steepening of $dN/dM$ for
high cluster masses, in fainter galaxies.

In \S\ref{sec:disc} we will further discuss the variations of $M_{\rm TO}$,
$M_C$, and $\Delta$ with galaxy luminosity, and how they relate to questions
of GC formation and dynamical evolution.

\subsection{GCLF Turnovers in the Faintest Galaxies}
\label{ssec:dwarfs}

In all of our galaxies there is evidence for the
presence of a peak in the GCLF. Recently, van den Bergh (2006)
claimed that the combined GCLF for a sample of local dwarf galaxies fainter
than $M_{V, {\rm gal}} > -16$ does {\it not} show a turnover, but
continues to increase
to GC masses as low as $\approx\! 10^4\,M_\odot$. These galaxy luminosities
translate to $B$-band
$M_{B, {\rm gal}} \ga -15.2$, which is essentially
the magnitude limit of our ACSVCS sample. 

Even though we do not probe down to the galaxy luminosities
where van den Bergh (2006) claims a drastically different GCLF behavior,
it is nonetheless worth noting that the turnover mass in our faintest galaxies
is still fairly close to the ``canonical''
$M_{\rm TO}\sim 2\times10^5\,M_\odot$. There is no hint of any
systematics that would cause $M_{\rm TO}$ to fall to $10^4\,M_\odot$ or less
in galaxies just 1 mag fainter than the smallest systems observed here (e.g.,
see Fig.~\ref{fig:MTO_GZ}). It is thus likely relevant that the results of van
den Bergh (2006) are based mostly on data from Sharina et~al. (2005), who
do not account for any potential contamination in their lists of candidate GCs
in the local dwarfs. Any GCLF derived from these data must therefore be
regarded as quite uncertain, at the faint end especially. Spectroscopic 
confirmation of the Sharina et al.~GC candidates is required.

\subsection{Variations with Galactocentric Radius}
\label{ssec:gradius}

To achieve a fuller understanding of the GCLF, and in particular the competing
influences of cluster formation and dynamical evolution on it, we would like
to know how it might vary in form as a function of position in its parent
galaxy. It has long been understood that the turnover of the Milky Way GCLF is
essentially invariant with Galactocentric radius (e.g., Harris 2001),
and multiple studies of the M87 GCLF have concluded that its overall shape is
basically the same from the center of the galaxy out to several 
effective radii (McLaughlin, Harris, \& Hanes 1994; Harris, Harris, \&
McLaughlin 1998; Kundu et al.~1999). Beyond this, however, little is known
about the generic situation in most galaxies.

For the most part, our data are not well-suited to address this question, due
to the 
small field of view of the ACS. However, we are afforded serendipitously
long baselines of galactocentric radius in M87 and M49, by the inclusion in the
ACSVCS of a number of low-luminosity galaxies that are projected close to
each of these large galaxies. We refer to these galaxies as ``companions,''
even though they might not be physically associated with their ``hosts.''
The majority of the GCs observed in the fields of these smaller systems
belong to the giants. While each companion does have some GCs
of its own, their numbers will be reduced to negligible levels,
compared to the M87 or M49 globulars, outside some sufficiently
large radius in the low-luminosity galaxy.
Thus, we take our original GC samples for the companions present in the survey
and consider 
only those cluster candidates that are found more than 6 effective radii from
the companion centers.\footnotemark
\footnotetext{Because the light profiles of the companion galaxies might have
been affected by an interaction with their giant host (in the case they
were physically associated), we use the median
effective radius of all VCS galaxies with magnitides within $0.5$~mag of each 
companion galaxy, instead of their measured one (the effective radii of all
ACSVCS galaxies have been measured by Ferrarese et al.~2006a).} 
Since the effective radii of the GC
spatial distributions are generally $\approx\! 2$ times larger than those of
the underlying galaxy light (Peng et~al. 2006, in preparation), 
this corresponds to excluding sources that are within about 3 GC-system scale
radii from the companion centers. This should effectively eliminate
$\approx\! 90\%$ of each companion's native GCs, leaving
us with fairly clean samples of extra M49 and M87 globulars,
located tens of kpc away from the giant galaxy centers.

We restrict our analysis to companions that have more than 50 GC candidates
left after this selection. These are VCC~1199 (companion to M49, projected
$4\farcm5$ away); VCC~1192 (M49, $4\farcm2$); VCC~1297 (M87, $7\farcm3$); and
VCC~1327 (M87, $7\farcm5$). Note that
$1\arcmin=4.8$~kpc for an average distance of $D=16.5$~Mpc to Virgo.

\begin{figure}
\epsscale{1}
\plotone{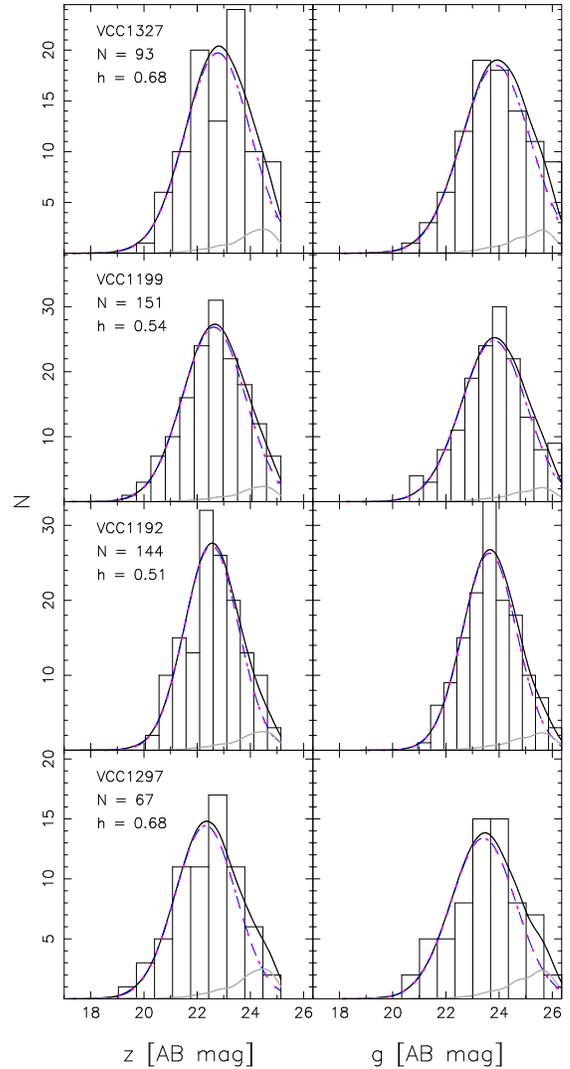}
\caption[]{Histograms and Gaussian fits for the GCLFs of GCs in the field
of view of four companions of M87 (=VCC~1316) and M49 (=VCC~1226) that lie
farther away than $6\,R_e$ from the companion galaxy, where $R_e$ is the
effective radius of the companion and is determined as described in
the text.
For each field we present the $z$-band and $g$-band GCLFs side by side.
The VCC name of the companion galaxy is indicated in the upper left corner of
the left panel, where we also indicate the total number $N$ of sources with
$p_{\rm GC}\ge 0.5$ and the bin-width $h$ used when constructing the
histograms. In each panel we show  the best fitting model (solid black curve), the
intrinsic Gaussian component (dashed curve), the Gaussian component
multiplied by the completeness fraction (dotted  curve),
and a kernel-density estimate of the expected contamination in the sample
(solid gray curve). The solid black curve is the sum of the solid gray and dotted  curves.
Details of the fits are given in Table \ref{tab:gclfpars_ore}.}
\label{fig:gclf_ore}
\end{figure}

\begin{figure*}
\epsscale{1}
\plottwo{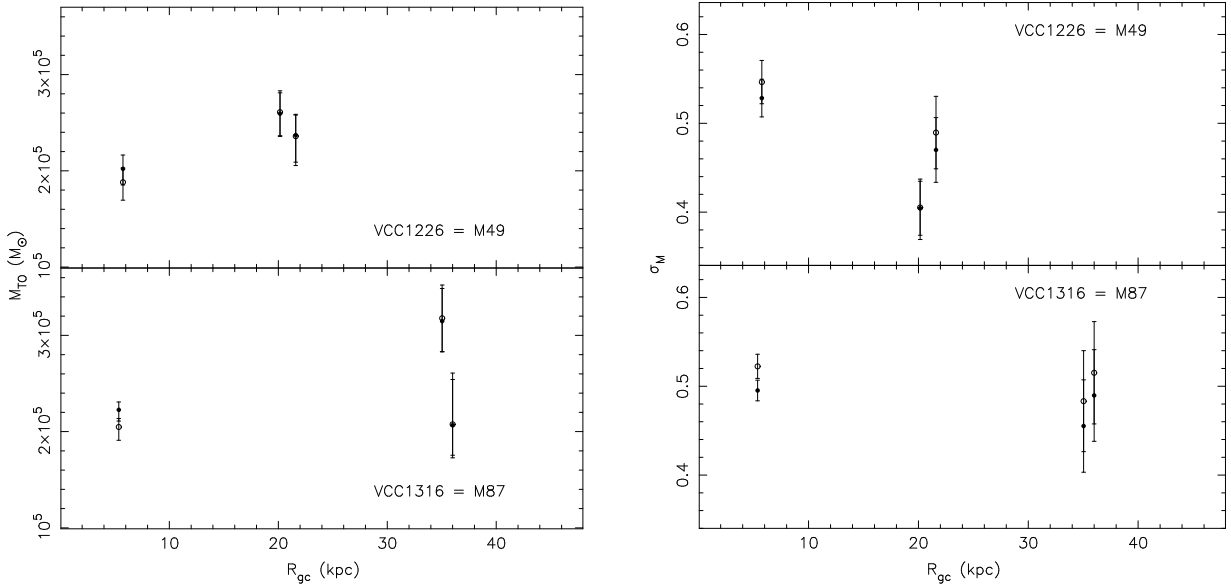}{f18b.eps}
\caption[]{({\it Left}) GCLF turnover mass $M_{\rm TO}$ as a function of
projected galactocentric distance $R_{\rm gc}$ in M49=VCC~1226 (top) and
M87=VCC~1316 (bottom). Filled symbols refer to $z$-band measurements of
$M_{\rm TO}$ while open symbols refer to $g$-band ones.
({\it Right}) Gaussian dispersion of logarithmic cluster masses, $\sigma_M$,
as a function of projected galactocentric distance $R_{\rm gc}$ in
M49=VCC1226 (top) and M87=VCC1316 (bottom). Filled symbols refer to $z$-band
measurements while open symbols refer to $g$-band ones.
In the M49 panels, the leftmost pairs of points refer to the GCLF parameters
derived from the central ACS pointings in the main body of the galaxy; the
next pair out correspond to the companion VCC~1192; and the rightmost pair
correspond to the companion VCC~1199.
In the M87 panels, the leftmost pairs of points refer to the GCLF parameters
derived from the central ACS pointings in the main body of the galaxy; the
next pair out correspond to the companion VCC~1297; and the rightmost pair
correspond to the companion VCC~1327.
}
\label{fig:Rgc}
\end{figure*}

In Figure~\ref{fig:gclf_ore} we show the luminosity functions and 
Gaussian fits for the resulting GC samples in the four fields neighbouring M87
and M49. In Table~\ref{tab:gclfpars_ore} we list the best-fit parameters
and the mean $(g-z)$ colors and mass-to-light ratios assumed to convert the
results to mass. The results are summarized in Figure~\ref{fig:Rgc}, where we
show the GCLF turnovers and dispersions as a function of galactocentric
distance in M87 and M49 separately. Evidently, none 
of the Gaussian GCLF parameters shows significant ($>3 \sigma$)
variation over the 20--35 kpc baselines probed. Fits of evolved Schechter
functions to these GCLFs confirm that $M_{\rm TO}$ in particular does not
change. As we discuss further in \S\ref{ssec:evaporation}, 
this lack of any significant radial trend in $M_{\rm TO}$ with galactocentric
distance is hard to reconcile with a picture in which 
the GCLF turnover is determined solely by dynamical effects (primarily
evaporation) acting on a universal power-law like initial cluster
mass function evolving in a fixed, time-independent galaxy potential.
(In fact, if it varies at all, $M_{\rm TO}$ may even get
slightly more massive with increasing radius in Fig.~\ref{fig:Rgc}. While we
do not claim that any such trend is in fact detected here, it would
be {\it opposite} to naive expectations.)

\placefigure{fig:gclf_ore}

\placetable{\ref{tab:gclfpars_ore}}

\placefigure{fig:Rgc}

\section{Discussion}
\label{sec:disc}

We have found interesting trends in three mass scales of physical
interest in connection with GC luminosity functions.

The GCLF turnover or peak mass takes a value of
$M_{\rm TO}=(2.2\pm0.4)\times 10^5\,M_\odot$ in most bright galaxies, but
shows some downward scatter in dwarfs fainter than
$M_{B, {\rm gal}} \ga -18$. In M87 and M49, the data are consistent with a
more or less constant $M_{\rm TO}$ to projected galactocentric radii of 20--35
kpc.

The higher-mass scale $M_c$ in an evolved Schechter function, which marks the
onset of an exponential cut-off in the number of clusters per unit mass, grows
steadily {\it smaller} in fainter galaxies. This drives a systematic
narrowing of the dispersion in more traditional Gaussian fits to the GCLF,
or equivalently a steepening of pure power-law fits to the mass function
$dN/dM$ at cluster masses $M\ga M_{\rm TO}$.

And the mass $\Delta$ in the evolved Schechter function,
which controls the shape of the low-mass end of the GC mass distribution and
is instrumental in setting $M_{\rm TO}$, varies by factors of
a few---although not entirely monotonically---as a function of galaxy
luminosity.

We now discuss these results in terms of their implications for GC formation
and dynamical evolution. We begin by focusing on $\Delta$ in the evolved
Schechter function, which, in the context of
Fall \& Zhang's (2001) dynamical theory for the GCLF,
is meant to measure the average amount of mass lost per globular cluster in a
galaxy, over a Hubble time of evolution. We then move on to $M_c$ and $M_{\rm
TO}$, asking specifically to 
what extent the observed variations in these high-mass characteristics of the
GCLF might be caused by dynamical friction rather than initial conditions.

\subsection{Evaporation and the Low-Mass Side of the GCLF}
\label{ssec:evaporation}

The defining feature of the evolved Schechter function in equation
(\ref{eq:esmass})---which we have found to fit the GC mass
distributions of galaxies in the ACSVCS just as well as the
traditional, but ad hoc,
lognormal form---is the flat shape of $dN/dM$ in the limit of low
masses. This asymptotic flatness always follows naturally from a
time-independent rate of cluster mass loss, regardless of the assumed 
initial form of $dN/dM_0$ (Fall \& Zhang 2001, and
\S\ref{sssec:esevolve} above). The exact values of the average
cumulative mass losses per GC for the galaxies in our sample
are, however, more specific to the assumption that
$dN/dM_0 \propto M_0^{-2}\,\exp(-M_0/M_c)$---a form chosen to match the
observed mass functions of young massive clusters in local mergers and
starbursts.

It is worth noting that, even though the average mass loss
$\Delta$ in an evolved Schechter function is key to setting the GCLF turnover
mass, $M_{\rm TO}$ does 
not vary as much or as systematically as $\Delta$ does in the ACSVCS
sample (cf.~Figures \ref{fig:mto_bd} and \ref{fig:MB_ES}). This is
because the value of the upper-mass cut-off $M_c$
also influences $M_{\rm TO}$ (see
\S\ref{ssec:esmod}), and $M_c$ varies in such a way as to largely
counteract the variation of $\Delta$, keeping $M_{\rm TO}$ more steady
as a function of $M_{B,{\rm gal}}$.

Since
$M_{\rm TO}$ is observed to be so nearly constant independently of any
functional fitting---at least in large galaxies---this balance
between variations in $\Delta$ and $M_c$ might be viewed simply as a
necessary condition to make evolved Schechter functions match the data at
all. But more interesting is that if the physical arguments behind the
fitting function are close to correct, our results imply
that the near-universality of the GCLF turnover in bright galaxies
($M_{B,{\rm gal}}\la -18$) is in some sense a coincidence resulting from
steeper initial $dN/dM$ (with lower $M_c$) in fainter systems being eroded by
faster mass-loss rates (yielding larger $\Delta$).

As we discussed in \S\ref{ssec:esmod}, some amount of cluster mass
loss may result from tidal shocks, but we expect that in general the
largest part comes from two-body relaxation and evaporation, at a rate
determined by the mean cluster density inside its half-mass radius:
$\mu_{\rm ev}\propto \rho_h^{1/2}$. This basic dependence holds independently
of any host galaxy properties, so if the cluster evaporation rate varies
systematically as a function of $M_{B,{\rm gal}}$, it presumably reflects
systematics
in the typical $\rho_h$ of the cluster systems. Then, if GCs are tidally
limited, such that their average densities are determined by the galaxy
density inside their orbits (e.g., King 1962), variations in their
characteristic $\rho_h$ should 
correspond in some way to variations in the host-galaxy densities. The easiest
way to quantify any such connection is to assume a spherical,
time-independent galaxy potential with a simple analytical form.
Thus, in their models of the Milky Way GC system, Fall \& Zhang (2001)
relate the $\rho_h$ of individual clusters to their orbital pericenters $r_p$
in a logarithmic potential with a fixed circular
speed, $V_c$, so that
$\rho_h \propto \rho_{\rm gal}(r_p)\propto V_c^2/r_p^2$.
We address the validity of these particular (strong) assumptions
about the host galaxy below; but for the moment we follow Fall \& Zhang and
most other authors (e.g., Vesperini 2000, 2001; Baumgardt \& Makino 2003) in
making them. What do our fitted $\Delta$ values for the ACSVCS galaxies then
imply for the distribution of GC densities and pericenters in these systems?

The evaporation rate of a cluster with observable, {\it projected}
half-mass radius $R_h$ depends on the density
$\rho_h\equiv 3M/(8\pi R_h^3)$ roughly as
\begin{equation}
\mu_{\rm ev}({\rm theo}) \simeq 345 \ M_\odot\,{\rm Gyr}^{-1} \ 
    \left(\frac{\rho_h}{M_\odot\,{\rm pc}^{-3}}\right)^{1/2} \ ,
\label{eq:rhohmuev}
\end{equation}
which again is independent of any assumptions on the host-galaxy
potential.\footnotemark
\footnotetext{Equation (\ref{eq:rhohmuev}) follows from
$\mu_{\rm ev}=0.045\,M/t_{\rm rh}$ (H\'enon 1961; see also Fall \& Zhang
2001), where the half-mass relaxation time $t_{\rm rh}$ is given by equation
(8-72) of Binney \& Tremaine (1987) with (1) an average stellar mass of
$0.7\,M_\odot$ and a Coulomb logarithm $\ln\,\Lambda = 12$ assumed
constant in time, as in Fall \& Zhang, and (2) a generic proportionality,
$R_h\simeq0.75\,r_h$, between the projected half-mass radius $R_h$ and its
unprojected counterpart $r_h$ (e.g., Spitzer 1987).}
However, if $\rho_h$ is taken to be set by a well defined $r_p$ in a
steady-state, singular isothermal sphere, then we also have (from
eqs.~[4] and [15] of Fall \& Zhang 2001)
\begin{eqnarray}
\mu_{\rm ev}({\rm theo}) & \simeq & 2.9\times10^4\ M_\odot\,{\rm Gyr}^{-1}\,
                            \left(r_p/{\rm kpc}\right)^{-1}
			    \left(V_c/220\ {\rm km~s}^{-1}\right) \nonumber \\
 & & \quad \times \left[1-\ln\,(r_p/r_c)\right]^{1/2} \ .
\label{eq:theorate}
\end{eqnarray}
In the last term on the right-hand side, which is derived by Innanen, Harris,
\& Webbink (1983), $r_c$ is the radius of a circular orbit with the same
energy as an arbitrary orbit with $r_p\le r_c$.

Now, for the Milky Way, recall from \S\ref{ssec:modcomp}
(eq. [\ref{eq:MWesmass}]), that we estimate 
\begin{equation}
\Delta({\rm MW}) = (2.5\pm0.5)\times10^5\,M_\odot
\label{eq:MWdelta}
\end{equation}
from our fit of an evolved Schechter function to the GCLF. For a GC
age of 13 Gyr, this implies a mass-loss rate 
(averaged over the distribution of cluster $\rho_h$ or, given the
assumptions behind eq.~[\ref{eq:theorate}], over all cluster orbits) of
\begin{equation}
\langle \mu_{\rm ev}\rangle ({\rm fit)}
         = \frac{\Delta({\rm MW})}{13\ {\rm Gyr}}
         = (1.9\pm0.4)\times10^4\,M_\odot\,{\rm Gyr}^{-1}\ .
\label{eq:MWrate}
\end{equation}
Comparing equation (\ref{eq:MWrate}) to equation (\ref{eq:rhohmuev})
implies an average
$\langle \rho_h \rangle \simeq
    (3000\pm 600)\,\,M_\odot\,{\rm pc}^{-3}$
for GCs in the Milky Way. This average falls towards the upper end of the
range of cluster $\rho_h$ observed today, but it is within a factor of
$\simeq\!2$--3 of the mean (e.g., see the data in Harris 1996). Equation
(\ref{eq:theorate}) further  
suggests an average pericenter of $\langle r_p\rangle\approx 2$~kpc.
This is roughly the same answer found by Fall \& Zhang (2001; see
their Figure 13), which shows that an evolved Schechter function is a
reasonable analytical approximation to their full numerical
theory. While such an $\langle r_p \rangle$ is slightly small---just as
$\langle \rho_h\rangle$ is slightly high---compared to more direct
pericenter estimates for Galactic globulars
(cf.~Innanen et al.~1983; van den Bergh 1995), it is again
within the range of standard values.

It is not at all obvious a priori that average cluster densities and
pericenters inferred strictly from fits to the Galactic GC mass function
should agree to within factors of 2 or 3 with values estimated by independent
methods. The fact 
that they do is an encouraging sign for the basic picture of
evaporation-dominated GCLF evolution. Some residual corrections---downward
in ``predicted'' $\langle \rho_h \rangle$ and up in $\langle r_p \rangle$---are
evidently required, but at a level that plausibly could come from
straightforward refinements in the various steps leading to 
equations (\ref{eq:rhohmuev}) and (\ref{eq:theorate}). For example,
there is some room for adjustment of the exact theoretical coefficients for
the evaporation rate $\mu_{\rm ev} \propto M/t_{\rm rh} \propto \rho_h^{1/2}$
and the pericenter $r_p\propto \rho_h^{-1/2}$
(see, e.g., the discussions in Fall \& Zhang 2001). In addition, we have
neglected here any additional mass loss caused by tidal shocks, and we have
adopted the idealization of a spherical and time-invariant Galactic potential.

To bring the ACSVCS data into this discussion, we focus on
the basic pattern of variation in $\Delta$ as a function of
galaxy luminosity, shown in the lower panel of Figure
\ref{fig:MB_ES}. First, $\Delta$ increases slightly from the brightest
$M_{B,{\rm gal}}\simeq-21.5$ to the fainter
$M_{B,{\rm gal}}\simeq-18$. The uncertainties and scatter in $\Delta$
are large, but the mean increase is perhaps a factor of
$\approx\!2$--5. Then, at fainter $M_{B,{\rm gal}}\ga -18$,
$\Delta$ holds more constant or even decreases again, possibly by as much as
a factor of $\approx\!2$--3 by the limiting $M_{B,{\rm gal}}\simeq -16$ of the
survey. 

If evaporation is responsible for these variations, then we should
expect them to be mirrored in the behavior of the average GC half-mass
radius as a function of galaxy luminosity: from equation
(\ref{eq:rhohmuev}), 
$\langle \mu_{\rm ev} \rangle \propto \langle \rho_h \rangle^{1/2}
                              \propto \langle R_h \rangle^{-3/2}$,
and by definition $\Delta\propto \langle \mu_{\rm ev} \rangle$ for coeval
clusters. Globulars in Virgo are marginally resolved with the ACS, and
Jord\'an et al.~(2005) have fit PSF-convolved King (1966) models to estimate
intrinsic $R_h$ values for individual sources (selected as described in
\S2 of Jord\'an et~al. 2005)
in most of the
galaxies that we have dealt with here. The behavior of mean
$\langle R_h \rangle$ versus $M_{B,{\rm gal}}$ is shown in Figure 5 of
Jord\'an et al.

A detailed comparison of $\langle R_h \rangle$ and $\Delta$ is not
straightforward, since these quantities were estimated separately for
GC samples defined differently by Jord\'an et al.~than in this paper.
Nevertheless, it is interesting that
$\langle R_h \rangle$ can be described as decreasing towards brighter
galaxy luminosity in the range $-21.5\la M_{B,{\rm gal}}\la -18$,
where $\Delta$ increases, and then turning around to increase somewhat 
at fainter $M_{B,{\rm gal}}\ga -18$, where $\Delta$ decreases again.
The changes in $\langle R_h \rangle$ are---as we would expect---smaller
and less clear than those in $\Delta$, but it is just plausible that
there is a net decrease of $\simeq\!35\%$ from
$M_{B,{\rm gal}}=-21.5$ to $M_{B,{\rm gal}}=-18$ and a slightly larger
increase from $M_{B,{\rm gal}}=-18$ to $M_{B,{\rm gal}}=-16$.
This would be consistent with the shallowest trends able to fit
$\Delta$ versus $M_{B,{\rm gal}}$ in Figure \ref{fig:MB_ES}.

We cannot use equation (\ref{eq:theorate}) to relate $\Delta$
to typical GC pericenters and average galaxy densities on a
case-by-case basis in the ACSVCS sample as in the Milky Way, since $V_c$
observations are not available for all systems.
However, scaling
relations can be used to some effect here. Large early-type galaxies with
$M_{B,{\rm gal}}\la -18$ generally obey
$V_c=\sqrt{2}\sigma \propto L_{\rm gal}^{0.25}$
(e.g., Faber \& Jackson 1976),
$(M/L)_{\rm gal}\propto L_{\rm gal}^{0.2-0.3}$
at optical wavelengths (van der Marel 1991; Cappellari et~al. 2006),
and thus
$R_{\rm eff}\propto L_{\rm gal}^{0.7-0.8}$
by the virial theorem (see also Ha\c{s}egan et al.~2005). Average mass
densities therefore increase towards lower $L_{\rm gal}$, such that
equation (\ref{eq:theorate}) implies 
\begin{equation}
\Delta \ \propto \ 
   \left(R_{\rm eff}/\langle r_p \rangle\right)\, R_{\rm eff}^{-1} V_c
   \ \propto \ 
   \left(R_{\rm eff}/\langle r_p \rangle\right)\, L_{\rm gal}^{-0.5 \pm 0.05}
\label{eq:deltabright}
\end{equation}
for bright galaxies. The situation is somewhat different for fainter
$M_{B,{\rm gal}}\ga -18$. For Coma Cluster galaxies in this regime,
Matkovi\'c \& Guzm\'an (2005) find
$V_c=\sqrt{2}\sigma \propto L_{\rm gal}^{0.5 \pm 0.1}$, while the data
in Graham \& Guzm\'an (2003) suggest
$R_{\rm eff}\propto L_{\rm gal}^{0.1-0.2}$.
If these systems are representative of those in Virgo, then their
average densities decrease towards lower $L_{\rm gal}$, and equation
(\ref{eq:theorate}) leads to
\begin{equation}
\Delta \ \propto \ 
   \left(R_{\rm eff}/\langle r_p \rangle\right)\, L_{\rm gal}^{0.35 \pm 0.1}
\label{eq:deltafaint}
\end{equation}
for faint dwarfs.

The major unknown in equations (\ref{eq:deltabright}) and
(\ref{eq:deltafaint}) is the ratio of galaxy $R_{\rm eff}$ to GC
$\langle r_p \rangle$, and how it might or might not vary systematically as a
function of galaxy luminosity. If the ratio is constant for all systems, then
the dotted lines drawn in the lower panel of Figure \ref{fig:MB_ES} show the
expected variation of the mass loss $\Delta$ versus $M_{B,{\rm gal}}$. These
lines are normalized to make $\Delta = 2.5 \times10^5\,M_\odot$ at
$M_{B, {\rm gal}}=-21$ and to make the bright- and faint-galaxy
scalings meet at $M_{B,{\rm gal}}=-18$.
The net increase of $L_{\rm gal}^{-0.5}$ from $M_{B,{\rm gal}}=-21.5$
to $M_{B,{\rm gal}}=-18$ is a factor of about 5, while the
decrease of $L_{\rm gal}^{0.35}$ from $M_{B,{\rm gal}}=-18$
to $M_{B,{\rm gal}}=-16$ is a factor of approximately 2.

These changes may be somewhat greater than suggested by the actual fitted
estimates of $\Delta$. Moreover, an increase of $\Delta$ by a factor
of 5 between $M_{B,{\rm gal}}=-21.5$ and $M_{B,{\rm gal}}=-18$
would imply a decrease in $\langle R_h \rangle$ by a
factor of $5^{2/3}\approx 3$, which is larger than the measurements of
Jord\'an et al.~(2005) support. However, this is clearly not an 
order-of-magnitude problem. It could easily be alleviated if
the galaxy total mass distributions are not isothermal spheres, or if
$R_{\rm eff}/\langle r_p \rangle$ depends even weakly 
on galaxy luminosity, or if uncertainties and scatter in the galaxy scalings
result in small deviations from the nominal exponents on
$L_{\rm gal}$ in equations (\ref{eq:deltabright}) and (\ref{eq:deltafaint}).
Tidal shocks may also contribute differently
to the net $\Delta$ in different galaxies, a complication that we have
entirely ignored. Again, then, it is encouraging that
these crude relations come as close as they do to explaining the systematics
in a cluster mass-loss parameter inferred only from the GCLF---accounting in
particular for the change in dependence of $\Delta$ on galaxy luminosity
around $M_{B,{\rm gal}}\simeq-18$.  

Obviously, more rigorous and detailed analyses of individual galaxies are
required to really make (or break) the case in general that the overall form
of an evolved Schechter function for the GC mass function, and the parameter
$\Delta$ especially, can be interpreted physically and self-consistently as
the result of evolution from an initial GC $dN/dM_0\propto M_0^{-2}$ with
individual cluster mass-loss rates that are constant in time. From our
discussion here, it does seem that this ``literal'' view of the simple fits to
the Milky Way and ACSVCS GCLFs is at least broadly compatible with
observations of the cluster densities or radii in these galaxies and with the
trends in $\Delta$ vs.~$L_{\rm gal}$, if
evaporation is the main disruptive process for clusters as massive as
$M_{\rm TO}\sim 2\times10^5\,M_\odot$.

Difficulties do arise, however, when considering the additional constraint
that the GCLF is invariant over wide ranges of galactocentric radius and GC
density in the Milky Way 
and other large galaxies. As described above, application of equation
(\ref{eq:theorate}) to the global Galactic GCLF ultimately implies an
average GC pericenter of $\langle r_p \rangle \simeq 2$~kpc,
corresponding to about half the effective radius of the
bulge. Similarly, our normalization of equation (\ref{eq:deltabright})
in Figure \ref{fig:MB_ES} implies
$\langle r_p \rangle < 0.5\, R_{\rm eff}$ for the brightest
early-type galaxies in Virgo.
But observationally, the GCLF turnover
$M_{\rm TO}$ (and thus $\Delta$) has the same, global value for clusters
currently found out to at least 10--15 effective radii in the Milky Way
(e.g., Harris 2001) and at least $\simeq 4\,R_{\rm eff}$ in M87 and
M49 (\S\ref{ssec:gradius}). This can only be consistent with
evaporation-dominated depletion of an intially steep GC
$dN/dM_0\propto M_0^{-2}$ at low masses, and with the additional
assumption that the mass loss $\Delta\propto r_p^{-1}$, if cluster
orbits are systematically much more elongated at larger galactocentric radius
in all these systems.

In fact, for the Milky Way and M87 respectively, Fall \& Zhang (2001) and
Vesperini et al.~(2003) have shown that following this chain of logic leads 
to the conclusion that globulars should {\it initially} have been on
predominantly radial orbits outside about one effective radius in each galaxy.
On the other hand, the {\it present} GC velocity distributions in the Galaxy,
in M87, and in M49 are all essentially isotropic---implying orbits with
typical axis ratios of only $r_a/r_p\simeq3$---out to the same spatial scales
of several $R_{\rm eff}$, over which the observed GCLF is invariant
(see, e.g., Dinescu, Girard, \& van Altena 1999; C\^ot\'e et al.~2001,
2003). Fall \& Zhang suggest that this difference between (presumed) initial
and (observed) present orbital properties might be explained by preferential
depletion of GCs on the most radial orbits. But while the idea remains to be
tested in detail for the Milky Way, Vesperini et al.~(2003) show that---again
if the galaxy potential is spherical and time-independent---it does not
suffice to account quantitatively for the combined GCLF and kinematics data
in M87.

Related to this is the average density,
$\langle \rho_h \rangle \simeq 3000\ M_\odot\,{\rm pc}^{-2}$, implied by the 
more general equation (\ref{eq:rhohmuev}) and the required total
$\Delta$ for Galactic globulars. A similar $\langle \rho_h \rangle$
is also suggested for GCs in the brightest Virgo galaxies by
the $\Delta$ values in Figure \ref{fig:MB_ES}. As we mentioned above, such
densities are observed for real clusters; but there is a broad distribution
of $\rho_h$, with an average slightly lower than
$3000\ M_\odot\,{\rm pc}^{-2}$ and a long tail to much smaller values of
$ <\! 100\ M_\odot\,{\rm pc}^{-2}$. More generally, the GCs in 
most large galaxies have half-mass radii that are largely uncorrelated with
cluster mass (e.g., van den Bergh, Morbey, \& Pazder 1991; Jord\'an
et al.~2005, and references therein), so that $\rho_h$ apparently
always ranges over more than two orders of magnitude. When
$\rho_h < 100\ M_\odot\,{\rm pc}^{-2}$, the total evaporative mass loss per
cluster over 13 Gyr is $< 5\times10^4\,M_\odot$, well below the typical
average $\Delta$ and global $M_{\rm TO}$ for entire GC systems. In the
Milky Way at least, the large majority of such low-density GCs are found at
Galactocentric distances $r_{\rm gc}\ga 10$~kpc, so in a sense the
problem is bound up with the weak radial variation of the GCLF.

These points are important, and they need to be resolved, but they should
not be taken as disproof of the idea that long-term dynamical
evolution alone might explain the difference between the mass
functions of old GCs and young massive clusters. Ultimately, the
near-flatness of $dN/dM$ at low masses, which is clearly seen in the
Milky Way and is entirely consistent with all of our Virgo GCLFs, only
demands that cluster masses decrease linearly in time if the
dynamical-evolution hypothesis is correct at all (see Fall \& Zhang
2001, and \S\ref{ssec:esmod} above). It is not absolutely necessary
that evaporation account for the full mass-loss rate of every
cluster, even though our discussion here has focused on exploring
this possibility (and shown that it does come remarkably
close, to within factors of 2--3 for the most part). For example,
globulars in the extreme low-density tails of $\rho_h$ distributions,
mentioned just above, might be much more strongly---and
differently---affected by tidal shocks than any previous GCLF
calculations have allowed. Such shock-dominated evolution could still
lead to a constant mass-loss rate of its own (see Dehnen et
al.~2004, and \S\ref{sssec:esevolve} above), which would add directly
to $\mu_{\rm ev}$ without otherwise changing any of the main arguments
here.

In more specific terms, the radial invariance of the GCLF might ultimately
be explained by modifying a single ancillary assumption in the
current dynamical-evolution models rather than discarding the idea altogether.
It is the notion of spherical and steady-state galaxy
potentials that prompts Fall \& Zhang (2001), Vesperini et
al.~(2003), and almost all other authors to 
use equations (\ref{eq:rhohmuev}) and
(\ref{eq:theorate}) to tie cluster densities to orbital pericenters in
these analyses. But, as Fall \& Zhang themselves point out, this is of course
an extreme simplification for galaxies that grow through hierarchical
merging.

Fall \& Zhang suggest, for example, that a major merger could obviate the
need for extremely radial orbits to distribute clusters with high mean
densities, fixed at small and well defined pericenters, over large
volumes in a galaxy. Instead, a merger may efficiently mix two 
globular cluster systems spatially and isotropize their velocity
distribution. This could then work to weaken any radial gradients in
the mass loss $\Delta$ and the GCLF turnover mass, which might have
resulted from realistic orbital distributions and $\mu_{\rm ev}(r_p)$
dependence like equation (\ref{eq:theorate}) in the progenitor galaxies.

In addition to this, multiple minor mergers---which are perhaps more
relevant than major mergers for a galaxy like our own---should
steadily bring in globulars formed with densities and evaporation
rates unrelated, at least initially, to their new orbits in the main
galaxy, making the use of equation (\ref{eq:theorate}) less than
straightforward. In fact, any use of it at all
could be questionable in this case, since all clusters would
constantly be sampling new pericenters in an evolving
potential. Again, then, weak spatial variations in 
$\Delta$ and $M_{\rm TO}$ need not imply highly radial GC orbits.
Prieto \& Gnedin (2006) have recently simulated the evolution of the
GCLF during the hierarchical growth of a Milky Way-sized galaxy.
Starting from an initial cluster mass function
$dN/dM_0\propto M_0^{-2}$, which is re-shaped primarily by
evaporation---but abandoning equation (\ref{eq:theorate}) and instead
adopting evaporation rates from GC densities fixed independently
of their orbits---they find that it is possible (even without a recent major
merger) to produce a final GC system with an isotropic velocity distribution
and a radially invariant GCLF similar to the observed Galactic distribution.

A caveat is that the hierarchical-growth simulations most favored by Prieto
\& Gnedin (2006) are ones in which they assume that all globular clusters have
a common mean density inside $R_h$ (just one that is not set by any orbital
pericenter). This is still incompatible with the wide range of $\rho_h$
observed for the GCs in many galaxies, and it is furthermore not obvious how
the cumulative mass loss $\Delta \propto \langle \rho_h \rangle^{1/2}$
should then vary as a function of galaxy luminosity. On the other hand,
Prieto \& Gnedin have 
also run some models allowing for an initial spread of GC densities followed
by evaporation at constant $\rho_h$. This is at least more reminiscent of real
$\rho_h$ distributions, and it still produces a GCLF that is not too
drastically different from the Galactic one. Clearly, more work is required
to clarify the dynamical evolution of initial power-law GC mass functions in
time-dependent galaxy potentials, with the totality of relevant observational
constraints taken into account: a flat $dN/dM$ at low masses; a weak or absent
correlation between GC radii and masses; radially invariant GCLFs; currently
isotropic velocity distributions; and mass losses $\Delta$ that vary with
galaxy luminosity as in Figure \ref{fig:MB_ES}.

Should all efforts along these lines fail to explain the combined data, the
only option left would seem to be that a peak in the GCLF was
established much earlier, by processes more related to cluster formation.
One possible scenario has been proposed by Vesperini \& Zepf (2003). They
suggest that low-mass globulars were initially less concentrated (with a
larger ratio of half-mass to tidal radius) than high-mass clusters. The
inevitable expansion of all clusters following mass loss driven by
stellar evolution 
would then cause many low-mass clusters preferentially to overflow their
tidal radii, leading ultimately to fast disruption times
of a few hundred Myr or less (Chernoff \& Weinberg 1990). This may
turn an initial
power-law $dN/dM_0$ at low masses into a roughly flat-topped or even
lognormal distribution, with $M_{\rm TO}$ near its current value, very
early on. Weaker long-term evaporation (i.e., lower cluster densities or
larger and more variable pericenters) could then suffice to explain
the residual difference between the initial, steep mass function and
the final, observed one, even in a static galaxy potential. 

Observations of the young massive clusters in the Antennae galaxies
already imply that early disruption is {\it independent} of cluster
mass, at least for clusters more massive than several $10^4\,M_\odot$
and younger than $\simeq\! 10^8$~yr (Zhang \& Fall 1999; Fall et 
al.~2005). Thus, if the disruption mechanism of Vesperini \& Zepf
(2003) is to work, the mass-selective aspect of it apparently 
must be restricted to timescales of $10^8$--$10^9$~yr or so.
In any case, the success of this or any similar picture further
relies on an appropriately tuned mass dependence in some key GC
property being built into cluster systems essentially as an initial
condition; but this still requires explanation in itself. 

\subsection{Dynamical Friction and the High-Mass Side of the GCLF}
\label{ssec:dynfric}

At GC mass scales $M\ga \Delta$, dynamical friction can in some cases
become more important than evaporation or shocks as a cluster destruction
mechanism. A point mass $M$ 
originally on a circular orbit of radius $r$ in a galaxy with a total-mass
distribution following a singular isothermal sphere will spiral in to the
galaxy center within a time (Binney \& Tremaine 1987)
\begin{equation}
\tau_{\rm df} \simeq
\frac{5.9\ {\rm Gyr}}{(\ln\Lambda)/10}\left(\frac{r}{{\rm kpc}}\right)^2 
\left(\frac{V_c}{220\ {\rm km\ s}^{-1}}\right)
\left(\frac{10^6\, M_\odot}{M}\right) \ ,
\label{eq:df_bt}
\end{equation}
where $V_c$ is the galaxy's circular speed and $\ln\Lambda \sim 10$ is
the usual Coulomb logarithm.

It is clear from equation (\ref{eq:df_bt}) that  dynamical friction cannot be
a major factor in deciding the evolution of all but the very most massive tip
of the GCLF in $\sim\! L_*$ and brighter galaxies with
$V_c \ga 200$~km~s$^{-1}$. However, the scaling $\tau_{\rm df}\propto V_c$
implies that the relevance of dynamical friction can increase significantly
for lower luminosity galaxies (e.g., Hernand\'ez \& Gilmore 1998; Lotz et
al.~2001). It is then reasonable
to ask whether a stronger depletion of massive GCs in dwarf
galaxies might be able to explain the systematic decrease of $M_c$ versus
$M_{B,{\rm gal}}$ in our fits of evolved Schechter functions for these
systems, and possibly even the slight decrease in average $M_{\rm TO}$ towards
the faintest $M_{B, {\rm gal}}$. 

\begin{figure}
\epsscale{1}
\plotone{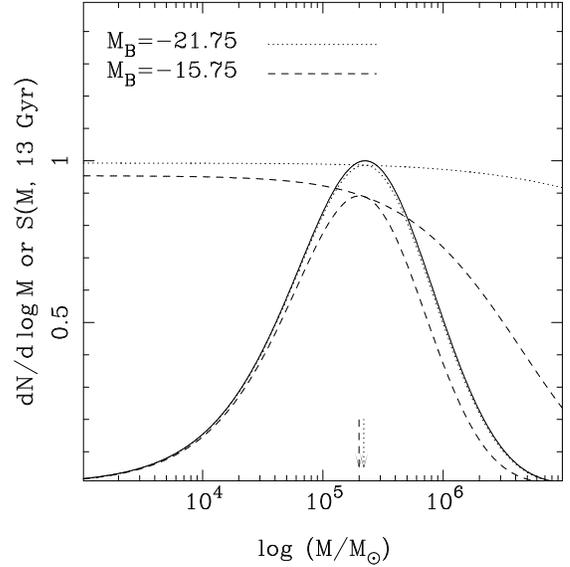}
\caption[]{A simple model for the effects of dynamical friction on the GC
mass function, illustrated in terms of the distribution $dN/d\,\log\,M$, which
is directly proportional to the GCLF. The solid curve shows this version of
an evolved Schechter function with $M_c=3.0\times10^6\,M_\odot$ and
$\Delta=2.6\times10^5\,M_\odot$ (hence $M_{\rm TO}=2.2\times10^5\,M_\odot$),
appropriate for a giant elliptical. This is assumed to be the mass function 
on which dynamical friction operates.
The monotonically decreasing {\it dotted} curve shows the function
$S(M,t=13\ {\rm Gyr})$ (eq.~[\ref{eq:S}]), calculated for a galaxy with
$M_{B,{\rm gal}}=-21.75$ as described in the text. The {\it dotted}
bell-shaped curve is the product of this $S$ times the solid curve; it
illustrates the cumulative effect of dynamical friction on the
GCLF in very massive galaxies.
The monotonically decreasing {\it dashed} curve shows the function
$S(M,t=13\ {\rm Gyr})$ as calculated for a galaxy with
$M_{B,{\rm gal}}=-15.75$. The {\it dashed} bell-shaped
curve is the product of this $S$ times the solid curve, illustrating the
net effect of dynamical friction on the GCLF in very faint galaxies.
The arrows indicate the position of the final turnover mass for each of the
resulting mass functions. There is a slight decrease in $M_{\rm TO}$ for the
low-luminosity galaxy as a consequence of dynamical friction, but not enough
to account fully for the observed behavior in Fig.~\ref{fig:mto_bd}
or Fig.~\ref{fig:binVSind}.}
\label{fig:dynfric}
\end{figure}

\placefigure{fig:dynfric}

We do not attempt here to find a definitive answer to this question, but only 
an indication of the ability of dynamical friction to produce the observed
trends.
One particular subtlety is that 
the expression for $\tau_{\rm df}$ in
equation (\ref{eq:df_bt}) does
not allow for clusters to evaporate. 
But a steadily decreasing cluster mass 
will lead to a longer total dynamical-friction timescale.
We deal with this complication in the simplest way possible: the
timescale $\tau_{\rm df}$ for a cluster with initial mass
$M_0$ and present mass $M=(M_0-\Delta)$ is approximated by evaluating equation
(\ref{eq:df_bt}) at the {\it average} mass, $(M+\Delta/2)$.

Let us denote by
$\widehat{\Psi}(M,t)$ the GC mass function that would be obtained after a time
$t$ of GC evolution in the absence of any dynamical friction. The effects of
dynamical friction are easily accounted for by subtracting from
$\widehat{\Psi}$ all clusters with instantaneous masses $M$ such that
\begin{equation}
(M+\Delta/2) > M_{\rm min}(r,t) \ ,
\label{eq:df_cond.1}
\end{equation}
where $M_{\rm min}$ follows from  equation (\ref{eq:df_bt}) by setting
$\tau_{\rm df}<t$:
\begin{equation}
M_{\rm min}(r, t) \simeq
\frac{4.5 \times 10^5\,M_\odot}{(\ln\Lambda)/10}
      \left(\frac{13\ {\rm Gyr}}{t}\right)
      \left(\frac{r}{{\rm kpc}}\right)^2
      \left(\frac{V_c}{220\,{\rm km\ s}^{-1}}\right) \ .
\label{eq:df_cond.2}
\end{equation}
The net, ``global'' GC mass function (averaged over all GC orbits, or
galactocentric radii) at any time $t$ is thus
$dN/dM = S(M,t) \times \widehat{\Psi}(M,t)$ where
\begin{equation}
S(M,t) = \frac{\int_0^{\infty}  \rho_{\rm GC}(r) \,
                                H[M_{\rm min}(r,t) - (M+\Delta/2)] \,
                                4\pi r^2 \, dr}
              {\int_0^{\infty}  \rho_{\rm GC}(r) \, 4\pi r^2 \, dr} \ .
\label{eq:S}
\end{equation}
Here $\rho_{\rm GC}(r)$ is the space density of 
GCs (assumed to be independent of cluster mass) and $H$ is the
Heaviside step function: $H(x) \equiv 1$ for $x>0$ and
$H(x) \equiv 0$ for $x < 0$.

This raises further points to be dealt with in more careful calculations
along these lines. First, dynamical friction will clearly affect also the
spatial distribution of GCs, so that $\rho_{\rm GC}(r)$ will have a
dependence on time, which we ignore. Second, the
effects of dynamical friction could introduce some dependence on
galactocentric position into the GC mass function, which in a complete
treatment would be contrasted with observational limits on any such
variations. Third, changing the assumed galaxy potential could significantly
affect the derived $\tau_{\rm df}$ (e.g., Hernandez \& Gilmore 1998; Read
et~al. 2006), as could relaxing the unrealistic assumption of strictly
circular orbits (e.g., Pesce, Capuzzo-Dolcetta, \& Vietri 1992; van den Bosch
et al.~1999). Finally, we do not take into account the fact that the ACS has a
fixed field of view, and thus we are not always observing truly globally
averaged GCLFs---although this point is relevant mainly for the most massive 
galaxies, where the effects of dynamical friction are expected to be
negligible in any case.

These issues notwithstanding, we proceed to estimate the effects of dynamical
friction 
by evaluating $S(M,t)$ as written in equation (\ref{eq:S}).
We  assume that the ``friction-free'' $\widehat{\Psi}(M,t)$
at the present day is well described by the GCLF
of bright ellipticals, where dynamical friction is negligible, and is
therefore given by equation
(\ref{eq:esmass}) with $\Delta=2.6\times 10^5 M_{\odot}$ and
$M_c = 3\times 10^6 M_{\odot}$
(see Figure \ref{fig:MB_ES} and Table \ref{tab:galbins}).
To obtain the final $dN/dM$ including dynamical friction, we then
multiply this by the function $S(M, t\equiv13\,{\rm Gyr})$.
In doing so, we always take the slowly varying Coulomb logarithm in equation
(\ref{eq:df_cond.2}) to be $\ln\Lambda=10$.

We assume that for giant galaxies with $M_B < -18$ we have 
$V_c \propto \sigma \propto L_{\rm gal}^{0.25}$ (Faber \& Jackson 1976),
with a zeropoint chosen to give $V_c=484$ km s$^{-1}$ at $M_B=-21.75$, 
based on the velocity dispersion of M87 (Bender, Saglia \& Gerhard 1994).
We impose a change in this scaling at $M_B>-18$, so that dwarfs follow
$V_c \propto \sigma \propto L_{\rm gal}^{0.5 \pm 0.1}$
(Matkovi\'c \& Guzm\'an 2005; cf.~\S\ref{ssec:evaporation} above).
We can then find $M_{\rm min}(r,t)$ from equation (\ref{eq:df_cond.2}) for any
GC in any galaxy.

To specify the spatial distribution of GCs and calculate
$S(M,t=13\ {\rm Gyr})$, we estimate the galaxy's effective radius
$R_{\rm eff}$ using the data from
Ferrarese et al.~(2006a); 
then we assume that the effective radius of the
GC system is just twice $R_{\rm eff}$ (Peng
et~al. 2006, in preparation). 
Finally, we assume that $\rho_{\rm GC}(r)$ is given
by the density profile of Prugniel \& Simien (1997;
see also Terzi\'c \& Graham 2005), which is an analytical approximation
to the deprojection of a Sersic profile ($R^{1/n}$ law),
and we let the Sersic index $n$ be determined by $M_{B, {\rm gal}}$ as per
equation (25) of Ferrarese et~al. (2006a).

The results of the calculations for two representative galaxy magnitudes,
$M_{B,{\rm gal}}=-21.75$ and $M_{B,{\rm gal}}=-15.75$, are illustrated in
Figure \ref{fig:dynfric}. The figure shows both $S(M,t=13\ {\rm Gyr})$ (the
monotonically decreasing curves)
and the function  $dN/d\log M$ (proportional to the GCLF and given 
by the peaked curves) that follows
from dynamical friction acting on the assumed evolved Schechter function.
The resulting turnover mass scales are indicated with arrows, which show that
the stronger dynamical friction in the fainter galaxy
leads to a slightly lower turnover mass scale.

\begin{figure}
\epsscale{1}
\plotone{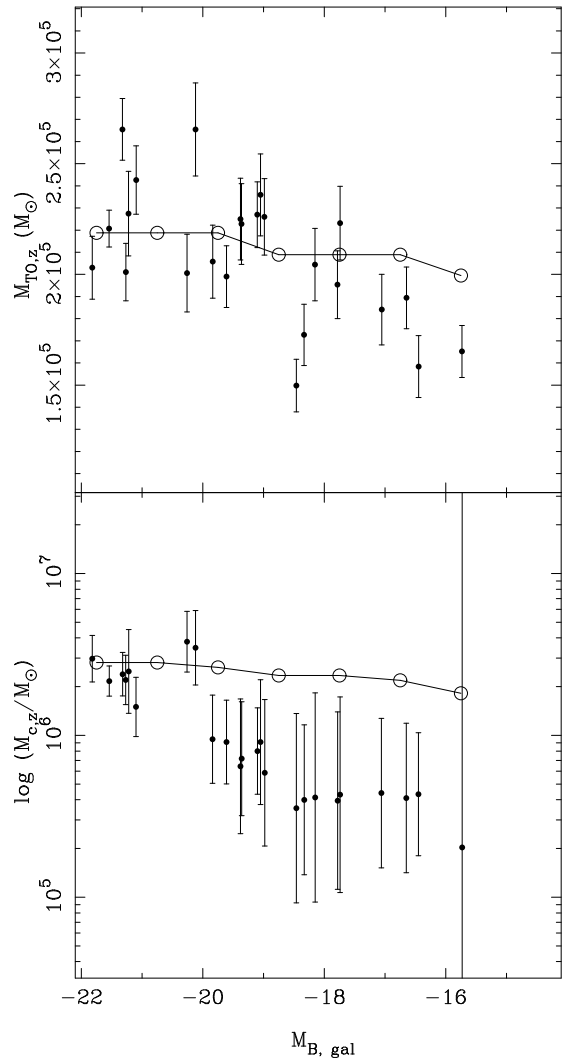}
\caption[]{({\it Upper panel}) GCLF turnover mass $M_{{\rm TO},z}$, inferred
from evolved Schechter fits to $z$-band data for the binned-galaxy GC samples,
versus $M_{B, {\rm gal}}$ (data are the same as the circles in
Fig.~\ref{fig:mto_bd}).
The open symbols connected by a solid line are the predicted change
in $M_{\rm TO}$ due to the increasing efficiency of dynamical friction
in fainter galaxies; see text. Dynamical friction might account for part of
the observed trend, but probably not all of it.
({\it Lower panel}) Evolved Schechter function cut-off mass $M_{c,z}$ versus
$M_{B, {\rm gal}}$ for binned-galaxy $z$-band samples.
The open symbols connected by a solid line are again our predictions for
the change in $M_c$ due to stronger dynamical friction in smaller galaxies.
Dynamical friction is not able to explain the observed behavior of $M_c$
as a function of galaxy luminosity.}
\label{fig:TO_MB_mod}
\end{figure}

We show the behavior of $M_{\rm TO}$ as a function of $M_{B,{\rm gal}}$ in
general, in the upper panel of Figure~\ref{fig:TO_MB_mod} (circles connected
by a solid line) and contrast it with the observed variation
in our binned-galaxy GC samples (Figure \ref{fig:mto_bd}).
The predicted $M_{\rm TO}$ varies quite slowly with $M_{B,{\rm gal}}$, but it
ultimately decreases by $\sim 10\%$ from our assumed $2.2\times10^5\,M_\odot$
in the brightest galaxies. This is comparable to the observed decrease of
$\approx\! 30\%$ in $M_{\rm TO}$. Thus, dynamical friction may be responsible
for some part of the the slow change in GCLF turnover mass with galaxy
magnitude. 

\placefigure{fig:TO_MB_mod}

In the lower panel of Figure~\ref{fig:TO_MB_mod} we show (again, as open
circles connected by a solid line) the $M_c$ values inferred by fitting
evolved Schechter functions to our model GC mass functions after calculating
the effects of dynamical friction. Evidently, we can expect dynamical
friction to cause perhaps a $\sim\! 30\%$--40\% decrease in the value of $M_c$
from the brightest to the faintest galaxies;
but this is altogether too little to account for the factor of
$\simeq\! 6$--7 decrease we actually observe. Similarly, if we fit
power laws to our dynamical-friction mass functions in the range
$3\times10^5\,M_\odot \le M \le 2\times10^6\,M_\odot$, we obtain rather
constant powers $\beta\simeq1.7$--1.8 for galaxy magnitudes
$-21.75 < M_{B,{\rm gal}} < -15.75$, which is far from being able to explain
the observational situation in Figure \ref{fig:power} above. We conclude
that {\it dynamical friction cannot account for more than a small fraction of
the observed steepening of the globular cluster mass function above the
GCLF turnover}. 

These results are essentially in agreement with those of Vesperini (2000), who
models the effects of evaporation and dynamical friction on the GCLF,
and predicts only slight decreases in the mean $\langle \log\,M \rangle$ and
the Gaussian dispersion $\sigma_M$ as galaxy luminosity decreases
(see his Figure 6), at levels much smaller than those seen in our data
(e.g., Figure \ref{fig:MTO_GZ}).
Thus, although we have emphasized the highly simplified nature of our
calculations, it nevertheless appears that galaxy-to-galaxy systematics in the
cluster formation processes, rather than dynamical evolution, must be largely
responsible for the observed variation in the detailed form of the GCLF at
high masses.

\subsubsection{Initial Conditions}
\label{sssec:init}

It seems inevitable from the discussion above that the observed steepening or
narrowing of the GCLF above the turnover point $M_{\rm TO}\sim
2\times10^5\,M_\odot$ in fainter galaxies---whether this is expressed in terms
of smaller Schechter-function mass scales $M_c$ or steeper power-law indices
$\beta$ or narrower Gaussian dispersions $\sigma_M$---must reflect
non-universal initial conditions in the cluster mass distribution, and
therefore some fundamental aspect of the star-formation process.

Observationally, it is known that the luminosity of the brightest young star
cluster in a star-forming galaxy scales with the global star formation rate
(Billet, Hunter \& Elmegreen 2002; Larsen 2002). There has been some
discussion as to whether this is just
a size-of-sample effect (if more clusters are formed, it is
statistically more likely to achieve higher masses by random sampling of an
underlying mass distribution that might still be universal) or indicative of
a real, physical limit to the initial cluster mass function (Larsen 2002;
Weidner, Kroupa \& Larsen 2004). 

Gieles et~al.~(2006a, b) argue that there is a physical upper limit, $M_{\rm
max}$, to cluster masses in each of NGC 6946, M51, and the Antennae galaxies
(though see Whitmore, Chandar \& Fall 2006 for a differing view).
The number of clusters found with $M>M_{\rm max}$ falls rapidly to zero in all
three cases, but the value of the upper limit is found by Gieles et al.~to vary
between the galaxies, from $M_{\rm max}\simeq 4$--$10\times10^5\,M_\odot$.
Qualitatively, a parameter like $M_{\rm max}$ can be identified with
$M_c$ in a Schechter-function description of (initial) GC mass
functions. Quantitatively, the range of $M_{\rm max}$ claimed by Gieles et
al.~for their young systems is very similar indeed to our fitted $M_c$ values
for the old GCs in early-type Virgo galaxies (see Figure \ref{fig:MB_ES}).

It will be interesting to explore this possible connection between
globulars and young massive clusters in more detail. 
Possibly one route to
take is suggested by the theory of the GCLF developed by Harris \& Pudritz
(1994), in which a distribution of cluster masses is built up by
collisions between gaseous protoclusters. 
McLaughlin \& Pudritz (1996) suggest that the total time required to  
produce very high-mass clusters may be longer for galaxies in 
lower-density environments, and this could perhaps be related to our
finding of a cut-off at lower $M_c$ (in our current notation) for the 
initial GC mass functions at fainter $M_{B,{\rm gal}}$.
If these types of ideas can be generalized, then
both our GCLF observations and the possible existence of an upper mass
``limit'' in young cluster systems could be reflecting a systematic variation
in gas-dynamical timescales as a function of galaxy mass and/or density.

In any case, the fact that dynamical friction
is unable to account for the steepening of an initially universal mass 
function as the mass of the host galaxy decreases, 
combined with the possible existence in young, relatively unevolved
cluster systems of a mass scale similar to $M_c$
in our old GC systems, leads us to favor the view that a significant part of
the observed morphology {\it at the highest-mass ends} of GCLFs is due to
systematics in the initial distributions. The precise extent to which this
part of the initial mass function is still reflected in the present-day one is
still something of an open question, the answer to which will be a crucial 
ingredient in our understanding of GC formation and evolution. 
A detailed understanding of the ``microscopic'' star-formation processes on
rather short timescales in very young clusters could well be key to making
much further progress in this direction.

\section{Summary and Conclusions}
\label{sec:conclusions}

We have presented the GCLFs of 89 early-type galaxies in the Virgo cluster
and determined maximum-likelihood estimates for model parameters
using fits of Gaussians and a simple ``evolved
Schechter function'' described in \S\ref{ssec:esmod}. The latter reflects the
effects of GC disruption (at a constant rate and presumably due mostly to
two-body relaxation and evaporation) on an initial cluster 
mass distribution that followed a Schechter function with a fixed
power-law index of $-2$ at low masses. The evolved mass function tends to
a flat shape at low $M$ and is an accurate analytical approximation to the
numerical distributions produced in the theory of Fall \& Zhang
(2001). We have tested the robustness of our results by simulations, by the
construction of GCLFs for galaxies binned together to contain a minimum
number of clusters, and by using alternate schemes to select GC candidates
from catalogues of observed sources. Our main results and conclusions are the
following: 

\begin{enumerate}

\item We find a remarkably regular decrease of the dispersion of the GCLF
as the luminosity of the host galaxy decreases
(\S\ref{sec:trends} and Jord\'an et al.~2006). Quantitatively,
the maximum-likelihood estimates of the dispersion $\sigma$ of Gaussian
fits to the $z$- and $g$-band data are well described by the linear
relations presented in equations (\ref{eq:sigmaz}) and (\ref{eq:sigmag}).
The dispersions for the GCLFs of the Milky Way and M31 fall in the midst
of our new data and thus the correlation of $\sigma$ with $M_{B,{\rm gal}}$ 
would appear to be more fundamental than the older view, that GCLF widths
depend on galaxy Hubble type.

This trend reflects a systematic steepening of the GC mass function for
{\it massive} clusters in particular ($M \ga 3\times10^5 M_{\odot}$, above the
peak of the GCLF) as the host galaxy luminosity decreases. When fitting
power-law mass functions to this upper cluster mass regime, the power-law
exponents in
a model of the form $dN/dM \propto M^{-\beta}$ increase from $\beta \la 2$ to
$\beta \ga 3 $ over the range of galaxy masses in our sample. This steepening
is in turn equivalent to a systematic decrease of the cut-off mass
$M_c$ in evolved-Schechter function fits to the GCLFs, from
$M_c\simeq 2$--$3\times 10^6\,M_\odot$ in the brightest galaxies to
$M_c\simeq\,3$--$4\times10^5\,M_\odot$ in the faintest systems.

\item The GCLF turnover mass $M_{\rm TO}$ is slightly smaller in
dwarf systems ($M_B \ga -18$), relative to the same quantity in more massive
galaxies. In the latter we have 
$M_{\rm TO} = (2.2\pm0.4) \times 10^5 M_{\odot}$, decreasing to
$M_{\rm TO} \simeq 1.6$--$1.7 \times 10^5 M_{\odot}$ on average for
the faintest galaxies in our sample---although individual dwarfs scatter
between $1\times 10^5\,M_\odot \la M_{\rm TO} \la 2\times10^5\,M_\odot$
(\S\ref{sec:trends}). We show that this
might be at least partly accounted for by the effects of dynamical
friction if all other processes shaping the mass function were to lead to
an invariant $M_{\rm TO}$ (\S\ref{ssec:dynfric}).

\item We explored radial variations of the GCLF over baselines of $20-35$ kpc
in M87 (VCC 1316) and M49 (VCC 1226) by studying GCs in the fields of
dwarf galaxies close in projection to these giant ellipticals
(\S\ref{ssec:gradius}). We find no evidence for a
variation of the turnover mass $M_{\rm TO}$ with galactocentric distance in
either galaxy, consistent with previous studies of M87 in particular. This
reinforces the importance of the radial invariance of GCLFs as
a constraint on models of GCLF formation and dynamical evolution.

\item Our success in fitting evolved Schechter functions to our data
(\S\ref{ssec:esfits}) means that the GC mass functions in early-type Virgo
galaxies are consistent with a universally flat shape,
$dN/dM \sim {\rm constant}$, in the limit of low masses---as is also found
in the Milky Way (\S\ref{ssec:modcomp} and Fall \& Zhang 2001). If this
feature is caused by dynamical evolution from a much steeper initial
distribution, it requires that cluster masses decrease linearly in
time. This can plausibly be expected if evaporation dominates the
cluster evolution, although tidal shocks may also lead to similar
behavior.

\item Fits of the evolved Schechter function imply that a narrow range of
average mass losses per GC---$\Delta \approx (2$--$10) \times 10^5 M_{\odot}$
at the outside---is required in all galaxies to 
account for our observed GCLFs. Such a range of $\Delta$ across a
factor of $\approx\! 400$ in galaxy luminosity is in rough agreement with
observed (small) variations in the mean half-mass radii of GCs in the ACSVCS
galaxies (Jord\'an et al.~2005), and with simple scalings of evaporation
rate as a function of host-galaxy luminosity
(\S\ref{ssec:evaporation}). However, more work is required to 
reconcile fully the main idea---that long-term dynamical evolution alone
transformed initial Schechter cluster mass functions into the presently
observed distributions---with the weak radial variation of GCLFs inside large
galaxies and with observations of the orbital distributions and range of mean
cluster densities in the same systems. 

\item The clear decrease of the GC cut-off mass $M_c$ with galaxy luminosity
in evolved-Schechter function descriptions of the GCLF (\S\ref{ssec:esfits})
is too pronounced to be explained by dynamical friction operating on a
universal $dN/dM$ with an initially constant $M_c$ in all galaxies
(\S\ref{ssec:dynfric}). It most likely reflects systematic variations at the
high-mass end of the initial GC mass function.

\end{enumerate}

The present-day mass functions of GCs were likely shaped by a variety of
processes acting on different timescales,
including systematic variations in the initial (proto-)cluster mass
function at the high-mass end; long-term dynamical erosion by
evaporation, tidal shocks, and dynamical friction; and global relaxation
effects in time-varying galaxy potentials (hierarchical merging). It is
further possible, though not yet entirely clear, that
mass-selective early dissolution of clusters due to stellar evolution
may have played some role in defining the observed mass distributions.
Future attempts to understand the whole of the GCLF will clearly
have to consider all of these processes, and their inevitable interplay, in
quite some detail. Such comprehensive modeling will also have to acknowledge
the increasingly complex and stringent empirical constraints that follow from
combining direct GCLF observations with other GC systematics---such as their
structural correlations, and the dynamics of cluster systems---for which
data are continually accumulating and improving in quality.

\acknowledgements

We thank Mike Fall for critical readings of earlier versions of this paper, and
for helpful discussions.
We also thank
M. Kissler-Patig, J. Liske, S. Mieske and S. Zepf 
for useful discussions and the referee, S{\o}ren Larsen, for a careful reading
of the manuscript.
Support for program GO-9401 was provided 
through a grant from the Space
Telescope Science Institute, which is operated by the Association of
Universities for Research in Astronomy, Inc., under NASA contract NAS5-26555.
S.M.  acknowledges additional support from NASA grant NAG5-7697 to the ACS
Team.
This research has made use of the NASA/IPAC Extragalactic Database (NED)
which is operated by the Jet Propulsion Laboratory, California Institute
of Technology, under contract with the National Aeronautics and Space 
Administration.

{\it Facility:} \facility{HST (ACS/WFC)}

%%%%%%%%%%%%%%%%%%%%%%%%%%%%%%%%%%%%%%%%%%%%%%%%%%%%%%%%%%%%%%%%%%%%%%%%%%%%%%%
%%%%%%%%%%%%%%%%%%%%%%%%%%%%% BIBLIOGRAPHY %%%%%%%%%%%%%%%%%%%%%%%%%%%%%%%%%%%%
%%%%%%%%%%%%%%%%%%%%%%%%%%%%%%%%%%%%%%%%%%%%%%%%%%%%%%%%%%%%%%%%%%%%%%%%%%%%%%%

%%%%%%%%%%%%%%%%%%%%%%%%%%%%%%%%%%%%%%%%%%%%%%%%%%%%%%%%%%%%%%%%%%%%%%%%%%%%%%%
%%%%%%%%%%%%%%%%%%%%%%%%%%%%%%%%%% TABLES %%%%%%%%%%%%%%%%%%%%%%%%%%%%%%%%%%%%%
%%%%%%%%%%%%%%%%%%%%%%%%%%%%%%%%%%%%%%%%%%%%%%%%%%%%%%%%%%%%%%%%%%%%%%%%%%%%%%%

\clearpage
\thispagestyle{empty}
\LongTables %C
\begin{landscape} %C
\begin{deluxetable}{cccccccccccccccccccccc}
\tablecaption{Luminosity Function Histograms for GCs and Expected Contaminants\tablenotemark{1}
\label{tab:gclf_hists}}
\tabletypesize{\tiny}
%\rotate%A
\tablewidth{0pt}
\tablehead{
\colhead{} &
\multicolumn{10}{c}{Sample with $p_{\rm GC} \ge 0.5$} &
\colhead{} & 
\multicolumn{10}{c}{Sample with $m_z < 25.15$ or $m_g < 26.35$, and $R_h < 
0\farcs064$}
\\
\cline{2-11} \cline{13-21}
\\
\colhead{VCC} &
\colhead{$m_z$} &
\colhead{$h_z$} &
\colhead{$N_{z, \rm tot}$} &
\colhead{$N_{z, \rm cont}$} &
\colhead{$f_z$} &
\colhead{$m_g$} &
\colhead{$h_g$} &
\colhead{$N_{g, \rm tot}$} &
\colhead{$N_{g, \rm cont}$} &
\colhead{$f_g$} &
\colhead{} &
\colhead{$m_z$} &
\colhead{$h_z$} & 
\colhead{$N_{z, \rm tot}$} &
\colhead{$N_{z, \rm cont}$} &
\colhead{$f_z$} &
\colhead{$m_g$} &
\colhead{$h_g$} &
\colhead{$N_{g, \rm tot}$} &
\colhead{$N_{g, \rm cont}$} &
\colhead{$f_g$}
\\
\colhead{(1)}  & \colhead{(2)}  & \colhead{(3)}  & \colhead{(4)}  &
\colhead{(5)}  & \colhead{(6)}  & \colhead{(7)}  & \colhead{(8)}  &
\colhead{(9)}  & \colhead{(10)} & \colhead{(11)} & \colhead{} &
\colhead{(12)} &
\colhead{(13)} & \colhead{(14)} & \colhead{(15)} & \colhead{(16)} &
\colhead{(17)} & \colhead{(18)} & \colhead{(19)} & \colhead{(20)} &
\colhead{(21)}
}
\startdata
1226 &   18.0 &   0.4 &    0 &    0.0 &  1.00  &    19.2 &   0.4 &    0 &    0.0 &  1.00 & ~  &    18.0 &   0.4 &    0 &    0.1 &  1.00  &    19.2 &   0.4 &    0 &    0.1 &  1.00 \\ 
1226 &   18.4 &   0.4 &    0 &    0.1 &  1.00  &    19.6 &   0.4 &    0 &    0.1 &  1.00 & ~  &    18.4 &   0.4 &    0 &    0.2 &  1.00  &    19.6 &   0.4 &    0 &    0.2 &  1.00 \\ 
1226 &   18.8 &   0.4 &    2 &    0.0 &  1.00  &    20.0 &   0.4 &    1 &    0.1 &  1.00 & ~  &    18.8 &   0.4 &    0 &    0.2 &  1.00  &    20.0 &   0.4 &    0 &    0.2 &  1.00 \\ 
1226 &   19.2 &   0.4 &    5 &    0.2 &  1.00  &    20.4 &   0.4 &    3 &    0.1 &  1.00 & ~  &    19.2 &   0.4 &    5 &    0.2 &  1.00  &    20.4 &   0.4 &    2 &    0.1 &  1.00 \\ 
1226 &   19.6 &   0.4 &    4 &    0.3 &  1.00  &    20.8 &   0.4 &    8 &    0.4 &  1.00 & ~  &    19.6 &   0.4 &    4 &    0.3 &  1.00  &    20.8 &   0.4 &    8 &    0.5 &  1.00 \\ 
1226 &   20.0 &   0.4 &   12 &    0.2 &  1.00  &    21.2 &   0.4 &   11 &    0.3 &  1.00 & ~  &    20.0 &   0.4 &   12 &    0.2 &  1.00  &    21.2 &   0.4 &   10 &    0.4 &  1.00 \\ 
1226 &   20.4 &   0.4 &   25 &    0.4 &  1.00  &    21.6 &   0.4 &   24 &    0.2 &  1.00 & ~  &    20.4 &   0.4 &   24 &    0.5 &  1.00  &    21.6 &   0.4 &   23 &    0.4 &  1.00 \\ 
1226 &   20.8 &   0.4 &   32 &    0.2 &  1.00  &    22.0 &   0.4 &   33 &    0.3 &  1.00 & ~  &    20.8 &   0.4 &   31 &    0.4 &  1.00  &    22.0 &   0.4 &   33 &    0.3 &  1.00 \\ 
1226 &   21.2 &   0.4 &   57 &    0.4 &  1.00  &    22.4 &   0.4 &   59 &    0.5 &  1.00 & ~  &    21.2 &   0.4 &   57 &    0.4 &  1.00  &    22.4 &   0.4 &   58 &    0.5 &  1.00 \\ 
1226 &   21.6 &   0.4 &   66 &    0.6 &  1.00  &    22.8 &   0.4 &   60 &    0.4 &  1.00 & ~  &    21.6 &   0.4 &   62 &    0.6 &  1.00  &    22.8 &   0.4 &   57 &    0.4 &  1.00 \\ 
1226 &   22.0 &   0.4 &   91 &    0.9 &  1.00  &    23.2 &   0.4 &   78 &    0.6 &  1.00 & ~  &    22.0 &   0.4 &   86 &    0.6 &  1.00  &    23.2 &   0.4 &   73 &    0.5 &  1.00 \\ 
1226 &   22.4 &   0.4 &   98 &    0.8 &  0.99  &    23.6 &   0.4 &  101 &    1.3 &  0.98 & ~  &    22.4 &   0.4 &   94 &    0.5 &  0.99  &    23.6 &   0.4 &   99 &    0.8 &  0.98 \\ 
1226 &   22.8 &   0.4 &   95 &    1.6 &  0.94  &    24.0 &   0.4 &  107 &    1.4 &  0.90 & ~  &    22.8 &   0.4 &   90 &    0.9 &  0.94  &    24.0 &   0.4 &  100 &    0.8 &  0.90 \\ 
1226 &   23.2 &   0.4 &   88 &    1.4 &  0.83  &    24.4 &   0.4 &   74 &    1.8 &  0.80 & ~  &    23.2 &   0.4 &   83 &    1.2 &  0.83  &    24.4 &   0.4 &   71 &    1.4 &  0.80 \\ 
1226 &   23.6 &   0.4 &   70 &    2.0 &  0.72  &    24.8 &   0.4 &   78 &    2.5 &  0.71 & ~  &    23.6 &   0.4 &   65 &    1.4 &  0.72  &    24.8 &   0.4 &   72 &    2.1 &  0.71 \\ 
1226 &   24.0 &   0.4 &   61 &    3.4 &  0.62  &    25.2 &   0.4 &   56 &    2.9 &  0.62 & ~  &    24.0 &   0.4 &   60 &    2.8 &  0.62  &    25.2 &   0.4 &   56 &    2.5 &  0.62 \\ 
1226 &   24.4 &   0.4 &   39 &    3.3 &  0.51  &    25.6 &   0.4 &   50 &    2.9 &  0.52 & ~  &    24.4 &   0.4 &   38 &    3.2 &  0.51  &    25.6 &   0.4 &   47 &    2.6 &  0.52 \\ 
1226 &   24.8 &   0.4 &   16 &    1.6 &  0.37  &    26.0 &   0.4 &   18 &    1.8 &  0.37 & ~  &    24.8 &   0.4 &   16 &    1.6 &  0.37  &    26.0 &   0.4 &   18 &    1.7 &  0.37 \\ 
1226 &   25.2 &   0.4 &    3 &    0.4 &  0.19  &    26.4 &   0.4 &    3 &    0.3 &  0.18 & ~  &    25.2 &   0.4 &    3 &    0.4 &  0.19  &    26.4 &   0.4 &    3 &    0.3 &  0.18 \\ 
1316 &   18.0 &   0.4 &    0 &    0.0 &  1.00  &    19.2 &   0.4 &    0 &    0.0 &  1.00 & ~  &    18.0 &   0.4 &    0 &    0.1 &  1.00  &    19.2 &   0.4 &    0 &    0.1 &  1.00 \\ 
\enddata
\tablenotetext{1}{Table \ref{tab:gclf_hists} is presented in its entirety in
the electronic version of this paper. A portion is shown here for guidance
regarding its form and content.}
\tablecomments{
Key to columns---(1) Galaxy VCC number; 
(2)--(3) Mean magnitude and width of bin in the $z$-band; 
(4) Total number of objects in bin with probability $p_{\rm GC} \ge 0.5$ of
 being a globular cluster; 
(5) Expected number of contaminants in bin;
(6) GC completeness fraction in bin;
(7)--(11) Same as (2)--(6) but for the $g$-band;
(12)--(21) Same as (2)--(11) but for GC samples constructed by applying
cuts in magnitude and half-light radius $R_h$, rather than by
selecting on the basis of $p_{\rm GC}$.
}
\end{deluxetable}
\clearpage
\end{landscape} %C

\LongTables %C
\begin{deluxetable}{ccccccccc}
\tablecaption{Gaussian GCLF Parameters for Individual ACSVCS Galaxies
\label{tab:gclfpars}}
\tabletypesize{\footnotesize}
\tablewidth{0pt}
\tablehead{
\colhead{VCC} &
\colhead{$B_{\rm gal}$} & 
\colhead{$\mu_g$} & 
\colhead{$\sigma_g$} &
\colhead{$\mu_z$} & 
\colhead{$\sigma_z$} &
\colhead{$\widehat{\cal B}$} &
\colhead{$N$} &
\colhead{Comments}
\\
\colhead{(1)}  & \colhead{(2)}  & \colhead{(3)}  & \colhead{(4)}  &
\colhead{(5)}  & \colhead{(6)}  & \colhead{(7)}  & \colhead{(8)}  &
\colhead{(9)}
}
\startdata
\object[VCC1226]{1226} &     9.31 & $   24.105 \pm    0.086  $ & $    1.366 \pm    0.061 $ &$   22.789 \pm    0.077 $ & $    1.321 \pm    0.053 $ &    0.023 &  764 & \nodata\\
\object[VCC1316]{1316} &     9.58 & $   24.018 \pm    0.049  $ & $    1.312 \pm    0.035 $ &$   22.689 \pm    0.041 $ & $    1.242 \pm    0.030 $ &    0.014 & 1745 & \nodata\\
\object[VCC1978]{1978} &     9.81 & $   24.062 \pm    0.077  $ & $    1.340 \pm    0.058 $ &$   22.747 \pm    0.070 $ & $    1.316 \pm    0.050 $ &    0.022 &  807 & \nodata\\
\object[VCC 881]{ 881} &    10.06 & $   23.950 \pm    0.097  $ & $    1.274 \pm    0.075 $ &$   22.834 \pm    0.093 $ & $    1.238 \pm    0.071 $ &    0.034 &  367 & \nodata\\
\object[VCC 798]{ 798} &    10.09 & $   25.120 \pm    0.232  $ & $    1.708 \pm    0.130 $ &$   23.722 \pm    0.179 $ & $    1.562 \pm    0.102 $ &    0.016 &  507 & Faint excess \\
\object[VCC 763]{ 763} &    10.26 & $   23.973 \pm    0.074  $ & $    1.178 \pm    0.055 $ &$   22.836 \pm    0.070 $ & $    1.159 \pm    0.052 $ &    0.035 &  506 & \nodata\\
\object[VCC 731]{ 731} &    10.51 & $   24.403 \pm    0.061  $ & $    1.207 \pm    0.046 $ &$   23.211 \pm    0.059 $ & $    1.199 \pm    0.044 $ &    0.021 &  907 & \nodata\\
\object[VCC1535]{1535} &    10.61 & $   23.685 \pm    0.097  $ & $    1.079 \pm    0.076 $ &$   22.512 \pm    0.092 $ & $    1.063 \pm    0.067 $ &    0.042 &  244 & \nodata\\
\object[VCC1903]{1903} &    10.76 & $   23.446 \pm    0.089  $ & $    1.192 \pm    0.071 $ &$   22.255 \pm    0.089 $ & $    1.215 \pm    0.073 $ &    0.046 &  308 & \nodata\\
\object[VCC1632]{1632} &    10.78 & $   23.951 \pm    0.103  $ & $    1.423 \pm    0.077 $ &$   22.717 \pm    0.095 $ & $    1.390 \pm    0.071 $ &    0.038 &  456 & \nodata\\
\object[VCC1231]{1231} &    11.10 & $   23.715 \pm    0.090  $ & $    1.103 \pm    0.072 $ &$   22.592 \pm    0.089 $ & $    1.106 \pm    0.069 $ &    0.058 &  254 & \nodata\\
\object[VCC2095]{2095} &    11.18 & $   24.429 \pm    0.296  $ & $    1.564 \pm    0.226 $ &$   23.503 \pm    0.333 $ & $    1.615 \pm    0.209 $ &    0.076 &  134 & Faint excess \\
\object[VCC1154]{1154} &    11.37 & $   23.902 \pm    0.092  $ & $    0.988 \pm    0.072 $ &$   22.813 \pm    0.094 $ & $    1.001 \pm    0.072 $ &    0.065 &  192 & \nodata\\
\object[VCC1062]{1062} &    11.40 & $   23.687 \pm    0.133  $ & $    1.218 \pm    0.110 $ &$   22.548 \pm    0.123 $ & $    1.203 \pm    0.097 $ &    0.066 &  179 & \nodata\\
\object[VCC2092]{2092} &    11.51 & $   24.009 \pm    0.198  $ & $    1.111 \pm    0.176 $ &$   22.882 \pm    0.186 $ & $    1.135 \pm    0.148 $ &    0.114 &   92 & \nodata\\
\object[VCC 369]{ 369} &    11.80 & $   23.622 \pm    0.117  $ & $    1.102 \pm    0.102 $ &$   22.447 \pm    0.108 $ & $    1.077 \pm    0.091 $ &    0.068 &  179 & Faint excess \\
\object[VCC 759]{ 759} &    11.80 & $   23.805 \pm    0.121  $ & $    1.120 \pm    0.098 $ &$   22.689 \pm    0.114 $ & $    1.084 \pm    0.090 $ &    0.067 &  172 & \nodata\\
\object[VCC1692]{1692} &    11.82 & $   23.872 \pm    0.146  $ & $    1.073 \pm    0.123 $ &$   22.831 \pm    0.153 $ & $    1.120 \pm    0.117 $ &    0.096 &  136 & \nodata\\
\object[VCC1030]{1030} &    11.84 & $   23.737 \pm    0.098  $ & $    0.980 \pm    0.078 $ &$   22.621 \pm    0.098 $ & $    1.021 \pm    0.076 $ &    0.072 &  176 & \nodata\\
\object[VCC2000]{2000} &    11.94 & $   23.482 \pm    0.119  $ & $    1.183 \pm    0.100 $ &$   22.471 \pm    0.109 $ & $    1.159 \pm    0.087 $ &    0.071 &  197 & \nodata\\
\object[VCC 685]{ 685} &    11.99 & $   23.692 \pm    0.135  $ & $    1.248 \pm    0.110 $ &$   22.584 \pm    0.127 $ & $    1.213 \pm    0.104 $ &    0.085 &  167 & \nodata\\
\object[VCC1664]{1664} &    12.02 & $   23.675 \pm    0.121  $ & $    1.049 \pm    0.094 $ &$   22.502 \pm    0.110 $ & $    1.009 \pm    0.086 $ &    0.092 &  146 & \nodata\\
\object[VCC 654]{ 654} &    12.03 & $   23.981 \pm    0.200  $ & $    0.911 \pm    0.192 $ &$   23.053 \pm    0.207 $ & $    0.930 \pm    0.166 $ &    0.194 &   48 & \nodata\\
\object[VCC 944]{ 944} &    12.08 & $   23.721 \pm    0.140  $ & $    0.868 \pm    0.121 $ &$   22.712 \pm    0.140 $ & $    0.893 \pm    0.114 $ &    0.132 &   91 & \nodata\\
\object[VCC1938]{1938} &    12.11 & $   23.798 \pm    0.145  $ & $    1.077 \pm    0.123 $ &$   22.830 \pm    0.140 $ & $    1.020 \pm    0.130 $ &    0.113 &  101 & \nodata\\
\object[VCC1279]{1279} &    12.15 & $   23.666 \pm    0.111  $ & $    1.031 \pm    0.088 $ &$   22.612 \pm    0.111 $ & $    1.035 \pm    0.086 $ &    0.097 &  138 & \nodata\\
\object[VCC1720]{1720} &    12.29 & $   23.672 \pm    0.159  $ & $    0.798 \pm    0.150 $ &$   22.615 \pm    0.161 $ & $    0.871 \pm    0.141 $ &    0.141 &   71 & \nodata\\
\object[VCC 355]{ 355} &    12.41 & $   24.618 \pm    0.364  $ & $    1.221 \pm    0.250 $ &$   23.406 \pm    0.239 $ & $    1.036 \pm    0.168 $ &    0.167 &   62 & \nodata\\
\object[VCC1619]{1619} &    12.50 & $   24.255 \pm    0.238  $ & $    1.050 \pm    0.207 $ &$   23.178 \pm    0.235 $ & $    1.061 \pm    0.175 $ &    0.165 &   66 & \nodata\\
\object[VCC1883]{1883} &    12.57 & $   24.135 \pm    0.217  $ & $    1.106 \pm    0.175 $ &$   23.066 \pm    0.184 $ & $    1.064 \pm    0.144 $ &    0.124 &   83 & \nodata\\
\object[VCC1242]{1242} &    12.60 & $   23.741 \pm    0.130  $ & $    0.919 \pm    0.115 $ &$   22.624 \pm    0.126 $ & $    0.963 \pm    0.101 $ &    0.105 &  116 & \nodata\\
\object[VCC 784]{ 784} &    12.67 & $   24.299 \pm    0.203  $ & $    0.870 \pm    0.188 $ &$   23.122 \pm    0.179 $ & $    0.813 \pm    0.164 $ &    0.178 &   64 & \nodata\\
\object[VCC1537]{1537} &    12.70 & $   23.688 \pm    0.279  $ & $    0.977 \pm    0.279 $ &$   22.789 \pm    0.328 $ & $    1.143 \pm    0.274 $ &    0.256 &   45 & \nodata\\
\object[VCC 778]{ 778} &    12.72 & $   24.197 \pm    0.215  $ & $    1.081 \pm    0.166 $ &$   23.120 \pm    0.204 $ & $    1.043 \pm    0.147 $ &    0.163 &   74 & \nodata\\
\object[VCC1321]{1321} &    12.84 & $   24.025 \pm    0.275  $ & $    0.831 \pm    0.276 $ &$   23.057 \pm    0.255 $ & $    0.849 \pm    0.223 $ &    0.198 &   50 & \nodata\\
\object[VCC 828]{ 828} &    12.84 & $   23.817 \pm    0.177  $ & $    1.042 \pm    0.159 $ &$   22.800 \pm    0.147 $ & $    0.902 \pm    0.126 $ &    0.143 &   80 & \nodata\\
\object[VCC1250]{1250} &    12.91 & $   23.585 \pm    0.163  $ & $    0.815 \pm    0.147 $ &$   22.611 \pm    0.165 $ & $    0.834 \pm    0.132 $ &    0.200 &   54 & \nodata\\
\object[VCC1630]{1630} &    12.91 & $   24.201 \pm    0.406  $ & $    1.310 \pm    0.313 $ &$   23.150 \pm    0.361 $ & $    1.316 \pm    0.252 $ &    0.217 &   57 & \nodata\\
\object[VCC1146]{1146} &    12.93 & $   23.895 \pm    0.153  $ & $    0.901 \pm    0.185 $ &$   22.757 \pm    0.180 $ & $    0.892 \pm    0.173 $ &    0.148 &   82 & \nodata\\
\object[VCC1025]{1025} &    13.06 & $   24.265 \pm    0.125  $ & $    0.844 \pm    0.117 $ &$   23.357 \pm    0.148 $ & $    0.933 \pm    0.116 $ &    0.143 &  104 & \nodata\\
\object[VCC1303]{1303} &    13.10 & $   23.640 \pm    0.150  $ & $    0.780 \pm    0.128 $ &$   22.836 \pm    0.158 $ & $    0.807 \pm    0.122 $ &    0.176 &   61 & \nodata\\
\object[VCC1913]{1913} &    13.22 & $   23.764 \pm    0.137  $ & $    0.750 \pm    0.128 $ &$   22.749 \pm    0.144 $ & $    0.759 \pm    0.120 $ &    0.180 &   65 & \nodata\\
\object[VCC1327]{1327} &    13.26 & $   23.686 \pm    0.133  $ & $    1.259 \pm    0.107 $ &$   22.624 \pm    0.118 $ & $    1.211 \pm    0.094 $ &    0.081 &  173 & VCC1316 Companion\\
\object[VCC1125]{1125} &    13.30 & $   23.701 \pm    0.144  $ & $    0.791 \pm    0.144 $ &$   22.650 \pm    0.146 $ & $    0.783 \pm    0.123 $ &    0.179 &   62 & \nodata\\
\object[VCC1475]{1475} &    13.36 & $   24.094 \pm    0.159  $ & $    0.999 \pm    0.155 $ &$   23.239 \pm    0.190 $ & $    1.112 \pm    0.146 $ &    0.138 &   85 & \nodata\\
\object[VCC1178]{1178} &    13.37 & $   23.621 \pm    0.148  $ & $    1.002 \pm    0.114 $ &$   22.574 \pm    0.129 $ & $    0.953 \pm    0.093 $ &    0.124 &   90 & \nodata\\
\object[VCC1283]{1283} &    13.45 & $   24.058 \pm    0.172  $ & $    0.880 \pm    0.155 $ &$   23.062 \pm    0.179 $ & $    0.930 \pm    0.139 $ &    0.170 &   66 & \nodata\\
\object[VCC1261]{1261} &    13.56 & $   24.038 \pm    0.350  $ & $    1.146 \pm    0.342 $ &$   23.058 \pm    0.358 $ & $    1.246 \pm    0.263 $ &    0.217 &   46 & \nodata\\
\object[VCC 698]{ 698} &    13.60 & $   23.793 \pm    0.096  $ & $    0.832 \pm    0.071 $ &$   22.799 \pm    0.089 $ & $    0.814 \pm    0.064 $ &    0.105 &  119 & \nodata\\
\object[VCC1422]{1422} &    13.64 & $   23.711 \pm    0.276  $ & $    0.703 \pm    0.261 $ &$   22.595 \pm    0.236 $ & $    0.694 \pm    0.220 $ &    0.256 &   37 & \nodata\\
\object[VCC2048]{2048} &    13.81 & $   23.481 \pm    0.508  $ & $    0.976 \pm    0.303 $ &$   22.444 \pm    0.340 $ & $    0.893 \pm    0.284 $ &    0.420 &   22 & \nodata\\
\object[VCC1871]{1871} &    13.86 & $   23.597 \pm    0.739  $ & $    1.194 \pm    0.588 $ &$   22.619 \pm    0.690 $ & $    1.190 \pm    0.581 $ &    0.516 &   18 & \nodata\\
\object[VCC   9]{   9} &    13.93 & $   23.863 \pm    0.547  $ & $    1.023 \pm    0.378 $ &$   22.833 \pm    0.371 $ & $    0.897 \pm    0.236 $ &    0.246 &   34 & \nodata\\
\object[VCC 575]{ 575} &    14.14 & $   24.952 \pm    0.263  $ & $    0.558 \pm    0.219 $ &$   23.881 \pm    0.333 $ & $    0.316 \pm    0.362 $ &    0.386 &   27 & \nodata\\
\object[VCC1910]{1910} &    14.17 & $   23.787 \pm    0.237  $ & $    1.181 \pm    0.198 $ &$   22.655 \pm    0.215 $ & $    1.141 \pm    0.185 $ &    0.180 &   60 & \nodata\\
\object[VCC1049]{1049} &    14.20 & $   24.110 \pm    0.564  $ & $    0.559 \pm    0.530 $ &$   23.221 \pm    0.463 $ & $    0.671 \pm    0.373 $ &    0.487 &   18 & \nodata\\
\object[VCC 856]{ 856} &    14.25 & $   23.886 \pm    0.263  $ & $    0.922 \pm    0.189 $ &$   22.797 \pm    0.193 $ & $    0.870 \pm    0.139 $ &    0.211 &   50 & \nodata\\
\object[VCC 140]{ 140} &    14.30 & $   24.029 \pm    0.321  $ & $    0.800 \pm    0.281 $ &$   23.027 \pm    0.300 $ & $    0.822 \pm    0.196 $ &    0.327 &   29 & \nodata\\
\object[VCC1355]{1355} &    14.31 & $   24.536 \pm    0.957  $ & $    1.260 \pm    0.714 $ &$   23.696 \pm    0.785 $ & $    1.168 \pm    0.675 $ &    0.468 &   20 & \nodata\\
\object[VCC1087]{1087} &    14.31 & $   23.741 \pm    0.151  $ & $    0.929 \pm    0.120 $ &$   22.722 \pm    0.139 $ & $    0.900 \pm    0.119 $ &    0.162 &   68 & \nodata\\
\object[VCC1297]{1297} &    14.33 & $   23.401 \pm    0.119  $ & $    1.140 \pm    0.097 $ &$   22.298 \pm    0.106 $ & $    1.083 \pm    0.086 $ &    0.092 &  152 & VCC1316 Companion\\
\object[VCC1861]{1861} &    14.37 & $   23.688 \pm    0.293  $ & $    1.042 \pm    0.243 $ &$   22.603 \pm    0.225 $ & $    0.953 \pm    0.187 $ &    0.233 &   49 & \nodata\\
\object[VCC 543]{ 543} &    14.39 & $   23.908 \pm    0.235  $ & $    0.701 \pm    0.177 $ &$   22.844 \pm    0.195 $ & $    0.646 \pm    0.139 $ &    0.330 &   28 & \nodata\\
\object[VCC1431]{1431} &    14.51 & $   24.132 \pm    0.190  $ & $    1.050 \pm    0.169 $ &$   23.112 \pm    0.199 $ & $    1.088 \pm    0.149 $ &    0.158 &   71 & \nodata\\
\object[VCC1528]{1528} &    14.51 & $   23.552 \pm    0.149  $ & $    0.717 \pm    0.119 $ &$   22.621 \pm    0.136 $ & $    0.702 \pm    0.115 $ &    0.221 &   49 & \nodata\\
\object[VCC1695]{1695} &    14.53 & $   24.408 \pm    0.461  $ & $    0.957 \pm    0.481 $ &$   23.462 \pm    0.558 $ & $    1.093 \pm    0.434 $ &    0.380 &   22 & \nodata\\
\object[VCC1833]{1833} &    14.54 & $   24.116 \pm    0.268  $ & $    0.700 \pm    0.246 $ &$   22.953 \pm    0.160 $ & $    0.500 \pm    0.209 $ &    0.332 &   28 & \nodata\\
\object[VCC 437]{ 437} &    14.54 & $   23.942 \pm    0.198  $ & $    0.782 \pm    0.169 $ &$   23.063 \pm    0.179 $ & $    0.845 \pm    0.140 $ &    0.229 &   50 & \nodata\\
\object[VCC2019]{2019} &    14.55 & $   23.543 \pm    0.255  $ & $    0.858 \pm    0.258 $ &$   22.612 \pm    0.236 $ & $    0.849 \pm    0.213 $ &    0.303 &   34 & \nodata\\
\object[VCC 200]{ 200} &    14.69 & $   24.471 \pm    0.326  $ & $    0.672 \pm    0.463 $ &$   23.578 \pm    0.402 $ & $    0.825 \pm    0.359 $ &    0.379 &   25 & \nodata\\
\object[VCC 571]{ 571} &    14.74 & $   24.362 \pm    0.684  $ & $    0.938 \pm    0.822 $ &$   24.366 \pm    1.728 $ & $    1.460 \pm    0.956 $ &    0.478 &   17 & \nodata\\
\object[VCC  21]{  21} &    14.75 & $   24.332 \pm    0.802  $ & $    1.427 \pm    0.802 $ &$   23.293 \pm    0.701 $ & $    1.350 \pm    0.478 $ &    0.351 &   26 & \nodata\\
\object[VCC1488]{1488} &    14.76 & $   24.137 \pm    0.421  $ & $    0.573 \pm    0.364 $ &$   23.030 \pm    0.539 $ & $    0.511 \pm    0.392 $ &    0.471 &   19 & \nodata\\
\object[VCC1499]{1499} &    14.94 & $   24.496 \pm    0.691  $ & $    1.352 \pm    0.608 $ &$   23.806 \pm    0.674 $ & $    1.325 \pm    0.387 $ &    0.271 &   35 & \nodata\\
\object[VCC1545]{1545} &    14.96 & $   24.079 \pm    0.178  $ & $    0.884 \pm    0.175 $ &$   23.148 \pm    0.178 $ & $    0.894 \pm    0.152 $ &    0.189 &   63 & \nodata\\
\object[VCC1192]{1192} &    15.04 & $   23.777 \pm    0.091  $ & $    1.070 \pm    0.072 $ &$   22.660 \pm    0.086 $ & $    1.049 \pm    0.067 $ &    0.064 &  213 & VCC1226 Companion\\
\object[VCC1075]{1075} &    15.08 & $   23.514 \pm    0.240  $ & $    0.553 \pm    0.247 $ &$   22.522 \pm    0.197 $ & $    0.515 \pm    0.226 $ &    0.378 &   26 & \nodata\\
\object[VCC1440]{1440} &    15.20 & $   24.280 \pm    0.278  $ & $    0.887 \pm    0.251 $ &$   23.298 \pm    0.228 $ & $    0.826 \pm    0.175 $ &    0.259 &   38 & \nodata\\
\object[VCC 230]{ 230} &    15.20 & $   23.957 \pm    0.218  $ & $    0.545 \pm    0.198 $ &$   23.099 \pm    0.336 $ & $    0.581 \pm    0.319 $ &    0.274 &   38 & \nodata\\
\object[VCC2050]{2050} &    15.20 & $   23.900 \pm    0.436  $ & $    0.281 \pm    0.536 $ &$   22.964 \pm    0.389 $ & $    0.304 \pm    0.370 $ &    0.459 &   20 & \nodata\\
\object[VCC 751]{ 751} &    15.30 & $   23.508 \pm    0.267  $ & $    0.493 \pm    0.212 $ &$   22.674 \pm    0.247 $ & $    0.501 \pm    0.177 $ &    0.495 &   17 & \nodata\\
\object[VCC1828]{1828} &    15.33 & $   23.806 \pm    0.250  $ & $    0.701 \pm    0.265 $ &$   22.757 \pm    0.283 $ & $    0.664 \pm    0.329 $ &    0.355 &   27 & \nodata\\
\object[VCC1407]{1407} &    15.49 & $   24.449 \pm    0.144  $ & $    0.666 \pm    0.145 $ &$   23.468 \pm    0.150 $ & $    0.747 \pm    0.120 $ &    0.186 &   60 & \nodata\\
\object[VCC1886]{1886} &    15.49 & $   23.027 \pm    0.984  $ & $    0.967 \pm    1.086 $ &$   21.565 \pm    0.520 $ & $    0.463 \pm    0.566 $ &    0.622 &   14 & \nodata\\
\object[VCC1199]{1199} &    15.50 & $   23.828 \pm    0.102  $ & $    1.163 \pm    0.084 $ &$   22.679 \pm    0.092 $ & $    1.123 \pm    0.072 $ &    0.060 &  228 & VCC1226 Companion\\
\object[VCC1539]{1539} &    15.68 & $   23.810 \pm    0.213  $ & $    0.826 \pm    0.214 $ &$   22.821 \pm    0.207 $ & $    0.901 \pm    0.163 $ &    0.275 &   43 & \nodata\\
\object[VCC1185]{1185} &    15.68 & $   23.840 \pm    0.197  $ & $    0.691 \pm    0.137 $ &$   22.910 \pm    0.159 $ & $    0.639 \pm    0.113 $ &    0.292 &   33 & \nodata\\
\object[VCC1489]{1489} &    15.89 & $   23.977 \pm    0.439  $ & $    0.378 \pm    0.260 $ &$   23.156 \pm    0.381 $ & $    0.482 \pm    0.526 $ &    0.417 &   22 & \nodata\\
\object[VCC1661]{1661} &    15.97 & $   24.177 \pm    0.154  $ & $    0.225 \pm    0.201 $ &$   23.059 \pm    0.417 $ & $    0.615 \pm    0.336 $ &    0.477 &   19 & \nodata\\
\enddata
\tablecomments{
Key to columns---(1) Galaxy VCC number; 
(2) Galaxy $B$ magnitude from Binggeli et~al. (1995); 
(3)--(4) Maximum-likelihood estimates of the 
Gaussian mean $\mu$ and dispersion $\sigma$ of the $g$-band GCLF;
(5)--(6) Same as (3)--(4) but for the $z$-band;
(7) Fraction $\widehat{\cal B}$ of the sample that is expected to be 
contamination;
(8) Total number $N$ of all objects (including contaminants
and uncorrected for incompleteness) with $p_{\rm GC}\ge 0.5$; 
(9) Comments on individual galaxies.
}
\end{deluxetable}

\clearpage
\thispagestyle{empty}
\setlength{\voffset}{15mm}
\LongTables %C
\begin{landscape} %C
\begin{deluxetable}{cccccccccccccc}
\tablecaption{Definition of Binned GC Samples and Best-Fit GCLF Parameters
\label{tab:galbins}}
\tabletypesize{\tiny}
%\rotate %A
\tablewidth{0pt}
\tablehead{
\colhead{} & \colhead{} & \colhead{} & \colhead{} & \colhead{} & 
\colhead{} &
\multicolumn{4}{c}{Gaussian Fits} &
\multicolumn{4}{c}{Evolved Schechter Function Fits}
\\ \cline{7-10} \cline{11-14} \\
\colhead{Group} &
\colhead{$N_{\rm gal}$} &
\colhead{$\langle M_{B,{\rm gal}} \rangle$} &
\colhead{$M_{B,{\rm gal}}^{\rm min}$} &
\colhead{$M_{B,{\rm gal}}^{\rm max}$} &
\colhead{$N_{\rm GC}$} &
\colhead{$\mu_g$} & \colhead{$\sigma_g$} & \colhead{$\mu_z$} & 
\colhead{$\sigma_z$} &
\colhead{$\delta_g$} & \colhead{$m_{c,g}$} &
\colhead{$\delta_z$} & \colhead{$m_{c,z}$}
\\
\colhead{(1)}  & \colhead{(2)}  & \colhead{(3)}  & \colhead{(4)}  &
\colhead{(5)}  & \colhead{(6)}  & \colhead{(7)}  & \colhead{(8)}  &
\colhead{(9)}  & \colhead{(10)} & \colhead{(11)} & \colhead{(12)} &
\colhead{(13)} & \colhead{(14)}
}
\startdata
 0 &  1 (1226) &    -21.8 &    -21.8 &    -21.8 & 746 & $   -7.025 \pm    0.086 $ &  $    1.366 \pm    0.061 $ & $   -8.341 \pm    0.077 $ & $    1.321 \pm    0.053 $ & $   -7.150 \pm    0.133 $ &  $  -10.045 \pm    0.362 $ & $   -8.465 \pm    0.132 $ & $  -11.257 \pm    0.360 $ \\ 
 1 &  1 (1316) &    -21.5 &    -21.5 &    -21.5 & 1721 & $   -7.104 \pm    0.049 $ &  $    1.312 \pm    0.035 $ & $   -8.433 \pm    0.041 $ & $    1.242 \pm    0.030 $ & $   -7.287 \pm    0.089 $ &  $   -9.850 \pm    0.232 $ & $   -8.690 \pm    0.092 $ & $  -10.911 \pm    0.232 $ \\ 
 2 &  1 (1978) &    -21.3 &    -21.3 &    -21.3 & 789 & $   -7.014 \pm    0.077 $ &  $    1.340 \pm    0.058 $ & $   -8.329 \pm    0.070 $ & $    1.316 \pm    0.050 $ & $   -7.265 \pm    0.137 $ &  $   -9.750 \pm    0.356 $ & $   -8.617 \pm    0.146 $ & $  -10.928 \pm    0.381 $ \\ 
 3 &  1 (881) &    -21.2 &    -21.2 &    -21.2 & 355 & $   -7.334 \pm    0.097 $ &  $    1.274 \pm    0.075 $ & $   -8.450 \pm    0.093 $ & $    1.238 \pm    0.071 $ & $   -7.533 \pm    0.198 $ &  $   -9.877 \pm    0.525 $ & $   -8.607 \pm    0.221 $ & $  -11.043 \pm    0.647 $ \\ 
 4 &  1 (763) &    -21.1 &    -21.1 &    -21.1 & 488 & $   -7.385 \pm    0.074 $ &  $    1.178 \pm    0.055 $ & $   -8.522 \pm    0.070 $ & $    1.159 \pm    0.052 $ & $   -7.786 \pm    0.201 $ &  $   -9.371 \pm    0.460 $ & $   -8.955 \pm    0.200 $ & $  -10.499 \pm    0.457 $ \\ 
 5 &  1 (731) &    -21.3 &    -21.3 &    -21.3 & 888 & $   -7.431 \pm    0.061 $ &  $    1.207 \pm    0.046 $ & $   -8.623 \pm    0.059 $ & $    1.199 \pm    0.044 $ & $   -7.651 \pm    0.137 $ &  $   -9.789 \pm    0.339 $ & $   -8.818 \pm    0.134 $ & $  -11.011 \pm    0.337 $ \\ 
 6 &  1 (1903) &    -20.1 &    -20.1 &    -20.1 & 294 & $   -7.434 \pm    0.089 $ &  $    1.192 \pm    0.071 $ & $   -8.625 \pm    0.089 $ & $    1.215 \pm    0.073 $ & $   -7.641 \pm    0.209 $ &  $   -9.971 \pm    0.611 $ & $   -8.776 \pm    0.188 $ & $  -11.418 \pm    0.576 $ \\ 
 7 &  1 (1632) &    -20.3 &    -20.3 &    -20.3 & 439 & $   -7.089 \pm    0.103 $ &  $    1.423 \pm    0.077 $ & $   -8.323 \pm    0.095 $ & $    1.390 \pm    0.071 $ & $   -7.198 \pm    0.152 $ &  $  -10.393 \pm    0.474 $ & $   -8.443 \pm    0.154 $ & $  -11.516 \pm    0.469 $ \\ 
 8 &  1 (1231) &    -19.8 &    -19.8 &    -19.8 & 239 & $   -7.221 \pm    0.090 $ &  $    1.103 \pm    0.072 $ & $   -8.344 \pm    0.089 $ & $    1.106 \pm    0.069 $ & $   -7.841 \pm    0.311 $ &  $   -8.888 \pm    0.669 $ & $   -9.013 \pm    0.319 $ & $  -10.002 \pm    0.680 $ \\ 
 9 &  2 &    -19.6 &    -19.7 &    -19.5 & 347 & $   -7.196 \pm    0.074 $ &  $    1.102 \pm    0.057 $ & $   -8.308 \pm    0.076 $ & $    1.103 \pm    0.057 $ & $   -7.749 \pm    0.279 $ &  $   -8.909 \pm    0.635 $ & $   -8.946 \pm    0.294 $ & $   -9.957 \pm    0.647 $ \\ 
10 &  2 &    -19.4 &    -19.5 &    -19.2 & 248 & $   -7.282 \pm    0.088 $ &  $    1.111 \pm    0.069 $ & $   -8.444 \pm    0.089 $ & $    1.103 \pm    0.069 $ & $   -8.276 \pm    0.424 $ &  $   -8.597 \pm    0.838 $ & $   -9.726 \pm    0.557 $ & $   -9.586 \pm    1.040 $ \\ 
11 &  2 &    -19.4 &    -19.4 &    -19.4 & 283 & $   -7.341 \pm    0.086 $ &  $    1.092 \pm    0.066 $ & $   -8.426 \pm    0.089 $ & $    1.096 \pm    0.068 $ & $   -8.485 \pm    0.517 $ &  $   -8.478 \pm    0.974 $ & $   -9.431 \pm    0.453 $ & $   -9.696 \pm    0.879 $ \\ 
12 &  2 &    -19.1 &    -19.3 &    -18.9 & 347 & $   -7.380 \pm    0.072 $ &  $    1.091 \pm    0.055 $ & $   -8.447 \pm    0.071 $ & $    1.092 \pm    0.055 $ & $   -8.262 \pm    0.323 $ &  $   -8.767 \pm    0.652 $ & $   -9.410 \pm    0.336 $ & $   -9.814 \pm    0.667 $ \\ 
13 &  2 &    -19.0 &    -19.0 &    -19.0 & 212 & $   -7.333 \pm    0.084 $ &  $    0.982 \pm    0.065 $ & $   -8.446 \pm    0.083 $ & $    0.968 \pm    0.066 $ & $   -8.321 \pm    0.519 $ &  $   -8.467 \pm    1.013 $ & $   -9.596 \pm    0.602 $ & $   -9.482 \pm    1.131 $ \\ 
14 &  2 &    -19.1 &    -19.1 &    -19.0 & 214 & $   -7.459 \pm    0.084 $ &  $    1.051 \pm    0.065 $ & $   -8.479 \pm    0.085 $ & $    1.042 \pm    0.068 $ & $   -8.028 \pm    0.335 $ &  $   -9.134 \pm    0.766 $ & $   -9.212 \pm    0.443 $ & $   -9.943 \pm    0.964 $ \\ 
15 &  4 &    -18.5 &    -18.6 &    -18.3 & 283 & $   -6.908 \pm    0.088 $ &  $    1.042 \pm    0.067 $ & $   -7.996 \pm    0.086 $ & $    1.032 \pm    0.066 $ & $   -8.117 \pm    0.643 $ &  $   -7.887 \pm    1.186 $ & $   -9.330 \pm    0.806 $ & $   -8.931 \pm    1.463 $ \\ 
16 &  5 &    -18.3 &    -18.5 &    -18.1 & 257 & $   -7.081 \pm    0.089 $ &  $    1.012 \pm    0.070 $ & $   -8.146 \pm    0.087 $ & $    0.969 \pm    0.068 $ & $   -8.399 \pm    0.648 $ &  $   -7.972 \pm    1.158 $ & $   -9.413 \pm    0.642 $ & $   -9.057 \pm    1.157 $ \\ 
17 &  4 &    -18.2 &    -18.3 &    -18.0 & 208 & $   -7.338 \pm    0.087 $ &  $    0.949 \pm    0.072 $ & $   -8.330 \pm    0.087 $ & $    0.945 \pm    0.068 $ & $   -9.619 \pm    2.113 $ &  $   -7.835 \pm    6.880 $ & $   -9.929 \pm    0.949 $ & $   -9.095 \pm    1.615 $ \\ 
18 &  3 &    -17.8 &    -18.0 &    -17.6 & 205 & $   -7.263 \pm    0.082 $ &  $    0.951 \pm    0.062 $ & $   -8.276 \pm    0.085 $ & $    0.961 \pm    0.065 $ & $   -8.610 \pm    0.675 $ &  $   -8.111 \pm    1.202 $ & $   -9.882 \pm    0.802 $ & $   -9.041 \pm    1.371 $ \\ 
19 &  3 &    -17.7 &    -17.8 &    -17.7 & 197 & $   -7.409 \pm    0.079 $ &  $    0.901 \pm    0.059 $ & $   -8.421 \pm    0.081 $ & $    0.905 \pm    0.060 $ & $  -10.101 \pm    1.986 $ &  $   -7.839 \pm    7.800 $ & $  -10.042 \pm    0.896 $ & $   -9.133 \pm    1.512 $ \\ 
20 &  8 &    -17.1 &    -17.5 &    -16.8 & 196 & $   -7.149 \pm    0.099 $ &  $    0.953 \pm    0.080 $ & $   -8.216 \pm    0.094 $ & $    0.927 \pm    0.072 $ & $   -8.247 \pm    0.680 $ &  $   -8.025 \pm    1.253 $ & $   -9.371 \pm    0.630 $ & $   -9.161 \pm    1.154 $ \\ 
21 &  6 &    -16.6 &    -16.8 &    -16.5 & 222 & $   -7.217 \pm    0.080 $ &  $    0.943 \pm    0.060 $ & $   -8.240 \pm    0.080 $ & $    0.916 \pm    0.060 $ & $   -8.343 \pm    0.591 $ &  $   -8.166 \pm    1.103 $ & $   -9.609 \pm    0.656 $ & $   -9.079 \pm    1.155 $ \\ 
22 &  9 &    -16.4 &    -16.7 &    -16.1 & 193 & $   -7.072 \pm    0.086 $ &  $    0.875 \pm    0.068 $ & $   -8.043 \pm    0.096 $ & $    0.921 \pm    0.071 $ & $   -9.383 \pm    1.795 $ &  $   -7.437 \pm    4.766 $ & $   -8.974 \pm    0.505 $ & $   -9.135 \pm    0.951 $ \\ 
23 & 10 &    -15.7 &    -16.0 &    -15.4 & 201 & $   -7.133 \pm    0.072 $ &  $    0.749 \pm    0.058 $ & $   -8.090 \pm    0.077 $ & $    0.762 \pm    0.062 $ & $   -9.792 \pm    0.088 $ &  $   -7.292 \pm    1.956 $ & $  -10.815 \pm    0.092 $ & $   -8.315 \pm    5.536 $ \\ 
\enddata
\tablecomments{
Key to columns---(1) Identification number of binned group; (2) Number of
galaxies that were used in the creation of this binned group. When only one
galaxy present its VCC identifier is indicated; 
(3)--(5) Average, minimum, and maximum $M_{B, \rm gal}$ 
of galaxies in this binned group;
(6) Number of expected GCs; 
(7)--(10) Best-fit Gaussian GCLF parameters $\mu$ and $\sigma$ 
in the $g$- and $z$-bands;
(11)--(14) Best-fit evolved Schechter GCLF parameters 
$\delta$ and $m_c$, in the $g$- and $z$-bands.
}
\end{deluxetable}
\clearpage
\setlength{\voffset}{0mm}
\end{landscape} %C

\begin{deluxetable}{rcccrcccrcc}
\tablecaption{Best-fit power law exponent $\beta$
\label{tab:beta}}
\tabletypesize{\small}
\tablewidth{0pt}
\tablehead{
\colhead{VCC} & \colhead{$\beta_z$} & \colhead{$\beta_g$} &\colhead{~} &
\colhead{VCC} & \colhead{$\beta_z$} & \colhead{$\beta_g$} &\colhead{~} &
\colhead{VCC} & \colhead{$\beta_z$} & \colhead{$\beta_g$} \\
\colhead{(1)} & \colhead{(2)} & \colhead{(3)} &  \colhead{~} &
\colhead{(1)} & \colhead{(2)} & \colhead{(3)} &  \colhead{~} &
\colhead{(1)} & \colhead{(2)} & \colhead{(3)}}
\startdata
1226 & $     1.80 \pm     0.11 $ & $     1.72 \pm     0.11 $ & ~ &  654 & $     2.58 \pm     0.76 $ & $     2.73 \pm     0.72 $ & ~ & 1178 & $     2.42 \pm     0.40 $ & $     1.82 \pm     0.38 $ \\ 
 1316 & $     1.75 \pm     0.07 $ & $     1.79 \pm     0.07 $ & ~ &  944 & $     2.50 \pm     0.44 $ & $     2.55 \pm     0.46 $ & ~ & 1283 & $     2.85 \pm     0.61 $ & $     2.90 \pm     0.61 $ \\ 
 1978 & $     1.84 \pm     0.11 $ & $     1.77 \pm     0.11 $ & ~ & 1938 & $     2.65 \pm     0.45 $ & $     2.83 \pm     0.45 $ & ~ & 1261 & $     1.63 \pm     0.55 $ & $     2.01 \pm     0.52 $ \\ 
  881 & $     1.80 \pm     0.16 $ & $     1.79 \pm     0.16 $ & ~ & 1279 & $     1.87 \pm     0.28 $ & $     1.85 \pm     0.28 $ & ~ &  698 & $     2.42 \pm     0.35 $ & $     2.38 \pm     0.33 $ \\ 
  798 & $     2.15 \pm     0.15 $ & $     1.95 \pm     0.15 $ & ~ & 1720 & $     2.69 \pm     0.47 $ & $     2.97 \pm     0.50 $ & ~ & 1422 & $     2.95 \pm     0.94 $ & $     4.07 \pm     1.17 $ \\ 
  763 & $     1.85 \pm     0.14 $ & $     1.87 \pm     0.14 $ & ~ &  355 & $     3.50 \pm     0.97 $ & $     2.75 \pm     0.79 $ & ~ & 2048 & $     1.44 \pm     0.87 $ & $     1.26 \pm     0.82 $ \\ 
  731 & $     1.71 \pm     0.10 $ & $     1.77 \pm     0.10 $ & ~ & 1619 & $     2.46 \pm     0.72 $ & $     2.25 \pm     0.71 $ & ~ &    9 & $     2.42 \pm     0.82 $ & $     2.20 \pm     0.77 $ \\ 
 1535 & $     2.03 \pm     0.20 $ & $     1.94 \pm     0.20 $ & ~ & 1883 & $     3.18 \pm     0.60 $ & $     2.85 \pm     0.56 $ & ~ & 1910 & $     2.35 \pm     0.52 $ & $     2.07 \pm     0.51 $ \\ 
 1903 & $     1.87 \pm     0.18 $ & $     2.03 \pm     0.18 $ & ~ & 1242 & $     3.25 \pm     0.45 $ & $     3.06 \pm     0.45 $ & ~ &  856 & $     1.84 \pm     0.63 $ & $     1.70 \pm     0.64 $ \\ 
 1632 & $     1.89 \pm     0.16 $ & $     1.83 \pm     0.16 $ & ~ &  784 & $     3.77 \pm     1.14 $ & $     3.23 \pm     1.01 $ & ~ &  140 & $     3.55 \pm     1.31 $ & $     2.51 \pm     1.21 $ \\ 
 1231 & $     2.22 \pm     0.23 $ & $     2.13 \pm     0.23 $ & ~ & 1537 & $     2.13 \pm     0.62 $ & $     2.48 \pm     0.67 $ & ~ & 1087 & $     2.73 \pm     0.54 $ & $     2.50 \pm     0.52 $ \\ 
 2095 & $     1.79 \pm     0.34 $ & $     1.85 \pm     0.33 $ & ~ &  778 & $     2.07 \pm     0.48 $ & $     2.00 \pm     0.48 $ & ~ & 1861 & $     2.52 \pm     0.71 $ & $     1.92 \pm     0.62 $ \\ 
 1154 & $     1.81 \pm     0.28 $ & $     2.02 \pm     0.28 $ & ~ & 1321 & $     3.67 \pm     1.29 $ & $     4.99 \pm     1.97 $ & ~ & 1431 & $     2.55 \pm     0.57 $ & $     2.69 \pm     0.59 $ \\ 
 1062 & $     2.13 \pm     0.26 $ & $     1.99 \pm     0.26 $ & ~ &  828 & $     2.28 \pm     0.45 $ & $     2.48 \pm     0.42 $ & ~ & 1528 & $     2.19 \pm     0.70 $ & $     2.56 \pm     0.67 $ \\ 
 2092 & $     2.30 \pm     0.41 $ & $     2.35 \pm     0.41 $ & ~ & 1250 & $     1.92 \pm     0.52 $ & $     2.09 \pm     0.49 $ & ~ &  437 & $     3.55 \pm     0.91 $ & $     4.04 \pm     1.10 $ \\ 
  369 & $     2.14 \pm     0.25 $ & $     2.13 \pm     0.25 $ & ~ & 1630 & $     1.91 \pm     0.52 $ & $     1.99 \pm     0.50 $ & ~ & 2019 & $     2.22 \pm     0.70 $ & $     3.17 \pm     0.81 $ \\ 
  759 & $     2.32 \pm     0.27 $ & $     2.19 \pm     0.27 $ & ~ & 1146 & $     2.33 \pm     0.52 $ & $     2.47 \pm     0.50 $ & ~ &   21 & $     2.32 \pm     0.88 $ & $     2.76 \pm     0.93 $ \\ 
 1692 & $     1.93 \pm     0.29 $ & $     2.40 \pm     0.29 $ & ~ & 1025 & $     3.09 \pm     0.75 $ & $     2.81 \pm     0.66 $ & ~ & 1499 & $     2.75 \pm     0.90 $ & $     2.69 \pm     0.88 $ \\ 
 1030 & $     1.93 \pm     0.25 $ & $     2.13 \pm     0.25 $ & ~ & 1303 & $     2.55 \pm     0.59 $ & $     2.48 \pm     0.54 $ & ~ & 1545 & $     2.61 \pm     0.74 $ & $     2.57 \pm     0.73 $ \\ 
 2000 & $     2.07 \pm     0.25 $ & $     2.16 \pm     0.24 $ & ~ & 1913 & $     3.03 \pm     0.58 $ & $     2.57 \pm     0.58 $ & ~ & 1075 & $     3.94 \pm     1.31 $ & $     3.85 \pm     1.28 $ \\ 
  685 & $     1.71 \pm     0.25 $ & $     1.71 \pm     0.24 $ & ~ & 1125 & $     2.78 \pm     0.57 $ & $     2.37 \pm     0.58 $ & ~ & 1539 & $     3.14 \pm     0.92 $ & $     2.69 \pm     0.85 $ \\ 
 1664 & $     1.85 \pm     0.29 $ & $     1.66 \pm     0.28 $ & ~ & 1475 & $     2.37 \pm     0.54 $ & $     2.55 \pm     0.55 $ & ~ & 1185 & $     5.63 \pm     1.62 $ & $     5.56 \pm     1.59 $ \\ 
 \enddata
\tablecomments{
Key to columns---(1) Galaxy VCC number; 
(2) Best-fit power-law exponent $\beta$ for the mass function of GCs
between $\simeq\! 0.5$--2.5 mag brighter than the turnover magnitude in
the $z$-band GCLF;
(3) As (2) but for the $g$-band.
}
\end{deluxetable}

\clearpage

\begin{deluxetable}{rccccrccccrccc}
\tablecaption{Average GC Colors and Mass-to-Light Ratios
\label{tab:Ups}}
\tabletypesize{\small}
\tablewidth{0pt}
\tablehead{
\colhead{VCC} & \colhead{$\langle(g-z)\rangle$} & \colhead{$\Upsilon_z$}
              & \colhead{$\Upsilon_g$} & \colhead{~} &
\colhead{VCC} & \colhead{$\langle(g-z)\rangle$} & \colhead{$\Upsilon_z$}
              & \colhead{$\Upsilon_g$} & \colhead{~} &
\colhead{VCC} & \colhead{$\langle(g-z)\rangle$} & \colhead{$\Upsilon_z$}
              & \colhead{$\Upsilon_g$} \\
\colhead{(1)} & \colhead{(2)} & \colhead{(3)} & \colhead{(4)} & \colhead{~} &
\colhead{(1)} & \colhead{(2)} & \colhead{(3)} & \colhead{(4)} & \colhead{~} &
\colhead{(1)} & \colhead{(2)} & \colhead{(3)} & \colhead{(4)}
}
\startdata
1226 &     1.24 &     1.47 &     2.69 & ~ &1242 &     1.11 &     1.49 &     2.44 & ~ &1297 &     1.05 &     1.50 &     2.33 \\
1316 &     1.23 &     1.47 &     2.67 & ~ & 784 &     1.14 &     1.48 &     2.50 & ~ &1861 &     1.00 &     1.50 &     2.24 \\
1978 &     1.25 &     1.47 &     2.72 & ~ &1537 &     1.00 &     1.50 &     2.24 & ~ & 543 &     0.94 &     1.51 &     2.12 \\
 881 &     1.09 &     1.49 &     2.41 & ~ & 778 &     1.04 &     1.50 &     2.31 & ~ &1431 &     1.00 &     1.50 &     2.24 \\
 798 &     1.14 &     1.48 &     2.50 & ~ &1321 &     1.04 &     1.50 &     2.31 & ~ &1528 &     0.95 &     1.51 &     2.14 \\
 763 &     1.11 &     1.49 &     2.44 & ~ & 828 &     1.00 &     1.50 &     2.24 & ~ &1695 &     1.01 &     1.50 &     2.26 \\
 731 &     1.19 &     1.47 &     2.59 & ~ &1250 &     0.98 &     1.51 &     2.20 & ~ &1833 &     1.01 &     1.50 &     2.26 \\
1535 &     1.18 &     1.48 &     2.57 & ~ &1630 &     1.10 &     1.49 &     2.42 & ~ & 437 &     0.90 &     1.52 &     2.05 \\
1903 &     1.18 &     1.48 &     2.57 & ~ &1146 &     1.20 &     1.47 &     2.61 & ~ &2019 &     0.90 &     1.52 &     2.05 \\
1632 &     1.21 &     1.47 &     2.63 & ~ &1025 &     0.97 &     1.51 &     2.18 & ~ & 200 &     0.82 &     1.54 &     1.91 \\
1231 &     1.12 &     1.48 &     2.46 & ~ &1303 &     0.94 &     1.51 &     2.12 & ~ & 571 &     0.92 &     1.52 &     2.09 \\
2095 &     1.07 &     1.49 &     2.37 & ~ &1913 &     1.02 &     1.50 &     2.27 & ~ &  21 &     0.88 &     1.52 &     2.01 \\
1154 &     1.12 &     1.48 &     2.46 & ~ &1327 &     1.06 &     1.49 &     2.35 & ~ &1488 &     0.87 &     1.52 &     1.99 \\
1062 &     1.14 &     1.48 &     2.50 & ~ &1125 &     0.93 &     1.51 &     2.11 & ~ &1499 &     0.93 &     1.51 &     2.11 \\
2092 &     1.13 &     1.48 &     2.48 & ~ &1475 &     0.94 &     1.51 &     2.12 & ~ &1545 &     0.93 &     1.51 &     2.11 \\
 369 &     1.15 &     1.48 &     2.52 & ~ &1178 &     1.06 &     1.49 &     2.35 & ~ &1192 &     1.10 &     1.49 &     2.42 \\
 759 &     1.10 &     1.49 &     2.42 & ~ &1283 &     1.03 &     1.50 &     2.29 & ~ &1075 &     0.93 &     1.51 &     2.11 \\
1692 &     1.08 &     1.49 &     2.39 & ~ &1261 &     1.05 &     1.50 &     2.33 & ~ &1440 &     0.98 &     1.51 &     2.20 \\
1030 &     1.14 &     1.48 &     2.50 & ~ & 698 &     1.00 &     1.50 &     2.24 & ~ & 230 &     0.92 &     1.52 &     2.09 \\
2000 &     1.05 &     1.50 &     2.33 & ~ &1422 &     1.09 &     1.49 &     2.41 & ~ &2050 &     0.89 &     1.52 &     2.03 \\
 685 &     1.07 &     1.49 &     2.37 & ~ &2048 &     1.01 &     1.50 &     2.26 & ~ & 751 &     0.85 &     1.53 &     1.96 \\
1664 &     1.18 &     1.48 &     2.57 & ~ &1871 &     0.96 &     1.51 &     2.16 & ~ &1828 &     0.88 &     1.52 &     2.01 \\
 654 &     0.99 &     1.50 &     2.22 & ~ &   9 &     1.01 &     1.50 &     2.26 & ~ &1407 &     1.02 &     1.50 &     2.27 \\
 944 &     1.06 &     1.49 &     2.35 & ~ & 575 &     1.00 &     1.50 &     2.24 & ~ &1886 &     0.80 &     1.55 &     1.90 \\
1938 &     0.99 &     1.50 &     2.22 & ~ &1910 &     1.06 &     1.49 &     2.35 & ~ &1199 &     1.13 &     1.48 &     2.48 \\
1279 &     1.04 &     1.50 &     2.31 & ~ &1049 &     0.97 &     1.51 &     2.18 & ~ &1539 &     0.97 &     1.51 &     2.18 \\
1720 &     1.08 &     1.49 &     2.39 & ~ & 856 &     1.02 &     1.50 &     2.27 & ~ &1185 &     0.92 &     1.52 &     2.09 \\
 355 &     1.09 &     1.49 &     2.41 & ~ & 140 &     1.00 &     1.50 &     2.24 & ~ &1489 &     0.98 &     1.51 &     2.20 \\
1619 &     1.06 &     1.49 &     2.35 & ~ &1355 &     0.92 &     1.52 &     2.09 & ~ &1661 &     0.95 &     1.51 &     2.14 \\
1883 &     1.06 &     1.49 &     2.35 & ~ &1087 &     0.94 &     1.51 &     2.12 & ~ &\nodata & \nodata & \nodata & \nodata \\
\enddata
\tablecomments{
Key to columns---(1) Galaxy VCC number; 
(2) Mean GC $(g-z)$ color (from data in Peng et al.~2006a);
(3)--(4) Average GC mass-to-light ratio in the $z$ and $g$ bands,
obtained as described in \S~\ref{sssec:masstrend}.
}
\end{deluxetable}

\begin{deluxetable}{ccccccccccc}
\tablecaption{Gaussian GCLF Parameters for Outer GCs of M87/M49 
Companions\tablenotemark{1}
\label{tab:gclfpars_ore}}
%\rotate%A
\tabletypesize{\small}
\tablewidth{0pt}
\tablehead{
\colhead{VCC} &
\colhead{$\mu_g$} & 
\colhead{$\sigma_g$} &
\colhead{$\mu_z$} & 
\colhead{$\sigma_z$} &
\colhead{$\widehat{\cal B}$} &
\colhead{$N$} &
\colhead{Host} & 
\colhead{$\langle(g-z)\rangle$} &
\colhead{$\Upsilon_g$} &
\colhead{$\Upsilon_z$}
\\
\colhead{(1)}  & \colhead{(2)}  & \colhead{(3)}  & \colhead{(4)}  &
\colhead{(5)}  & \colhead{(6)}  & \colhead{(7)}  & \colhead{(8)}
& \colhead{(9)}  & \colhead{(10)} & \colhead{(11)} }
\startdata
1327 & $   23.886 \pm    0.182  $ & $    1.288 \pm    0.144 $ &$   22.777 \pm    0.164 $ & $    1.224 \pm    0.129 $ &    0.070 &   93 & VCC1316  &    1.104 &    2.43 &    1.49 \\
1199 & $   23.805 \pm    0.127  $ & $    1.224 \pm    0.102 $ &$   22.631 \pm    0.117 $ & $    1.175 \pm    0.091 $ &    0.053 &  151 & VCC1226  &    1.100 &    2.42 &    1.49 \\
1192 & $   23.643 \pm    0.102  $ & $    1.013 \pm    0.080 $ &$   22.540 \pm    0.099 $ & $    1.011 \pm    0.076 $ &    0.064 &  144 & VCC1226  &    1.155 &    2.54 &    1.48 \\
1297 & $   23.410 \pm    0.183  $ & $    1.208 \pm    0.142 $ &$   22.320 \pm    0.166 $ & $    1.138 \pm    0.130 $ &    0.103 &   67 & VCC1316  &    1.090 &    2.40 &    1.49 \\
\enddata
\tablenotetext{1}{All reported numbers refer {\it only} to those GC candidates
that are more than $6\,\langle R_e \rangle$ away from the centers of the
galaxies indicated, where $\langle R_e \rangle$ is the median effective radius
of other VCS galaxies that have magnitudes within 0.5 mag of the target
galaxy.} 
\tablecomments{
Key to columns---(1) Galaxy VCC number;
(2)--(3) Maximum-likelihood estimates of the
Gaussian mean $\mu$ and dispersion $\sigma$ of the $g$-band GCLF;
(4)--(5) Same as (2)--(3) but  for the $z$-band;
(6) Fraction $\widehat{\cal B}$ of the sample that is expected to be 
contamination;
(7) Total number $N$ of all objects with $p_{\rm GC}\ge 0.5$
(including contaminants and uncorrected for incompleteness);
(8) VCC number of giant elliptical galaxy close in projection;
(9) Mean GC $(g-z)$ color;
(10)--(11) Average GC mass-to-light ratio in the $g$ and $z$ bands, obtained
as described in \S~\ref{sssec:masstrend}.
}
\end{deluxetable}

%%%%%%%%%%%%%%%%%%%%%%%%%%%%%%%%%%%%%%%%%%%%%%%%%%%%%%%%%%%%%%%%%%%%%%%%%%%%%%%
%%%%%%%%%%%%%%%%%%%%%%%%%%%%%%%%% FIGURES %%%%%%%%%%%%%%%%%%%%%%%%%%%%%%%%%%%%%
%%%%%%%%%%%%%%%%%%%%%%%%%%%%%%%%%%%%%%%%%%%%%%%%%%%%%%%%%%%%%%%%%%%%%%%%%%%%%%%

\end{document}